%% file: photodet.tex
\documentclass[10pt,twocolumn]{article}
\usepackage[paperwidth=210mm,paperheight=297mm,centering,hmargin=1.5cm,vmargin=1.5cm]{geometry}

\usepackage{caption,subfig}

\usepackage{graphicx}
\usepackage[utf8]{inputenc}
\usepackage{amsmath}
\usepackage{amssymb}
\usepackage{appendix}
\usepackage[dvipsnames]{xcolor}

\usepackage{hyperref}
\usepackage{comment}
\usepackage[normalem]{ulem}

\usepackage{pdfpages}

\def\@maketitle{%
  \newpage\spacing{1}\setlength{\parskip}{12pt}%
    {\Large\bfseries\noindent\sloppy \textsf{\@title} \par}%
    {\noindent\sloppy \@author}%
}

\makeatletter
\def\blfootnote{\xdef\@thefnmark{}\@footnotetext}
\makeatother

\usepackage{cleveref}

\crefname{equation}{Eq.}{Eqs.}
\Crefname{equation}{Equation}{Equations}
\crefname{figure}{Fig.}{Figs.}
\Crefname{figure}{Figure}{Figures}
\crefname{section}{Sec.}{Secs.}
\Crefname{section}{Section}{Sections}

\newcommand{\ket}[1]{|#1\rangle}
\newcommand{\braket}[1]{\langle #1 \rangle}

\newcommand{\tr}{\text{tr}}
\newcommand{\e}{e}

\newcommand{\dd}{d}

\newcommand{\half}{\frac{1}{2}}

\newcommand{\ha}{\hat{a}}

\newcommand{\hb}{\hat{b}}

\newcommand{\hH}{\hat{H}}

\newcommand{\bomega}{\bar{\omega}}

\newcommand{\Hc}{\text{H.c.}}

\newcommand{\sm}{\text{Supplementary Materials}}
\newcommand{\mm}{\text{Methods}}

\newcommand{\tml}{\text{tml}}

\newcommand{\dg}{^\dagger}

\newcommand{\SNAIL}{Q}

\newcommand{\soutpar}[1]{\expandafter\sout\expandafter{#1}}

\begin{document}


\title{Quantum metamaterial for nondestructive microwave photon counting}

\author{\normalsize Arne L. Grimsmo$^{*1,2}$, Baptiste Royer$^{1,3}$, John Mark Kreikebaum$^{4,5}$, Yufeng Ye$^{6,7}$, Kevin O'Brien$^{6,7}$, Irfan Siddiqi$^{4,5}$, 
\\ \normalsize Alexandre Blais$^{1,8}$\\
{\footnotesize$^{1}$ Institut quantique and D\'{e}partment de Physique, Universit\'{e} de Sherbrooke, Sherbrooke, Qu\'{e}bec, Canada J1K 2R1}\\
{\footnotesize$^{2}$ Centre for Engineered Quantum Systems, School of Physics, The University of Sydney, Sydney, NSW 2006, Australia}\\
{\footnotesize$^{3}$ Department of Physics, Yale University, New Haven, CT 06520, USA}\\
{\footnotesize$^{4}$ Materials Science Division, Lawrence Berkeley National Laboratory, Berkeley, California 94720, USA}\\
{\footnotesize$^{5}$ Quantum Nanoelectronics Laboratory, Department of Physics, University of California, Berkeley, California 94720, USA}\\
{\footnotesize$^{6}$
Research Laboratory of Electronics, Massachusetts Institute of Technology, Cambridge, MA 02139, USA}\\
{\footnotesize$^{7}$
Department of Electrical Engineering and Computer Science,Massachusetts Institute of Technology, Cambridge, MA 02139, USA 
}\\
{\footnotesize$^{9}$ Computational Research Division, Lawrence Berkeley National Laboratory, Berkeley, California 94720, USA}\\
{\footnotesize$^{8}$ Canadian Institute for Advanced Research, Toronto, Canada}\\
}

\date{}

\maketitle

{\bf 
Detecting traveling photons is an essential primitive for many quantum information processing tasks.  We introduce a single-photon detector design operating in the microwave domain, based on a weakly nonlinear metamaterial where the nonlinearity is provided by a large number of Josephson junctions. The combination of weak nonlinearity and large spatial extent circumvents well-known obstacles limiting approaches based on a localized Kerr medium. Using numerical many-body simulations we show that the single-photon detection fidelity increases with the length of the metamaterial to approach one at experimentally realistic lengths. A remarkable feature of the detector is that the metamaterial approach
allows for a large detection bandwidth. In stark contrast to conventional photon detectors operating in the optical domain, the photon is not destroyed by the detection and the photon wavepacket is minimally disturbed. The detector design we introduce offers new possibilities for quantum information processing, quantum optics and metrology in the microwave frequency domain.
}

\blfootnote{$^*$Corresponding author. Email: arne.grimsmo@sydney.edu.au}

\section*{Introduction}

In contrast to infrared, optical and ultraviolet frequencies where single-photon detectors are a cornerstone of experimental quantum optics,
the realization of a detector with similar performance at microwave frequencies is far more challenging~\cite{Helmer09,Chen11,Sathyamoorthy14,Fan14,Koshino16,Inomata16,Narla16,Kyriienko16,leppakangas2018multiplying,Kono:18a,Besse:18a,Royer18,lescanne2019detecting}. The interest in realizing such a detector is intimately linked to the emergence of engineered quantum systems whose natural domain of operations is in the microwaves, including superconducting quantum circuits~\cite{blais2020quantum}, spin ensembles~\cite{Kubo10}, and semiconductor quantum dots~\cite{Burkard2020}.
The continuing improvement in coherence and control over these quantum systems offers a wide range of new applications for microwave single-photon detection, such as
photon-based quantum computing \cite{Spring13}, modular quantum computing architectures~\cite{Nickerson14}, high-precision sensing~\cite{Lloyd08}, and the detection of dark matter axions~\cite{Lamoreaux:13}.

A number of theoretical proposals and experimental demonstrations of microwave single-photon detectors have emerged recently. These  schemes can broadly be divided into two categories: 
Time-gated schemes where accurate information about the photon's arrival time is needed a priori~\cite{Chen11,Inomata16,Narla16,Kono:18a,Besse:18a,lescanne2019detecting}, and 
detectors that operate continuously in time and attempt to accurately record the photon arrival time~\cite{Helmer09,Sathyamoorthy14,Fan14,Koshino16,Kyriienko16,leppakangas2018multiplying,Royer18,lescanne2019detecting}.
In this work, we are concerned with the last category, which is simultaneously the most challenging to realize and finds the widest range of applications.

Depending on the intended application, there are several metrics characterizing the usefulness of single-photon detectors. Not only is high single-photon detection fidelity required for many quantum information applications, but large bandwidth, fast detection and short dead times are also desirable~\cite{Hadfield09}. Moreover, nondestructive photon counting is of fundamental interest and offers new possibilities for quantum measurement and control.
In this article, we introduce the Josephson Traveling-Wave Photodetector (JTWPD), a non-destructive single-photon detector which we predict to have remarkably high performance across the mentioned metrics.
In particular, this detector can have detection fidelities approaching unity
without sacrificing detector bandwidth.

The JTWPD exploits a weakly nonlinear, one-dimensional metamaterial, designed to respond to the presence of a single photon. The nonlinearity is provided by a large number of Josephson junctions, inspired by the Josephson traveling wave parametric amplifier~\cite{Macklin15}.
Because the detector response does not rely on any resonant interaction, the detector bandwidth can be designed to range from tens of MHz to the GHz range. The detection and reset times are predicted to be in the range of tens of $\mu$s for typical parameters.
Moreover, the signal-to-noise ratio (SNR) grows linearly with the length of the metamaterial which can be made large, leading to single-photon detection fidelities approaching unity. By interrogating the nonlinear medium with
a ``giant probe''~\cite{Guo17}---a probe system that couples to the medium over a spatial extent that is large compared to the length of the signal photons---this approach bypasses previous no-go results for photon counting based on localized cross-Kerr interactions~\cite{Shapiro06a,Shapiro07a,Gea-Banacloche10a,Fan13}.

\section*{Results}

\begin{figure}
  \centering
  \includegraphics[scale=0.85]{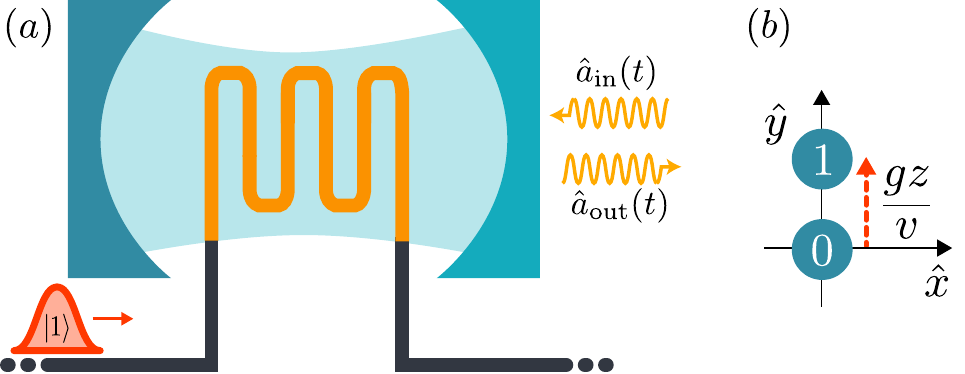}
  \caption{\label{fig:cartoon}  a) Sketch of the JTWPD. Standard transmission lines (black) are coupled to both ends of a one-dimensional metamaterial (orange) of length $z$ and linear dispersion relation, $\omega = v k$. A cross-Kerr interaction $\chi$ between the metamaterial and the giant probe mode (blue) leads to a phase shift in the strong measurement tone (yellow) while the signal photon (red) travels through the metamaterial. b) Phase space picture of the probe mode. With respect to the idle coherent state $\ket{\alpha}$, the presence of a signal photon displaces the states by $gz/v$, with $g = \chi \alpha$.}
\end{figure}

\begin{figure*}
  \centering
  \includegraphics[scale=1.0]{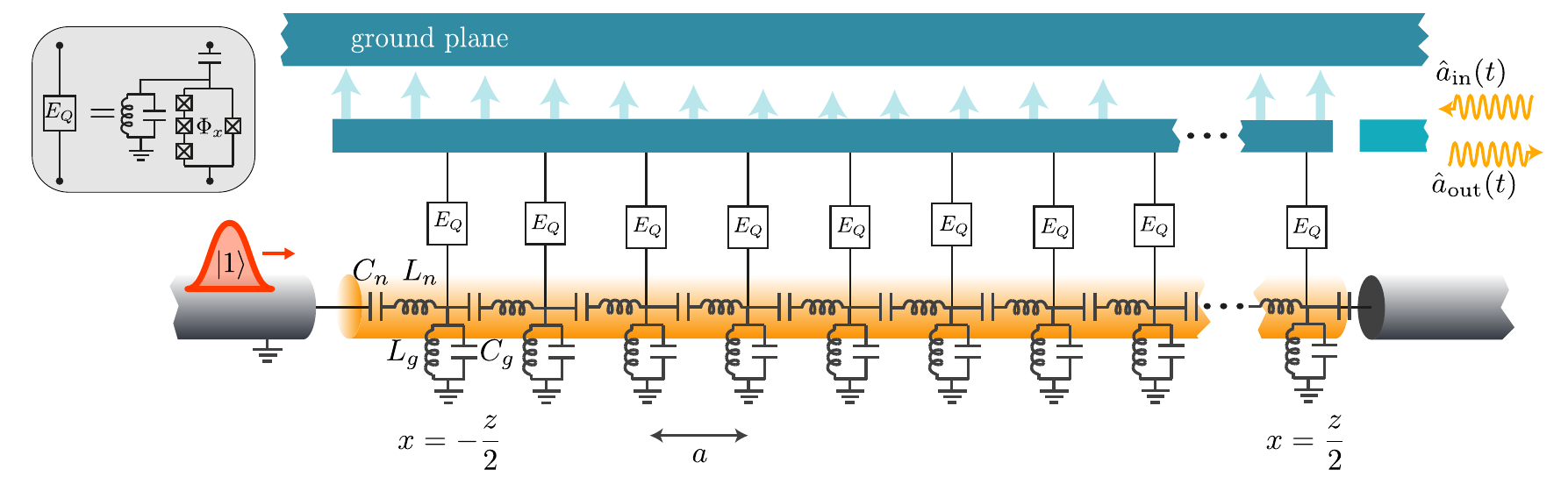}
  \caption{\label{fig:metamaterial}Schematic representation of the JTWPD. The probe resonator with ground plane on top and the center conductor below (blue), as well as a readout port on the right, acts as a giant probe. The light blue arrows illustrate the fundamental mode function of a $\lambda/2$ resonator. This probe is coupled via a position dependent cross-Kerr interaction $\chi(x)$, mediated by an array non-linear couplers (inset), to a metamaterial waveguide (orange). The metamaterial is coupled to impedance matched input/output transmission lines at $x=- z/2$ and $x=z/2$ (grey). An incoming photon of Gaussian shape $\xi(x,t)$ is illustrated (red).}
\end{figure*}

Many proposals for itinerant microwave photon detection rely on capturing the incoming photon in a localized absorber mode that is interrogated using heterodyne detection~\cite{Helmer09,Sathyamoorthy14,Fan14,Koshino16,Royer18,lescanne2019detecting}. A first challenge associated with this approach is linked to a version of the quantum Zeno-effect: continuously and strongly monitoring the absorber will prevent the incoming photon from being absorbed~\cite{Helmer09,Royer18}, limiting the detector's quantum efficiency. Another difficulty concerns the tradeoff between efficiency and bandwidth. A large detector response to a single photon requires a sufficiently long interaction time with the photon. In principle, this can be achieved by making the absorber mode long-lived. However, as the mode linewidth is inversely proportional to the photon lifetime, this imposes a serious constraint on the detector bandwidth.

Our solution to overcome these obstacles is illustrated schematically in \cref{fig:cartoon}: In place of a localized absorber, we use a long and weakly nonlinear metamaterial. Backscattering is avoided by using a nonlinearity that is locally weak, yet a large response is made possible by having a long photon time-of-flight through the metamaterial. The presence of a photon is recorded using a continuously monitored probe mode that is coupled to the metamaterial along the full extent of its length. Thanks to a nonlinear cross-Kerr coupling, in the presence of the measurement tone $\hat a_\mathrm{in}(t)$, a single photon in the metamaterial induces a displacement of the output field $\hat a_\mathrm{out}(t)$ relative to its idle state.  While the interaction between the metamaterial and the probe mode is locally too weak to cause any noticeable change in $\hat a_\mathrm{out}(t)$, the displacement accumulates as the photon travels through the metamaterial leading to a large enough signal to be recorded using homodyne detection.

\paragraph*{JTWPD design and working principle}

As illustrated in \cref{fig:metamaterial}, the backbone of the metamaterial is a waveguide of length $z$ (orange) realized as a linear chain of coupled LC oscillators, in a configuration known as composite right/left handed (CRLH) metamaterial~\cite{Caloz2004novel}. The LC oscillators are coupled via an array of nonlinear couplers (inset) to a readout resonator acting as a giant probe~(blue).
With the metamaterial coupled at $x=\pm z/2$ to impedance matched linear transmission lines, the interaction time between the photon and the giant probe is $\tau = z/v$ where $v$ is the speed of light in the metamaterial. As an alternative to this transmission mode, the interaction time can be doubled by terminating the metamaterial at $x=+ z/2$ with an open where the photon wavepacket is reflected. To simplify the analysis, we consider the transmission mode
in most of the treatment below, but return to a discussion of reflection mode when discussing potential experimental implementation and parameters.

The full detector Hamiltonian can be expressed as $\hH = \hH_{0} + \hH_r + \hH_{\rm int}$, where $\hH_0$ contains the linear part of the waveguide including the metamaterial as well as the input and output linear waveguides, $\hH_r$ is the probe resonator Hamiltonian and $\hH_{\rm int}$ describes the nonlinear coupling between the probe and the metamaterial. As shown in the \sm, in the continuum limit where the size $a$ of a unit cell of the metamaterial is small with respect to the extent of the photon wavepacket, $\hH_0$ takes the form
\begin{equation}\label{eq:H_0}
  \hH_0 = \sum_{\nu=\pm} \int_\Omega \dd \omega \hbar\omega \hb_{\nu\omega}^\dagger \hb_{\nu\omega}.
\end{equation}
In this expression, $\hb_{\pm\omega}^\dagger$ creates a delocalized right/left- moving photon with energy $\hbar\omega$ and satisfies the canonical commutation relation $[\hb_{\nu\omega},\hb^\dagger_{\mu\omega'}] = \delta_{\nu\mu}\delta(\omega-\omega')$. The subscript $\Omega$ in \cref{eq:H_0} is used to indicate that we only consider a band of frequencies around which the metamaterial's dispersion relation is approximately linear. The probe resonator Hamiltonian $\hat H_r$ can be written in a displaced and rotating frame with respect to the coherent drive field as
\begin{equation}\label{eq:H_r_displaced}
  \hH_r' = \frac{\hbar K}{2} \ha^{\dagger 2} \ha^2,
\end{equation}
where $K$ is a self-Kerr nonlinearity induced by the nonlinear couplers (see \mm).

The coupling elements also lead to cross-Kerr interaction between the array of oscillators and the probe mode. As mentioned above, this coupling is chosen to be locally weak such that the nonlinearity is only activated by the presence of a strong coherent drive $\hat a_\text{in}(t)$ on the probe. In this limit, the nonlinear interaction Hamiltonian $\hat H_{\rm int}$ is in the same rotating and displaced frame given by
\begin{equation}\label{eq:H_int1}
  \begin{aligned}
    \hat H_{\rm int}' ={}& \hbar \sum_{\nu\mu}\int_{-z/2}^{z/2}\dd x \chi(x) \hat b_{\nu}^\dagger(x) \hat b_{\mu}(x)\left( \hat a^\dagger \hat a + \alpha^2 \right)\\
    +\,& \hbar \sum_{\nu\mu}\int_{-z/2}^{z/2}\dd x g(x) \hat b_{\nu}^\dagger(x) \hat b_{\mu}(x)\left( \hat a^\dagger + \hat a\right),
  \end{aligned}
\end{equation}
where we have defined the $x$-dependent photon annihilation operators
\begin{equation}\label{eq:bx}
  \hb_{\nu}(x) = \sqrt{\frac{\bomega}{2\pi v}} \int_{\Omega} \frac{\dd\omega}{\sqrt{\omega}} \hb_{\nu\omega}\e^{\nu i\omega x/v},
\end{equation}
with $\bar\omega$ a nominal center frequency for the incoming photon and which is introduced here for later convenience. The parameter $\chi(x)$ is a dispersive shift per unit length given in~\cref{eq:chi},
while $g(x) = \alpha \chi(x)$ with $\alpha$ the displacement of the probe resonator field under the strong drive $\hat a_\mathrm{in}$. The expression for $\alpha$, which we take to be real without loss of generality, can be found in \cref{eq:alpha} of the \mm.

As can be seen from the second term of \cref{eq:H_int1} which dominates for small $\chi(x)$ and large $\alpha$, the combined effect of the cross-Kerr coupling and the strong drive results in a longitudinal-like interaction between the metamaterial and the probe mode~\cite{Didier15}. This corresponds to a photon-number dependent displacement of the probe field relative to the idle state displacement $\alpha$, which accumulates when a photon travels along the metamaterial. By continuously monitoring the output field of the probe mode, a photon is registered when the integrated homodyne signal exceeds a predetermined threshold. This approach shares similarities with the photodetector design introduced in Ref.~\cite{Royer18}, with the important distinction that here the photon is probed \emph{in-flight} as it travels through the metamaterial rather than after interaction with a localized absorber mode. 
This distinction is the key to achieving large detection fidelities without sacrificing bandwidth.

An important feature of this detector design is that although the detection bandwidth is large, the CRLH metamaterial can be engineered such as to have frequency cutoffs~\cite{Caloz2004novel}. The low-frequency cutoff avoids the detector from being overwhelmed by low-frequency thermal photons. Decay of the probe mode via the metamaterial to the input and output waveguides 
is minimized by choosing the probe mode resonance frequency to be outside of the metamaterial's bandwidth. In this situation, the metameterial effectively acts as a Purcell filter for the probe mode, thereby avoiding degradation of the probe mode quality factor.
Hybridization of the probe resonator and the metamaterial is further minimized by using a nonlinear coupler, illustrated in the inset of~\cref{fig:metamaterial}. As discussed in \mm, the coupler is operated at a point where the quadratic coupling vanishes leaving a quartic potential of strength $E_Q$ as the dominant contribution.

\paragraph*{Backaction and detector noise}

In the JTWPD, backaction on the incoming photon's wavevector, and therefore photon backscattering, is minimized by working with a giant probe which, optimally, does not acquire information about the photon's position. Focusing first on the ideal case where the probe mode self-Kerr nonlinearity $K$ and the dispersive shift $\chi(x)$ can be neglected compared to $g(x)=\alpha\chi(x)$, we clarify the dominant noise process for the probe resonator and the associated backaction on the photon by deriving a perturbative master equation.
In the subsequent section, we turn to full numerical analysis including the effect of the nonlinearities $K$ and $\chi(x)$.

Considering the ideal case for the moment and ignoring the spatial dependence of $g(x)$, the interaction Hamiltonian takes the simple longitudinal-coupling form
\begin{equation}\label{eq:H_ideal}
    \hat H_{\rm ideal} =
    \hbar g \sum_{\nu\mu}\int_{-z/2}^{z/2}\dd x \hat b_{\nu}^\dagger(x) \hat b_{\mu}(x)\left( \hat a^\dagger + \hat a\right).
\end{equation}
We model the incoming photon by an emitter system with annihilation operator $\hat c$, $[\hat c,\hat c\dg]=1$, located at $x_0 < -z/2$ and initialized in Fock state $\ket 1$. The decay rate $\kappa_c(t)$ of the emitter to the transmission line is chosen such as to have a Gaussian wavepacket with center frequency $\bar\omega$ and full width at half maximum (FWHM) $\gamma$ propagating towards the detector [see \cref{eq:wavepacket} of \mm]. Using Keldysh path integrals, we trace out the waveguide to find a perturbative master equation for the joint emitter-probe system. As discussed in the \mm, to second order in the interaction, this master equation takes a remarkably simple form
\begin{equation}\label{eq:effMEkeldysh}
\begin{aligned}
\dot {\hat \rho} ={}& -i\left[ gn_\text{det}(t)(\ha + \ha^\dag), \hat \rho_c \right]
 + \Gamma(t)\mathcal D[\ha + \ha^\dag] \hat \rho_c\\
 &+ \kappa_c(t)\mathcal D[\hat c]\hat \rho + \kappa_a \mathcal D[\ha]\hat \rho.
\end{aligned}
\end{equation}
In this expression, $\mathcal D[\hat o]\bullet = \hat o \bullet \hat o^\dag - 1/2\{\hat o^\dag \hat o,\bullet\}$ is the usual Lindblad-form dissipator
and we have defined $\hat \rho_c(t) = \hat c \hat \rho(t) \hat c^\dag/\braket{\hat c\dg \hat c}(t)$,
\begin{align}
    n_\text{det}(t) ={}& \frac{1}{v} \int_{-z/2}^{z/2}\dd x\left|\xi\left(x,t\right)\right|^2,\label{eq:etadet}\\
    \Gamma(t) ={}& \frac{4g^2}{\kappa_a v}
    \int_{-z/2}^{z/2}\dd x
    \left[1-\e^{-\frac{\kappa_a}{2v}\left(x + \frac{z}{2}\right)}\right]
    \left|\xi\left(x,t\right)\right|^2,\label{eq:Gamma}
\end{align}
with $\xi(x,t) = \xi(t-x/v)$ the incoming photon envelop and $n_\text{det}(t)$ the fraction of the photon that is in the metamaterial at time $t$. 
A term of order $g/\bar\omega$ describing back-scattering of the photon into the left-moving field has been dropped from~\cref{eq:effMEkeldysh}. With $\bar\omega$ the carrier frequency of the incoming photon, this contribution is negligible.

In~\cref{eq:effMEkeldysh}, $\hat \rho_c$ is the state of the system \emph{conditioned} on a photon having been emitted. The first term of~\cref{eq:effMEkeldysh} consequently has an intuitive interpretation that is consistent with the form of $ \hat H_{\rm ideal}$: The probe resonator is conditionally displaced by a drive equal to the longitudinal coupling amplitude times the photon fraction in the metamaterial, $g \times n_\text{det}(t)$. Indeed, while the $x$-quadrature of the probe, $\hat x = (\hat a\dg + \hat a)/\sqrt 2$, is a constant of motion under~\cref{eq:effMEkeldysh}, the $y$-quadrature, $\hat y = i(\hat a\dg - \hat a)/\sqrt 2$, is displaced.

\begin{figure}
  \centering
  \includegraphics[scale=1.0]{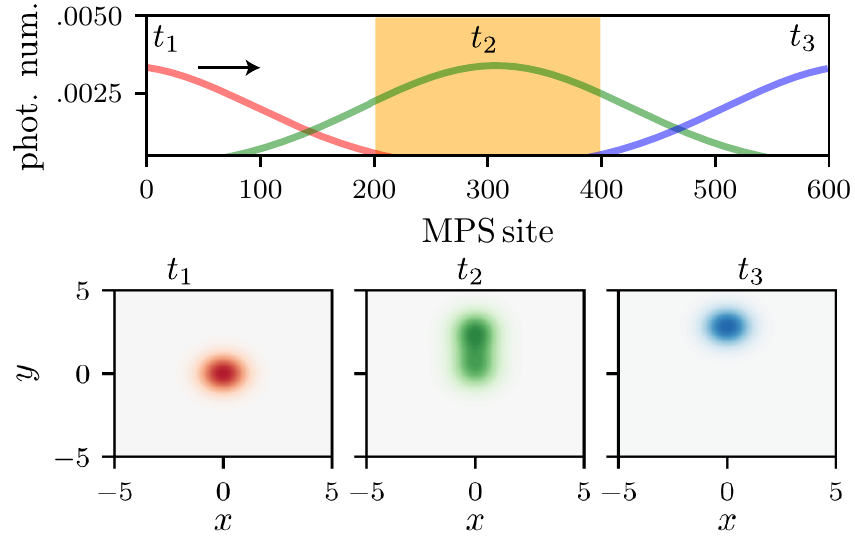}
  \caption{\label{fig:mps}The top panel shows snapshots of the photon number population along the MPS sites at three different times $t_1$ (red) $<$ $t_2$ (green) $<$ $t_3$ (blue). The white region corresponds to the linear waveguide and the orange region to the metamaterial with its coupling to the probe resonator. The lower three panels show the Wigner function $W(x,y)$ of the intracavity probe field at the three respective times. When the photon is only partially inside the metamaterial, the probe is in a superposition of displaced states (middle lower panel).
 Parameters are $\kappa_a=\chi(x)=K=0$, $g\tau = 2$ and $\gamma\tau=2$.
  }
\end{figure}

The second term of~\cref{eq:effMEkeldysh}, proportional to the rate $\Gamma(t)$, is the dominant process contributing to noise also along the $y$-quadrature.
The origin of the noise term can be understood as follows. When the photon first enters the detector and is only partially inside the metamaterial, the probe mode field evolves to a superposition of being displaced to different average values of $\hat y$, leading to enhanced fluctuations in this quadrature. This effect can be seen clearly in the numerical results of~\cref{fig:mps}, which are described in more detail below.
Finally, the last line of~\cref{eq:effMEkeldysh} describes the usual decay of the emitter and probe at respective rates $\kappa_c(t)$ and $\kappa_a$.

As the increased fluctuations in the $y$-quadrature arise due to uncertainty in the photon's position, a spatially longer photon is expected to lead to larger fluctuations. A measurement of the probe's $y$-quadrature will collapse the superposition of displaced states and thus lead to a backaction effect localizing the photon and randomizing its wavevector. This effect can be minimized by decreasing the interaction strength $g$ while keeping $gz/v$ constant by increasing $z$. In other words, backaction can be minimized by increasing the detector length relative to the spatial extent of the photon. This intuitive reasoning is confirmed by numerical results in the next section. 

\paragraph*{Numerical Matrix Product State simulations}

We now turn to numerical simulations of the JTWPD including the self- and cross-Kerr nonlinearities $K$ and $\chi$ that were dropped from the above discussion.
To go beyond the perturbative results of~\cref{eq:effMEkeldysh}, it is no longer possible to integrate out the waveguide degrees of freedom.
A brute-force numerical integration of the dynamics is, however, intractable, as the JTWPD is an open quantum many-body system with thousands of modes. We overcome this obstacle by using a numerical approach where the systems is represented as a stochastically evolving Matrix Product State (MPS) conditioned on the homodyne measurement record of the probe output field.

Our approach is based on trotterizing the time evolution and discretizing the photon waveguide, including the nonlinear metamaterial, along the $x$ axis. Building upon and extending recent developments of MPS in the context of waveguide QED~\cite{Grimsmo15,Pichler16},  this leads to a picture where the waveguide is represented by a ``conveyor belt'' of harmonic oscillators (referred to as MPS sites below) interacting with the probe resonator (see \mm). 
Measurement backaction under continuous homodyne detection of the probe resonator is included by representing the state as a quantum trajectory conditioned on the measurement record~\cite{Wiseman09}. With our approach this is simulated using a stochastic MPS algorithm.
Further details on this numerical technique can be found in \mm~and the \sm.

As in the previous section, we consider a Gaussian photon wavepacket with FWHM $\gamma$ propagating towards the detector by an emitter initialized in the state $\ket 1$ localized to the left of the detector. The interaction strength is quantified by the dimensionless quantity $g\tau$ where $\tau = z/v$ is the interaction time as before, and the photon width by the dimensionless quantity $\gamma\tau$.
Example snapshots of the photon number distribution along the MPS sites at three different times $t_1 < t_2 < t_3$ are shown in~\cref{fig:mps}, along with the corresponding Wigner functions of the probe mode field.
Because of the impedance match and negligible backaction, the photon wavepacket travels without deformation along the waveguide.

\begin{figure}
  \centering
  \includegraphics[scale=1.0]{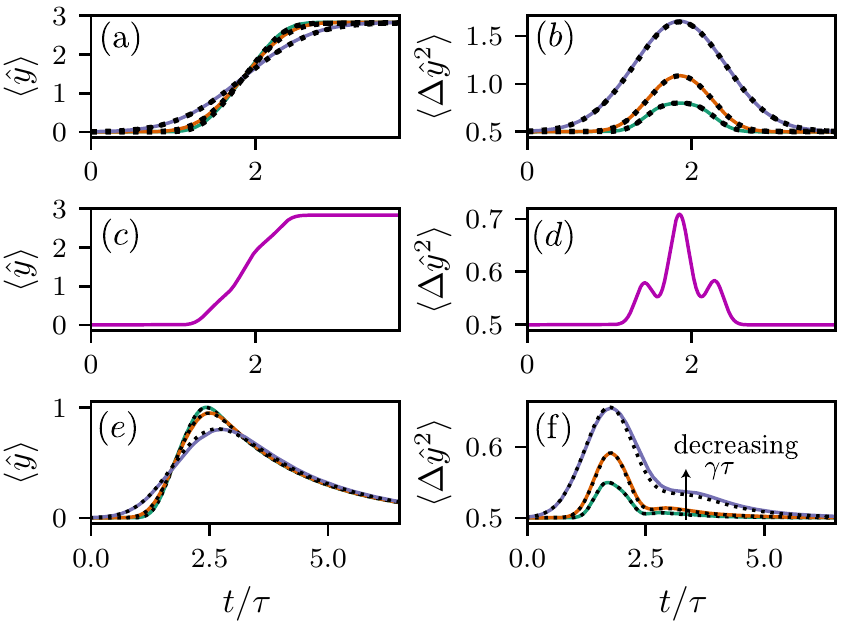}
  \caption{\label{fig:displacement}
  Time evolution of the intra-cavity probe displacement $\braket{\hat y}$ [$(a,c,e)$] and fluctuations $\braket{\Delta \hat y^2}$ [$(b,d,f)$], in the idealized case $\chi(x)=K=0$. Top row: $\kappa_a=0$ and $g\tau = 2$. Middle row: $\kappa_a=0$ and spatially varying $g(x)$ with average value $\bar g\tau = 2$. Bottom row: $\kappa_a\tau = 1.0$ and $g\tau = 2$. The solid lines correspond to MPS simulations with different photon widths $\gamma\tau=2$ (blue), 4 (orange), 6 (green) and 10 (bright purple), while the dotted lines are from integrating~\cref{eq:effMEkeldysh}.
  }
\end{figure}

We start by comparing numerical results from MPS simulations to the perturbative master equation obtained in~\cref{eq:effMEkeldysh}.  To help in directly comparing the simulation results, we first consider the idealized situation where $\chi(x) = K = 0$.
In~\cref{fig:displacement}, we show the average probe resonator displacement $\braket{\hat y}$ whose integrated value is linked to the detector signal and the noise $\braket{\Delta\hat y^2}$ as a function of time. To verify the prediction
that fluctuations in $\hat y$ increase for spatially longer photons, we compare Gaussian wavepackets of different spectral widths $\gamma$. 
Recall that a \emph{smaller} $\gamma\tau$ implies a \emph{longer} photon relative to the detector length. The solid lines in~\cref{fig:displacement} are obtained using MPS simulations with $\gamma\tau=2$ (blue), 4 (orange), 6 (green) and 10 (bright purple). The dotted lines are obtained from~\cref{eq:effMEkeldysh} for the same parameters. The agreement between the approximate analytical results and the full non-perturbative MPS results is remarkable. 

In panels $(c, d)$ of~\cref{fig:displacement} we use a spatially varying $g(x)$, and we consequently only show MPS results in these panels. In practice, the probe will be realized from a resonator whose vacuum fluctuations vary in space. To confirm the robustness of the detector to this variation, \cref{fig:displacement}~$(b,c)$ shows $\braket{\hat y}$ and $\braket{\Delta\hat y^2}$ versus time as obtained from MPS simulations for $g(x) = 2 \bar g \cos^2(2\pi x/z) + \mu(x)$. The cosine models the dependence on the mode function of a $\lambda/2$ resonator while $\mu(x)$ is added to take into account potential random variations in the coupling strength which we take here to be as large as 10\%. Moreover, to show the effect of a non-uniform $g(x)$ more clearly, we use $\gamma\tau=10$ corresponding to spatially shorter photons than in the other panels. Although additional structures can now be seen, the long-time average displacement remains unchanged confirming that the detector is robust against spatial variations of the metamaterial-probe coupling. 

Panels $(e, f)$ of~\cref{fig:displacement}
show results fo $\kappa_a > 0$. In this situation the MPS evolves stochastically with each trajectory resulting in a measured current $J_\text{hom}(t) = \sqrt{\kappa_a}\braket{\hat y}_\text{traj} + \xi(t)$, where $\xi(t) = \dd W_t/\dd t$ with $\dd W_t$ a Wiener process representing white noise~\cite{Wiseman09}. 
We compare $\braket{\hat y}$ and $\braket{\Delta \hat y^2}$ averaged over one thousand stochastic trajectories to the results obtained by integrating the Keldysh master equation~\cref{eq:effMEkeldysh}. The agreement is excellent for large $\gamma\tau$, but small deviations are observed when this parameter is decreased. We attribute this to terms of higher than second order in the interaction Hamiltonian, which are neglected in~\cref{eq:effMEkeldysh}. The exponential decay of $\braket{\hat y}$ at long time  observed in panel $(e)$ simply results from the finite damping rate $\kappa_a$. Indeed, the photon-induced displacement stops once the photon has travelled past the metamaterial at which point the probe mode relaxes back to its idle state.

For a given trajectory, we infer that a photon is detected if the homodyne current convolved with a filter~\cite{Fan14}
\begin{equation}\label{eq:Jhomconv}
  \bar J_\text{hom}(t) = \int_{0}^{\tau_m} \dd t' J_\text{hom}(t') f(t'-t).
\end{equation}
is larger than a threshold $y_\text{thr}$, i.e.~$\max_t \bar J_\text{hom}(t) > y_\text{thr}$. The filter $f(t)\propto\braket{\hat y(t)}$ is obtained from averaging over a large number of trajectories and is chosen such as to give more weight to times where the signal is on average larger. We maximize $t$ over the time window $[-\tau_m,\tau_m]$ and chose the threshold to optimize between quantum efficiency and dark counts. The quantum efficiency $\eta$ is defined as the probability of detecting a photon given that one was present. From the above procedure, it can be estimated as $\eta = N_{\mathrm{click}|1}/N_{\mathrm{traj}|1}$, with $N_{\mathrm{click}|1}$ the number of reported ``clicks'' and $N_{\mathrm{traj}|1}$ the number of simulated trajectories with a photon. On the other hand, the dark count probability is estimated similarly as the fraction of reported clicks $p_D = N_{\mathrm{click}|0}/N_{\mathrm{traj}|0}$ in a simulation with no incoming photon. In these simulations, the dark count rate is set by the threshold and the vacuum fluctuations of the probe resonator. A number that incorporates both $\eta$ and $p_D$, and is thus a good measure of the performance of a photodetector, is the assignment fidelity~\cite{Fan14}
\begin{equation}\label{eq:fidelity}
  \mathcal{F} = \half\left(\eta + 1-p_D\right).
\end{equation}
In practice, if the arrival time of the photon is known to lie within some time window, one can optimize $t$ in~\cref{eq:Jhomconv} over this window in a post-processing step~\cite{Royer18}. In our numerical simulations, the arrival time is known such that this optimization is not necessary and we can therefore simply evaluate $\bar J_\text{hom}(t)$ at $t=0$.

\begin{figure}
  \centering
  \includegraphics{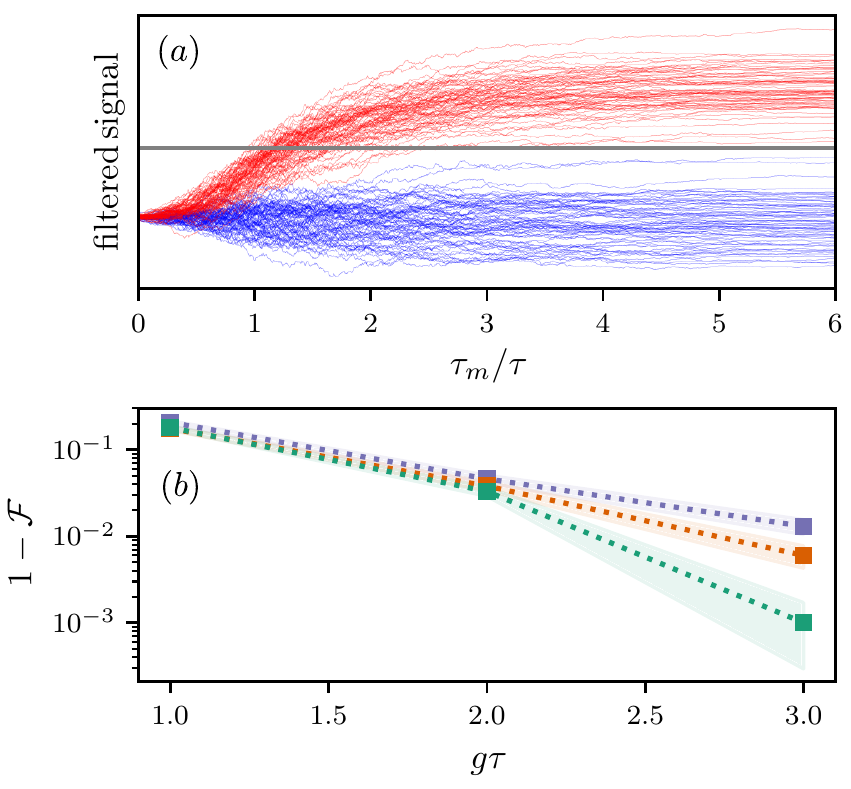}
  \caption{\label{fig:fidelity} 
  $(a)$~75 filtered homodyne currents (arbitrary units) for $g\tau=3$, $\kappa_a\tau=1.0$, $|K|/\kappa_a=10^{-2}$ and $g/\chi=5$. Red traces are obtained with an incoming Gaussian photon of unitless width $\gamma\tau = 6$, and blue traces for vacuum. The horizontal gray line is the threshold chosen to maximize the assignment fidelity.
  $(b)$~Infidelity versus $g\tau$ for $\gamma\tau = 2$ (blue), $4$ (orange), and $6$ (green), found by averaging over $N_\text{traj} = 2000$ trajectories. Other parameters as in $(a)$. The shaded regions indicate the standard error defined as $\pm \sqrt{\mathcal F(1-\mathcal F)/N_\text{traj}}$.
  }
\end{figure}

\cref{fig:fidelity} shows 75 typical filtered output records, $\bar J_\text{hom}(t=0)$, as a function of the measurement window $\tau_m$. These results are obtained from stochastic MPS simulations with $\gamma\tau=6$, $g\tau = 3$, $\kappa_a\tau = 1.0$, and include self- and cross-Kerr couplings with $|K|/\kappa_a = 10^{-2}$ and $g/\chi = 5$. The red traces correspond to simulations where a photon was present, while the blue traces are for incoming vacuum. The horizontal gray line is the threshold chosen to optimize the assignment fidelity.  At $\tau_m/\tau \gtrsim 3$, most traces are correctly identified. Panel $(b)$ shows the assignment fidelity for $\gamma\tau=$ 2 (blue), 4 (orange) and 6 (green) as as function of $g\tau$ but fixed $g/\chi = 5$. The measurement time $\tau_m$ is chosen sufficiently large to maximize $\mathcal{F}$. 
As expected from \cref{fig:displacement}, the fidelity is reduced for smaller $\gamma\tau$ because spatially longer photons (smaller $\gamma\tau$) lead to more noise in the measurement.

A remarkable feature of~\cref{fig:fidelity} is the clear trend of the assignment fidelity approaching unity with increasing $g\tau$. This number can be increased at fixed interaction strength $g$ by increasing the detector length. In the next section we show that values of $g\tau$ in the range $1$--$3$ used in~\cref{fig:fidelity} are within reach for experimentally realistic parameters and metamaterial lengths.

\paragraph*{Towards experimental realization}

The JTWPD shares similarities with the Josephson Traveling Wave Parametric Amplifier (JTWPA)~\cite{Macklin15,White15,Planat2020}. State of the art JTWPAs consists of a metamaterial with up to tens of thousands of unit cells, each comprised of a large Josephson junction and a shunt capacitance to ground. In addition, LC oscillators used to engineer the dispersion relation are placed every few unit cells. We envision a JTWPD with a similar number of unit cells, albeit with an increase in complexity for each unit cell. A significant design difference is that in the JTWPD every unit cell is coupled to the same probe resonator. In practice, this resonator can be a coplanar waveguide resonator or a 3D cavity.

As shown in the \sm, the number of unit cells necessary to reach a given value of $g\tau$ can be approximated by
\begin{equation}\label{eq:unitcells}
  N_\text{cells} \simeq \half \left(\frac{g\tau}{\alpha} \frac{R_K}{8\pi Z_\tml} \right)^2 \frac{\bar\omega^2}{K_QE_Q/\hbar}.
\end{equation}
where we neglect spatial dependence of the parameters for simplicity. In contrast to the simulation results presented above, we assume here that the detector is operated in reflection mode, effectively halving the number of unit cells needed for a given value of $\tau$. In this expression, $\alpha$ is the displacement of the probe resonator as before,
$R_K = h/e^2$ is the quantum of resistance, $Z_\tml$ the characteristic impedance of the metamaterial at the center frequency $\bar\omega$, 
and $E_Q$ the nonlinear energy of the coupling elements, discussed in more detail in \mm.

The parameter $K_Q$ appearing in~\cref{eq:unitcells} is the self-Kerr nonlinearity of the resonator [see~\cref{eq:selfKerr}] due to the nonlinear couplers in~\cref{fig:metamaterial}.
An interesting feature of the the coupling element we make use of is that the self-Kerr is always positive $K_Q>0$, in contrast to a more conventional Josephson junction element~\cite{Nigg12}.
The total Kerr non-linearity of the resonator can be adjusted by introducing another nonlinear element such as one or more Josephson junctions galvanically or capacitively coupled to the resonator.
We can then write the total Kerr non-linearity as $K = K_Q + K_J$, where $K_Q>0$ is the contribution from the couplers in~\cref{fig:metamaterial}, and $K_J<0$ comes from one or more Josephson junctions.
The latter elements can moreover be made tunable, allowing an in-situ tuning of $K_J<0$.
Following this approach, we can allow for a detector with a larger $K_Q$ contributing to reducing $N_\mathrm{cells}$, yet still have a total Kerr nonlinearity $K \simeq 0$ to avoid nonlinear response of the probe mode.
Similar ideas have recently been used to cancel unwanted cross-Kerr nonlinearities~\cite{Zhang2018}.

\begin{figure}
  \centering
  \includegraphics[scale=1.0]{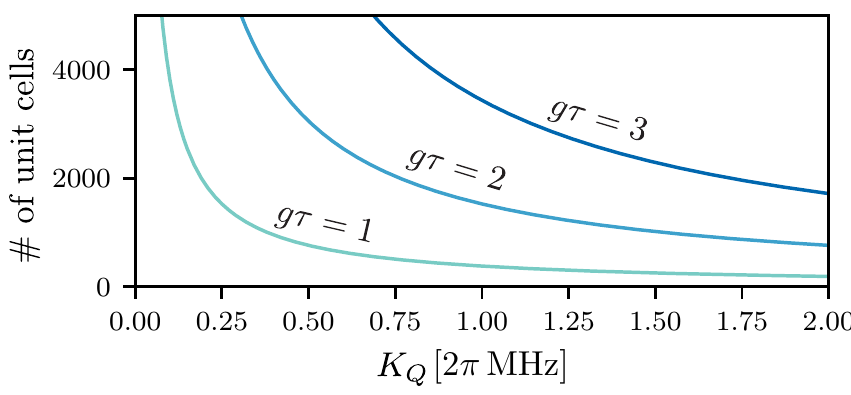}
  \caption{\label{fig:unitcells}Number of unit cells needed to reach $g\tau$ in the range $1$--$3$ as a function of self-Kerr non-linearity $K_Q$, for
  $\alpha=5$, $\bar\omega/(2\pi) = 5$ GHz, $I_s=E_Q/\varphi_0 = 1.1\,\mu$A and $Z_\tml = 50\,\Omega$. The total Kerr non-linearity of the resonator $K = K_Q + K_J$ can be tuned close to zero by introducing another non-linearity with $K_J<0$.
  }
\end{figure}

\cref{fig:unitcells} shows $N_\mathrm{cells}$ as a function of the 
self-Kerr $K_Q$
to reach $g\tau$ in the range $1$--$3$, for a photon center frequency of $\bar\omega/(2\pi) = 5$ GHz. In these plots we use a nonlinearity $I_s=E_Q/\varphi_0=1.1\,\mu$A for the coupling element, c.f. \mm, and the other parameters are $\alpha=5$ and $Z_\tml = 50\,\Omega$.
 Crucially, it is possible to reach $g\tau$ in the range $1$--$3$, as in our numerical simulations above, using a few thousand unit cells without an excessively large $K_Q$.
Alternatively, the same value of $g\tau$ can be reached for a smaller $K_Q$ by increasing the transmission line characteristic impedance, $Z_\tml$, as is clear from~\cref{eq:unitcells}.
As discussed in more detail in the \sm, $K_Q$ can be tuned by varying the coupling capacitance between the junctions and the probe resonator, or by tuning the characteristic impedance of the coupler mode.

The CRLH metamaterial has a frequency-independent characteristic impdeance $Z_\tml = \sqrt{L_n/C_g}$ given that $\sqrt{L_n/C_g}=\sqrt{L_g/C_n}$, referred to as a balanced CRLH~\cite{Caloz2004novel}. Close to the center of the CLRH frequency band, the dispersion relation is approximately linear, with a speed of light given by $v = 1/\sqrt{4 L_n C_g}$. For typical parameters, discussed in more detail in the \sm, we expect detection times in the range $\tau=1$--$10\,\mu$s.
To have $\kappa_a\tau=1$ as in the simulations above, this then
suggests a probe decay rate in the range $\kappa_a/2\pi \simeq 0.015$--$0.15$ MHz.
Larger values of $\kappa_a\tau$ might be preferable in practice, but we found this regime too demanding for numerical simulations due to the prohibitively small time steps needed. A larger $\kappa_a$ relaxes the constraint on reducing the total self-Kerr nonlinearity $|K|$.

Based on the numerical results in the previous section, the detection time is of the order $\tau_m \simeq 3\tau$, and thus expected to be in the $\mu$s to tens of $\mu$s range for the above value of $\tau$. The detector reset time is naturally of the order $1/\kappa_a$, but can likely be made faster using active reset protocols.
To avoid significant backaction effects, the photon's spectral width must not be too small as we have shown in the preceding sections. A value for the dimensionless photon width of $\gamma\tau = 2$ corresponds to a FWHM of $\gamma/(2\pi) = 0.25$ MHz, for the value $\tau=1\,\mu$s.
We emphasize that the detection fidelity increases with increasing $\gamma$,
and from our numerical results we thus expect photons of spectral width in the MHz range or larger to be detectable with very high fidelity.

The bandwidth of the detector is set by appropriately choosing the 
parameters of the CRLH metamaterial. In the \sm{} we show example parameter sets with bandwidths ranging from several GHz to 100s of MHz. For some applications that require very low dark count rates, lowering the bandwidth might be desirable. In principle the CLRH bandwidth can be made arbitrarily small, but the circuit parameters required might become challenging to realize. Another option is to replace the coupling element shown in the inset of~\cref{fig:metamaterial} by a floating coupler, such that the bandwidth is controlled by a coupling capacitance. All of these various options are discussed in more detail in the \sm.

\section*{Discussion}

Previous work have questioned whether cross-Kerr interaction can be used for high-fidelity single photon counting~\cite{Fan13}, seemingly in contradiction with our results.
There is, however, a fundamental difference between our proposal and the approach of Ref.~\cite{Fan13}. There, a number of nonlinear absorbers \emph{independently} couple to a traveling control field. This is similar to an alternative version of our proposal where each unit cell of the metematerial couples to an independent probe resonator. More generally, we can consider a situation where we partition the $N_\text{cells}$ unit cells of the detector into $M$ blocks,
with each block coupled to an independent readout probe resonator. With $M=N_\text{cells}$ we have a setup similar to Ref.~\cite{Fan13}, while $M=1$ corresponds to the JTWPD. 
However, as shown in \mm, such a setup gives a $\sqrt M$ \emph{reduction} in the probe resonator's displacement.
Our proposal thus has an $\sqrt{N_\text{cells}}$ improvement in the SNR scaling. This improvement comes from using what we referred to in the introduction as a giant probe, i.e.~a probe resonator that has a significant length compared to the photon. This contrasts with conventional circuit QED-based photodectors relying on point-like probe systems. 
Such a setup does not have any obvious analog in the optical domain, demonstrating the potential of  using metamaterials based on superconducting quantum circuits to explore fundamentally new domains of quantum optics.

In summary, we have introduced the JTWPD, a microwave single-photon detector based on a weakly nonlinear metamaterial coupled to a giant probe. This detector is unconditional in the sense that no apriori information about the photon arrival time or detailed knowledge of the photon shape is needed for its operation.
Detection fidelities approaching unity are predicted for metamaterial length that are compatible with state-of-the-art experiments.
Moreover, because the JTWPD does not rely on absorption in a resonant mode, large detection bandwidths are possible. 

A remarkable feature of the JTWPD, which distinguishes this detector from photodetectors operating in the optical regime, is the nondestructive nature of the interaction. Our numerical simulations clearly show that the shape of the photon population wavepacket is minimally disturbed by the detection. Together with the large bandwith and high detection fidelity,
this opens new possibilities for single-photon measurement and control, including feedback of photons after measurement, weak single-photon measurement, and cascading photon detection with other measurement schemes or coherent interactions.

\section*{Methods}

\paragraph*{Nonlinear coupling element}

We make use of a circuit identical to the SNAIL element introduced in Ref.~\cite{Frattini17}, but used at a different operating point.
The coupler consists of a loop of $n_s$ large junctions with Josephson energy $E_J$ and a single smaller junction with energy $\beta E_J$, leading to a nonlinear potential
\begin{equation}
  \hat U_\SNAIL(\hat\varphi) = - \beta E_J \cos(\hat\varphi - \varphi_x) - n_s E_J \cos\left(\frac{\hat\varphi}{n_s}\right),
\end{equation}
where $\varphi_x$ is the dimensionless flux encircled by the loop. The coupler is operated at the point $\varphi_x = \pi$ and $\beta = 1/n_s$ where the potential becomes
\begin{equation}\label{eq:U_SNAIL}
  \hat U_\SNAIL(\hat\varphi) = \frac{E_Q}{24} \hat\varphi^4 + \dots,
\end{equation}
and we have introduced $E_Q = E_J(n_s^2-1)/n_s^3$. Here we have expanded the nonlinear potential around $\hat\varphi \simeq 0$, which is valid based on the fact that each end of the element is coupled to harmonic modes with small zero-point flux fluctuations. The crucial property of this coupler is that it provides a purely nonlinear quartic potential while the quadratic contribution cancels out. This minimizes hybridization between the metamaterial and the probe resonator in the JTWPD, and is a very useful tool for generating non-linear interaction in general~\cite{Ye2020}.
In practice there will be small deviations from the ideal operation point $\phi_x=\pi$, $\beta=1/n_s$, but as we shown in the \sm, the JTWPD is robust to such imperfections.

The positive quartic potential in~\cref{eq:U_SNAIL} leads to positive self- and cross-Kerr nonlinearities for the probe-metamaterial system, in contrast to more conventional Josephson junction nonlinearities.
In the \sm{} we use a black-box quantization approach~\cite{Nigg12} to estimate the Kerr nonlinearities.
In particular, the self-Kerr nonlinearity of the probe mode induced by $N_\text{cells}$ coupler elements takes the form
\begin{equation}\label{eq:selfKerr}
\hbar K_Q =
\sum_{n=0}^{N_\text{cells}} E_{Q,n} |\varphi_r(x_n)|^4,
\end{equation}
with
$E_{Q,n}$ the energy of the $n$th nonlinear coupler and $\varphi_r(x_n)$ the dimensionless zero-point flux fluctuations of the probe mode biasing the $n$th coupling element.

\paragraph*{Dynamics of the probe resonator}

The probe resonator Hamiltonian can be written as
\begin{equation}\label{eq:H_r}
  \hH_{r} = \hbar \omega_r \ha^\dagger \ha + \frac{\hbar K}{2} \ha^{\dagger 2} \ha^2 + \hbar \left(i\varepsilon\e^{-i\omega_d t} \hat a^\dagger + \Hc\right),
\end{equation}
with $\ha$ the annihilation operator for the probe mode satisfying $[\ha,\ha^\dagger] = 1$. 
The resonator frequency $\omega_r$ includes significant frequency shifts due to the nonlinear couplers.
Moreover, the Kerr-nonlinearity $K=K_Q+K_J$ includes both a contribution $K_Q>0$ coming from the $N_\text{cells}$ coupler elements and a contribution $K_J<0$ which can be used to cancel out $K\simeq 0$, as discussed in the main text.
The last term of $\hH_{r}$ describes a resonator drive with amplitude $\varepsilon$ and frequency $\omega_d$. Taking damping of the probe resonator into account, the dynamics of the system is described  by the master equation
\begin{equation}
  \dot \rho = -\frac{i}{\hbar}[\hat H, \rho] + \kappa_a \mathcal{D}[\hat a]\rho.
\end{equation}

Moving to a frame rotating at the drive frequency and then displacing the field such that $\hat a \to \hat a + \alpha$, $\hH_{r}$ takes the form
\begin{equation}
  \hH'_{r}/\hbar = (\delta + 2K|\alpha|^2) \ha^\dagger \ha + \frac{K}{2} \ha^{\dagger 2} \ha^2,
\end{equation}
where $\delta = \omega_r-\omega_d$ and with $\alpha$ chosen such as to satisfy the steady-state equation
\begin{equation}
  (\delta + K|\alpha|^2)\alpha - \frac{i\kappa_a}{2}\alpha + i \varepsilon = 0.
\end{equation}

To drive the probe mode on resonance despite the Kerr nonlinearity, we chose $\omega_d$ such that $\delta = - 2K|\alpha|^2$. With this choice, the transformed probe Hamiltonian reduces to
\begin{equation}
\hH_r' = \hbar K/2 \ha^{\dagger2} \ha^2
\end{equation}
while the nonlinear equation for $\alpha$ becomes
\begin{equation}\label{eq:alpha}
  K|\alpha|^2\alpha + \frac{i\kappa_a}{2}\alpha = i\varepsilon.
\end{equation}
For $|K\alpha|^2 \ll \kappa_a$, the solution is approximately $\alpha = 2\varepsilon/\kappa_a$ and the steady-state of the resonator is to a good approximation the coherent state $\ket{\alpha}$. As discussed further in the \sm, in the opposite limit, the steady-state becomes non-Gaussian something which can reduce the signal-to-noise ratio of the detector.
To remain in the linear regime for sizeable $\alpha$, we require $|K|/\kappa_a$ to be small.

\paragraph*{Metamaterial-probe cross-Kerr coupling}

In the laboratory frame, the cross-Kerr interaction between the probe resonator and the waveguide takes the form
\begin{equation}\label{eq:H_int0}
    \hat H_{\rm int} = \hbar \sum_{\nu\mu}\int_{-z/2}^{z/2}\dd x \chi(x) \hat b_{\nu}^\dagger(x) \hat b_{\mu}(x)\hat a^\dagger \hat a,
\end{equation}
to fourth order in the Josephson nonlinear potentials~\cref{eq:U_SNAIL} and where $\nu = \pm$ refers to the direction of propagation of the photon. In this expression, we have defined the dispersive shift per unit length
\begin{equation}\label{eq:chi}
    \hbar \chi(x_n) =
    \frac{v E_{Q,n}}{a} \frac{4\pi Z_\tml}{R_K\bar\omega} |\varphi_r(x_n)|^2
\end{equation}
with $\bar\omega$ the photon center frequency, $Z_\tml$ the characteristic impedance of the transmission line at frequency $\bomega$, and we recall that $a$ is the unit cell length. Because we are only interested in small photon number in the waveguide, we have safely dropped fast-rotating terms and higher-order terms in $\hat b_{\nu\omega}$ from \cref{eq:H_int0}. Moving to the rotating and displaced frame introduced for the probe resonator above, \cref{eq:H_int0} leads to~\cref{eq:H_int1} where $g(x) = \alpha \chi(x)$ with $\alpha$ given by~\cref{eq:alpha} and where we take $\alpha$ to be real without loss of generality. 

The integral in $\hat H_{\rm int}$ should be interpreted as a Riemann sum, and the continuum limit is valid as long as all relevant wavelengths are much longer than $a$. Moreover, the expression for $\hat b_\nu(x)$ in~\cref{eq:bx} and $\chi(x)$ in~\cref{eq:chi} are derived under the assumption that dispersion is negligible over a relevant frequency band around $\bomega$, where the photon number is non-zero. In other words, we are working under the assumption that the incoming photon is sufficiently narrow. Nevertheless, we expect that photons with large spread of frequency components compared to previous proposals can be detected.

\paragraph*{Effective Keldysh master equation}

We describe the main steps of the derivation leading to \cref{eq:effMEkeldysh} and refer the reader to the \sm\, for more details. We model the incoming photon using an emitter located at position $x_0$ to the left of the metamaterial and of annihilation operator $\hat c$. After initializing the emitter in the  state $\ket{1}$, the emitter decay rate, $\kappa_c(t)$, is chosen such as to model the desired single-photon wavepacket.
Here, we choose a Gaussian wavepacket $\xi(t)$ of variance $\sigma^2$
\begin{equation}\label{eq:wavepacket}
\begin{aligned}
\xi(t) = \left(\frac{2\sigma^2}{\pi}\right)^{1/4}\e^{-i\bar \omega t} \e^{-\sigma^2 (t + x_0/v)^2},
\end{aligned}
\end{equation}
by using~\cite{Gough:12a}
\begin{equation}\label{eq:metohds:kappac}
\kappa_c(t) = \sqrt{\frac{8\sigma^2}{\pi}}\frac{\e^{-2\sigma^2 t^2}}{1 - \text{erf}[\sqrt 2\sigma t]},
\end{equation}
with $\text{erf}(x)$ the error function. The FWHM $\gamma$ used in the main text is related to the variance as $\gamma = 2\sqrt{2\ln 2}\sigma$.

The ideal Hamiltonian for the detector, emitter, and waveguide is given by
\begin{equation}
\begin{aligned}
\hat H &= \hat H_0 + \hat H_{\mathrm{ideal}} + \hH_c,\\
\hat H_0 &= \sum_{\nu} \int_\Omega d\omega\, \hbar\omega \hat b^\dag_{\nu, \omega} \hat b_{\nu,\omega},\\
\hat H_{\mathrm{ideal}} &= \hbar g\sum_{\nu\mu}\int_{-z/2}^{z/2} dx\, \hat b^\dag_{\nu}(x) \hat b_{\mu}(x)(\hat a^\dag + \hat a),\\
\hH_c &= \bar \omega \hat c^\dag \hat c + \sqrt{\kappa_c(t) v} \left[ \hat b^\dag_{+}(x_0)\hat c + \Hc\right].
\end{aligned}
\end{equation}
Using this Hamiltonian and adding decay of the probe resonator, we write the corresponding Keldysh action following Ref.~\cite{Sieberer:16a}. 
As explained in the \sm, to do this we take advantage of the fact that the action is quadratic in the fields $\hat b_\pm(x)$ and integrate out the waveguide degrees of freedom. The result is then expanded in a Taylor series in the interaction strength, which yields an effective Keldysh action for the emitter-resonator system. Finally, from that effective action, we find the equivalent master equation~\cref{eq:effMEkeldysh}.

\subsubsection*{Detector response neglecting backaction}

To help build intuition for the detector's response to a single photon, it is useful to neglect backaction effects and any correlations between the emitter and detector. Under these approximations, upon tracing out the emitter from~\cref{eq:effMEkeldysh}, we can replace the term
$\tr_C\left[\hat c \hat \rho \hat c\dg\right]$
by the approximate expression $\braket{\hat c\dg \hat c}\otimes \hat \rho_A$,
where $\tr_C\bullet$ is a partial trace over the emitter and $\hat \rho_A$ is the reduced state of the probe resonator. In this way, the reduced master equation for the probe resonator takes the form 
\begin{equation}
\dot {\hat \rho}_A \simeq -i\left[ g n_\text{det}(t)(\ha + \ha^\dag),\hat \rho_A \right]
+ \kappa_a \mathcal D[\ha]\hat \rho_A.
\end{equation}
The associated quantum Langevin equation is
\begin{equation}\label{eq:methods:inout}
    \dot{\hat a} \simeq -ign_\text{det}(t) - \frac{\kappa_a}{2}\hat a + \sqrt{\kappa_a}\hat a_\text{in}(t),
\end{equation}
with $\hat a_\text{in}(t)$ the input field which is in the vacuum state in the displaced frame, i.e. $\braket{\hat a_\text{in}(t)}=0$.
The solution for the expectation value $\braket{\hat a(t)}$ is then given by
\begin{equation}\label{eq:aout_avg}
    \braket{\ha(t)} \simeq -ig \int_{t_0}^t \dd t' \e^{-\kappa_a(t-t')/2}n_\text{det}(t).
\end{equation}
As expected, the number of photon in the metamaterial, $n_\text{det}(t)$, leads to a displacement of the probe field. 
We have confirmed that for the parameters used in~\cref{fig:displacement}, the above approximate expression is indistinguishable from the solution found from the full Keldysh master equation [dotted lines in \cref{fig:displacement}~$(a,e)$].

\subsubsection*{Detectors in series}

We can generalize the above discussion to a situation where the metamaterial
is divided into $M$ equal subsections, individually coupled to a set of $M$ independent and identical probe resonators. The interaction Hamiltonian then takes the form
\begin{equation}
    \begin{aligned}
    \hat H_\text{ideal} = \hbar g \sum_{m=0}^{M-1} \sum_{\nu\mu} & \int_{x_m-\Delta x/2}^{x_{m+\Delta x/2}}\dd x\\
    &\times \hat b_{\nu}^\dagger(x) \hat b_{\mu}(x) \left(\hat a_m^\dagger + \hat a_m\right),
    \end{aligned}
\end{equation}
with $x_m = -z/2 + \left(m + \half\right) \Delta x$, $\Delta x = z/M$, and $[\hat a_m,\hat a_n\dg] = \delta_{mn}$. 
Defining the collective mode
\begin{equation}
    \hat a_\Sigma = \frac{1}{\sqrt M} \sum_{m=0}^{M-1} \hat a_m,
\end{equation}
satisfying $[\hat a_\Sigma,\hat a_\Sigma\dg]=1$,
and assuming that each probe resonator labeled by $m$ couples identically with rate $\kappa$ to a common input-output waveguide, leads to the quantum Langevin equation for the collective mode
\begin{equation}
    \begin{aligned}
    \dot{\hat a}_\Sigma ={}& \frac{i}{\hbar}[\hat H_\text{ideal},\hat a_\Sigma] - \frac{\kappa_\Sigma}{2}\hat a_\Sigma + \sqrt{\kappa_\Sigma}\hat a_\text{in}(t),
    \end{aligned}
\end{equation}
where $\kappa_\Sigma = M\kappa$ and where we have taken the resonator frequencies to be identical. Under a similar set of approximations as above, we find
\begin{equation}
    \dot{\hat a}_\Sigma \simeq -\frac{ig}{\sqrt M} n_\text{det}(t) - \frac{\kappa_\Sigma}{2}\hat a_\Sigma + \sqrt{\kappa_\Sigma}\hat a_\text{in}(t),
\end{equation}
Comparing to \cref{eq:methods:inout} which was obtained for $M=1$, we find a $\sqrt{M}$ reduction in the displacement. To compensate one could increase $g\to g\sqrt{M}$, but this leads to a breakdown of the assumption of negligible backaction.
In summary the JTWPD limit $M=1$ is ideal.

\paragraph*{Matrix Product State simulations}

The JTWPD is an open quantum many-body system with nonlocal interactions, and numerically simulating its time evolution poses a significant challenge. Recently, approaches based on Matrix Product States (MPS) have been developed to simulate point-like scatterers interacting with one-dimensional waveguides~\cite{Grimsmo15,Pichler16}. Applying these ideas to the JTWPD, however, requires nontrivial extensions of the techniques in order to deal with the nonlocal interaction and the stochastic nature of the evolution in the presence of continuous homodyne detection. We outline here the main ideas behind the method we have developed, leaving further details to the \sm.

To represent the system as an MPS, we discretize both time and space. In the following we only consider the right moving field in the waveguide. As long as the different parts of the waveguide are impedance matched and $g/\bar\omega \ll 1$, back scattering into the left-moving field is negligible and it can safely be dropped.
Following~\cite{Grimsmo15,Pichler16}, we trotterize the time evolution operator
\begin{equation}
U(T) = \mathcal{T} \e^{-i \int_0^T \dd t \hat H(t)} = \lim_{N_t\to\infty} \hat U_{N_t-1} \dots \hat U_1 \hat U_0,
\end{equation}
where $\hat H(t)$ is the Hamiltonian in the interaction picture, and $\hat U_i$ evolves the system for a small time $t_i$ to $t_i+\Delta t$.
We moreover similarly discretize the spatial integral for each $\hat U_i$
\begin{equation}
  \hat U_i = \lim_{N_x\to\infty} \hat U_{i,N_x-1}\dots \hat U_{i,1}\hat U_{i,0},
\end{equation}
where
\begin{equation}\label{eq:methods:mps:timestep}
  \hat U_{i,n} = \e^{-\frac{i}{\hbar}\int_{t_i}^{t_i+\Delta t} \dd t \int_{x_n}^{x_n+\Delta x} \dd x \hat{\mathcal H}_\text{int}(x,t) - \frac{i}{\hbar} \Delta t \hat H_r/N_x},
\end{equation}
and $\Delta x = v\Delta t$. 
We next make the approximations
\begin{equation}
\begin{aligned}
&\int_{t_i}^{t_i+\Delta t} \dd t \int_{x_n}^{x_n+\Delta x} \dd x \hat b_{+}^\dagger(x-vt) \hat b_{+}(x-vt) \hat{A}(x) \\
\simeq{}& \int_{t_i}^{t_i+\Delta t} \dd t \hat b_{+}^\dagger(x_n-vt) \int_{x_n}^{x_n+\Delta x} \dd x \hat b_{+}(x-vt_i) \hat{A}(x_n) \\
=& -\Delta t \hat b_{n-i} \hat b_{n-i} \hat{A}(x_{n}),
\end{aligned}
\end{equation}
with $\hat{A}(x) = \chi(x)\left( \hat a^\dagger \hat a + \alpha^2 \right) + g(x) \left( \hat a^\dagger + \hat a\right)$ and where,
in the last line, we have defined
\begin{equation}
  \hat b_n = \frac{1}{\sqrt{\Delta x}} \int_{x_n}^{x_n+\Delta x} \dd x \hat b_+(x).
\end{equation}
For a photon that is not too broad in frequency, we can extend the integration limits in~\cref{eq:bx} and approximate
\begin{equation}
  \hb_{+}(x) \simeq \sqrt{\frac{1}{2\pi v}} \int_{-\infty}^\infty \dd\omega \hb_{+\omega}\e^{i\omega x/v}.
\end{equation}
Since $[\hat b_{+\omega},\hat b_{+\omega'}\dg] = \delta(\omega-\omega')$ this leads to $[\hat b_n,\hat b_m\dg] \simeq \delta_{nm}$, such that these discrete modes can be interpreted as harmonic oscillators. 

\begin{figure}
\centering
\includegraphics{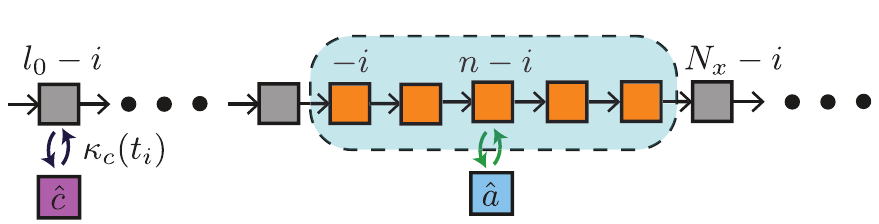}
\caption{\label{fig:conveyorbelt}At the $i$th time step, the probe resonator ($\hat a$) interacts with oscillators labeled $-i \le j < N_x-i$, as indicated by the dashed box. An emitter ($\hat c$) decays with rate $\sigma$ far to the left of the detector.}
\end{figure}

As illustrated in~\cref{fig:conveyorbelt}, 
\cref{eq:methods:mps:timestep} suggests the following picture: In the $i$th time step, the probe resonator interacts with waveguide modes $\hat b_j$ with $-i \le j < N_x-i$. In the next time step, the waveguide modes are shifted one unit cell to the right relative to the probe, such that interaction is now with $-i-1 \le j < N_x-i-1$, and so on. To model an incoming photon, we also include an emitter decaying at rate $\kappa_c(t_i)$ into the waveguide at site $l_0-i$ with $l_0 < 0$ to the left of the detector.

This discretized system can be evolved using methods described in Ref.~\cite{Pichler16}, with two important changes: 1)~Within each time step the probe resonator interacts with multiple waveguide oscillators, represented by the blue region in \cref{fig:conveyorbelt}. 
We therefore perform a single time step by swapping~\cite{Wall16} the MPS site corresponding to the probe resonator along the MPS, letting it interact with the waveguide modes one by one. 2)~For $\kappa_a > 0$, the probe resonator is coupled to an additional bath describing the input-output fields $\hat a_\text{in/out}(t)$, with $\hat a_\text{out}(t)$ being continuously monitored by homodyne detection. To avoid representing these bath degrees of freedom explicitly, we replace the unitary evolution $\e^{-i\hat H_r \Delta t}$ with a stochastic Schr\"odinger equation for the MPS integrated from $t_i$ to $t_i+\Delta t$. For this, we use the usual stochastic Schr\"odinger equation for homodyne detection which can be integrated using standard numerical solvers for stochastic differential equations~\cite{Kloeden92}. Note that only a single site of the MPS is changed during this step. Further details are given in the \sm.

\section*{Data Availability}
All relevant data to support the conclusions are within the paper and its Supplementary
Materials. Raw data and numerical code generated during the current study are available from the corresponding author on reasonable request.



\nocite{Grimsmo17,wang2018mode,Vool17,Bhat06,Carmichael2013,Schollwock11,Gardiner:00}

\bibliography{refs}
\input



\section*{Acknowledgements}
We thank J. Bourassa, T. Stace, J. Combes and B. Plourde for valuable discussions.
\textbf{Funding:}
This work is supported by the Australian Research Council (ARC) via Centre of Excellence in Engineered Quantum Systems (EQUS) Project No. CE170100009 and a Discovery Early Career Researcher Award (DE190100380).
Part of this work was supported by the Army Research Office under Grant no.~W911NF-15-1-0421, NSERC, the Vanier Canada Graduate Scholarship and by the Canada First Research Excellence Fund.
This research was funded in part by the MIT Center for Quantum Engineering via support from the Laboratory for Physical Sciences under contract number H98230-19-C-0292.
\textbf{Author contributions:}
ALG and BR devised the project and performed the numerical and analytical analysis. 
All authors contributed to analyzing and interpreting the results, and to writing the manuscript.
\textbf{Competing interests:}
The authors declare that they have no competing interests.

\input{inputsm}

\end{document}

%% file: inputsm.tex
\clearpage


\section*{Supplementary material}

# Supplementary Materials: Quantum metamaterial for non-destructive microwave photon counting

May 13, 2020

# Contents

# 1 Composite right/left handed (CRLH) metamaterial

## 1.1 Characteristic impedance and dispersion relation

Before considering the nonlinear coupling of the metamaterial to the probe resonator, it is useful to consider the properties of the metamaterial by itself, as illustrated in Fig. S1. This type of metamaterial is referred to as a composite right/left-handed (CRLH) transmission line metamaterial and is attractive due to its potential for dispersion and band-structure engineering [1, 2].

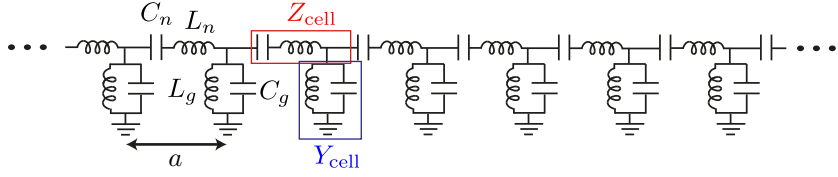

Figure S1: Array of identical LC resonators with series inductive and capacitive nearest neighbor couplings. For wavelengths large compared to the unit cell distance $a$, the metamaterial behaves as a transmission line. The transmission line in general has both a right-handed and a left-handed frequency band, and is referred to as a composite right/left-handed (CRLH) metamaterial. Right (left)-handed here means that the wavevector and group velocity have the same (oppsite) sign.

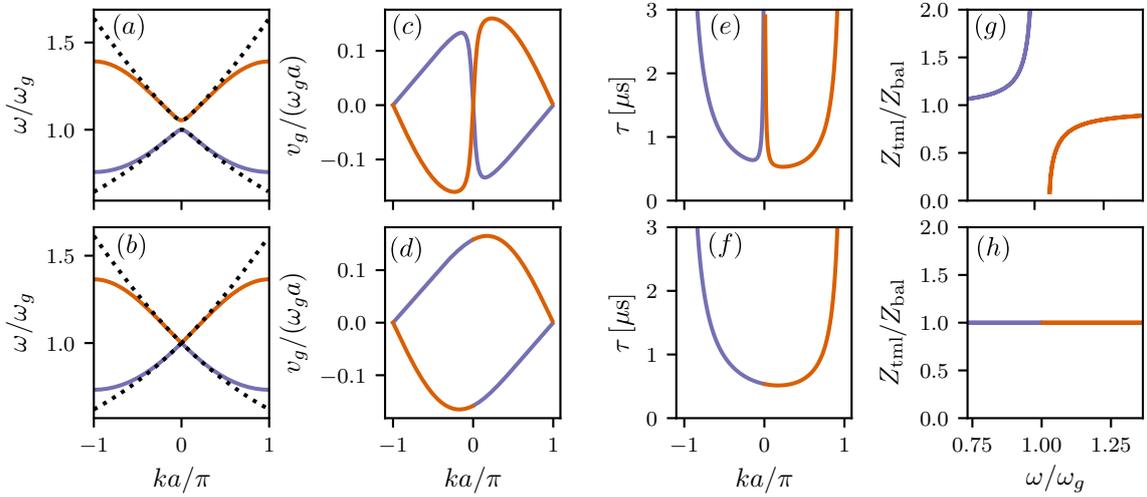

Figure S2: Characteristics of a CRLH transmission line metamaterial. $(a, b)$ Two positive solutions to Eq. (1) (solid) and small $ka$ approximation (dotted), $(c, d)$ group velocity $v_g = \partial \omega / \partial k$, $(e, f)$ time of flight through metamaterial $\tau = 2z/v_g$, for $N = 2000$ unit cells and $\omega_g/(2\pi) = 7.5$ GHz, $(g, h)$ Characteristic impedance $Z_{\text{tml}}(\omega)$. **Top row:** $cl = 0.9$. **Bottom row:** $cl = 1$.

As shown in Refs. [1, 2], the dispersion relation for waves propagating through the metamaterial is given by

$$\cos(ka) = 1 + \frac{Z_{\text{cell}}(\omega)Y_{\text{cell}}(\omega)}{2} = 1 - \frac{1}{2}\left(\omega L_n - \frac{1}{\omega C_n}\right)\left(\omega C_g - \frac{1}{\omega L_g}\right), \quad (1)$$

where $a$ is the unit cell distance, $k$ is the wavevector and

$$Z_{\text{cell}}(\omega) = i\left(\omega L_n - \frac{1}{\omega C_n}\right), \quad (2)$$

$$Y_{\text{cell}}(\omega) = i\left(\omega C_g - \frac{1}{\omega L_g}\right), \quad (3)$$

are, respectively, the series impedance and parallel admittance to ground for a unit cell, as illustrated in Fig. S1. In the limit of small $ka$, which corresponds to a long wavelength limit relative to the unit cell distance, the dispersion relation is approximately given by

$$ka = -i\sqrt{Z_{\text{cell}}(\omega)Y_{\text{cell}}(\omega)} \quad (|ka| \ll 1). \quad (4)$$

Eq. (1) has two positive solutions for $\omega$

$$\omega_{\uparrow\downarrow}(ka) = \frac{1}{\sqrt{L_n C_n}}\sqrt{\frac{x(ka) \pm \sqrt{x(ka)^2 - 4cl}}{2}}, \quad (5)$$

where $x(ka) = 1 + cl + 2c[1 - \cos(ka)]$, with $c = C_n/C_g$, $l = L_n/L_g$. Examples solutions for $\omega_\uparrow$ (orange) and $\omega_\downarrow$ (purple) are shown in Fig. S2 $(a)$ and $(b)$, together with the small $ka$ approximations.

Note that for the lower frequency band, the group velocity and the wave vector $k$ have opposite signs, such that the metamaterial behaves as a left-handed transmission line in this frequency range, while it behaves as a right-handed transmission line in the upper band [1]. The upper (lower) band has a minimum (maximum) at $ka = 0$, where

$$\omega_\uparrow(ka = 0) = \max(\omega_g, \omega_n), \quad (6)$$

$$\omega_\downarrow(ka = 0) = \min(\omega_g, \omega_n). \quad (7)$$

where $\omega_g = 1/\sqrt{C_g L_g}$ and $\omega_n = 1/\sqrt{C_n L_n}$. Unless $\omega_g = \omega_n$ ($cl = 1$) there is a bandgap between these two frequencies. In general the upper (lower) band has a high (low) frequency cut-off at

$$\Omega_{H/L} \equiv \omega_{\uparrow\downarrow}(ka = \pi) = \omega_n\sqrt{\frac{1 + cl + 4c \pm \sqrt{(1 + cl + 4c)^2 - 4cl}}{2}}. \quad (8)$$

The characteristic impedance of the metamaterial transmision line can be found by relating the current through a unit cell to the voltage to ground [2]. Consider the voltage across a single unit cell as illustrated in the following figure:

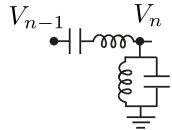

We have that the current across the cell is $I_n = (V_{n-1} - V_n)/Z_{\text{cell}}$. Assuming a traveling wave form for the voltage $V_n = V(\omega)e^{-ikan}$ we have that $I_n = I(\omega)e^{-ikan}$ with $I(\omega) = e^{ika/2}2i\sin\left(\frac{ka}{2}\right)\frac{V(\omega)}{Z_{\text{cell}}(\omega)}$. We define the characteristic impedance to be [2]

$$Z_{\text{tml}}(\omega) = \left|\frac{V(\omega)}{I(\omega)}\right| = \left|\frac{Z_{\text{cell}}(\omega)}{2i\sin(ka/2)}\right| = \sqrt{\frac{Z_{\text{cell}}(\omega)}{Y_{\text{cell}}(\omega)}}, \quad (9)$$

3

where in the last equality we used $|\sin(x/2)| = \sqrt{[1 - \cos(x)]/2}$ and Eq. (1). Of particular interest is the so-called balanced case $\omega_g = \omega_n$ ($cl = 1$) [1], for which the bandgap closes, $\omega_\uparrow(ka = 0) = \omega_\downarrow(ka = 0)$, and the characteristic impedance becomes independent of frequency

$$Z_{\text{tml}} = Z_{\text{bal}} = \sqrt{\frac{L_g}{C_n}} = \sqrt{\frac{L_n}{C_g}} \qquad (\omega_g = \omega_n). \tag{10}$$

This fact has been exploited to impedance match CRLH metamaterials to conventional 50 Ω microwave transmission lines over many GHz of bandwidth [1]. If $\omega_g \simeq \omega_n$ is only approximately satisfied, there will be small bandgap between these frequencies where $Z_{\text{tml}}$ diverges, while $Z_{\text{tml}}$ is weakly frequency dependent away from the bandgap.

Targeting a balanced CRLH metamaterial, for which $cl = 1$, the frequency cutoffs $\Omega_{H/L}$ can be controlled by varying $c/l$, since for $c \to 0$ we have that $\Omega_L \to \Omega_H$, i.e., the bandwidth vanishes, while for $l \to 0$, $\Omega_L \to 0$ and $\Omega_H \to \infty$.

## 1.2 Quantum treatment of a CRLH metamaterial

Based on the above classical arguments leading to analytical expressions for the wavevector and characteristic impedance, we expect that the quantized flux field along the metamaterial can be written in the standard form in a long wavelength approximation [3]

$$\hat{\phi}(x) = \sum_{\nu=\pm} \int_\Omega \phi(\omega) e^{\nu i k(\omega) x} \hat{b}_\nu(\omega) + \text{H.c.}, \tag{11}$$

where $[\hat{b}_\nu(\omega), \hat{b}_\mu^\dagger(\omega')] = \delta_{\nu\mu}\delta(\omega - \omega')$, the range of integration $\Omega$ refers to the frequency band supporting traveling waves in the metamaterial, and

$$\phi(\omega) = \sqrt{\frac{\hbar Z_{\text{tml}}(\omega)}{4\pi\omega}}. \tag{12}$$

quantify the magnitude of the flux field zero-point fluctuations. We here confirm this expectation by a quantized circuit treatment of the CRLH metamaterial illustrated in Fig. S1.

We start by defining two flux node variables $\{\theta_n, \phi_n\}$ for each unit cell, as illustrated in the following figure:

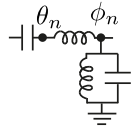

Following the standard circuit node approach [4], the Lagrangian for a CRLH metamaterial with $N$ unit cells is

$$L = \sum_{n=0}^{N-1} \frac{1}{2}\dot{\vec{\phi}}_n^T C \dot{\vec{\phi}}_n - \frac{1}{2}\vec{\phi}_n^T L_1^{-1} \vec{\phi}_n + \frac{1}{2}\vec{\phi}_n^T L_2^{-1} \vec{\phi}_{n-1} + \frac{1}{2}\vec{\phi}_n^T (L_2^{-1})^T \vec{\phi}_{n+1}, \tag{13}$$

where we have defined a vector $\vec{\phi}_n = [\theta_n, \phi_n]^T$ as well as capacitance and inductance matrices

$$C = \begin{pmatrix} C_n & -C_n \\ -C_n & C_g + C_n \end{pmatrix}, \quad L_1^{-1} = \begin{pmatrix} \frac{1}{L_n} & 0 \\ 0 & \frac{1}{L_g} + \frac{1}{L_n} \end{pmatrix}, \quad L_2^{-1} = \begin{pmatrix} 0 & \frac{1}{L_n} \\ 0 & 0 \end{pmatrix}. \tag{14}$$

It is convenient to introduce new rescaled variables $\vec{\phi}_n = C^{-1/2}\vec{\psi}_n$ such that the Lagrangian can be written

$$L = \sum_{n=0}^{N-1} \frac{1}{2}\dot{\vec{\psi}}_n^T \dot{\vec{\psi}}_n - \frac{1}{2}\vec{\psi}_n^T M_1 \vec{\psi}_n + \frac{1}{2}\vec{\psi}_n^T M_2 \vec{\psi}_{n-1} + \frac{1}{2}\vec{\psi}_n^T M_2^T \vec{\psi}_{n+1}, \tag{15}$$

where $M_i = C^{-1/2} L_i^{-1} C^{-1/2}$. Assuming periodic boundary conditions, we introduce a Fourier transform

$$\vec{\psi}_n = \frac{1}{\sqrt{N}} \sum_{l=0}^{N-1} \vec{\psi}_l e^{ik_l x_n} + \text{c.c.}, \qquad (16)$$

where we have defined $k_l = 2\pi l/Na$ and $x_n = na$. This leads to

$$L = \sum_{l=0}^{N-1} \left[ \dot{\vec{\psi}}_l^\dagger \dot{\vec{\psi}}_l - \vec{\psi}_l^\dagger M(k_l a) \vec{\psi}_l \right], \qquad (17)$$

with

$$M(ka) = M_1 - M_2 e^{-ika} - M_2^T e^{ika}. \qquad (18)$$

The matrix $M(ka)$ is Hermitian and can thus be diagonalized by a unitary: $U(ka)^\dagger M(ka) U(ka) = \Omega^2(ka) = \text{diag}[\omega_\uparrow^2(ka), \omega_\downarrow^2(ka)]$. Defining new variables via $\psi_l = U(k_l a) \xi_l$ then leads to

$$L = \sum_{l=0}^{N-1} \left[ \dot{\vec{\xi}}_l^\dagger \dot{\vec{\xi}}_l - \vec{\xi}_l^\dagger \Omega_l^2 \vec{\xi}_l \right], \qquad (19)$$

with $\Omega_l = \Omega(k_l a)$. Transforming to a Hamiltonian is also straight forward

$$H = \sum_{l=0}^{N-1} \left[ \vec{q}_l^\dagger \vec{q}_l + \vec{\xi}_l^\dagger \Omega_l^2 \vec{\xi}_l \right], \qquad (20)$$

where $\vec{q} = \left[ \frac{\partial L}{\partial \xi_1}, \frac{\partial L}{\partial \xi_2} \right]^T$ is a vector of canonical momenta corresponding to $\vec{\xi}$. We next promote the classical variables to operators

$$\vec{\xi}_l = \begin{pmatrix} \sqrt{\frac{\hbar}{2\omega_\uparrow(k_l a)}} \hat{c}_l \\ \sqrt{\frac{\hbar}{2\omega_\downarrow(k_l a)}} \hat{a}_l \end{pmatrix}, \qquad (21)$$

$$\vec{q}_l = i \begin{pmatrix} \sqrt{\frac{\hbar \omega_\uparrow(k_l a)}{2}} \hat{c}_l^\dagger \\ \sqrt{\frac{\hbar \omega_\downarrow(k_l a)}{2}} \hat{a}_l^\dagger \end{pmatrix}, \qquad (22)$$

where $[\hat{a}_l, \hat{a}_l^\dagger] = [\hat{c}_l, \hat{c}_l^\dagger] = 1$. This immediately gives

$$H = \sum_{l=0}^{N-1} \hbar \omega_\downarrow(k_l a) \hat{a}_l^\dagger \hat{a}_l + \hbar \omega_\uparrow(k_l a) \hat{c}_l^\dagger \hat{c}_l. \qquad (23)$$

Since $M(ka)$ is a two-by-two matrix, the frequencies can be found analytically. We have confirmed that the result for $\omega_{\uparrow\downarrow}(ka)$ agrees with that give in Eq. (5). The annihilation operators $\hat{a}_l$ thus refer to the lower, left-handed band, and the annihilation operators $\hat{c}_l$ refer to the upper, right-handed band.

We can now furthermore express the original position dependent variables in terms of the normal mode annihilation and creation operators:

$$\vec{\phi}_n = \begin{bmatrix} \hat{\theta}_n \\ \hat{\phi}_n \end{bmatrix} = \sum_{l=0}^{N-1} T(k_l a) \begin{bmatrix} \sqrt{\frac{\hbar}{2\omega_\uparrow(k_l a)}} \hat{c}_l \\ \sqrt{\frac{\hbar}{2\omega_\downarrow(k_l a)}} \hat{a}_l \end{bmatrix} e^{ik_l x_n} + \text{c.c.} \qquad (24)$$

where $T(k_l a) \equiv C^{-1/2} U(k_l a)/\sqrt{N}$. Let us focus specifically on the $\hat{\phi}_n$ component. Here $k$ is a Bloch vector such that $ka$ can be identified with $ka + \pi$, and it is therefore convenient to write

$$\hat{\phi}_n = \sum_{\nu=\pm} \sum_{l=0}^{N/2-1} \Delta ka \left[ \phi_\downarrow(k_l a) \frac{\hat{a}_{\nu l}}{\sqrt{\Delta ka}} e^{\nu i k_l x_n} + \phi_\uparrow(k_l a) \frac{\hat{c}_{\nu l}}{\sqrt{\Delta ka}} e^{\nu i k_l x_n} \right] + \text{H.c.}, \qquad (25)$$

5

where $\Delta ka = 2\pi/N$, and

$$\phi_\downarrow(ka) = \sqrt{\frac{1}{\Delta ka}}\sqrt{\frac{\hbar|T_{10}(ka)|^2}{2\omega_\downarrow(ka)}}, \tag{26a}$$

$$\phi_\uparrow(ka) = \sqrt{\frac{1}{\Delta ka}}\sqrt{\frac{\hbar|T_{10}(ka)|^2}{2\omega_\uparrow(ka)}}, \tag{26b}$$

where $T_{ij}(ka)$ with $i,j \in \{0,1\}$ refers to the matrix elements of $T(ka)$, and we made use of a slight abuse of notation based on the fact that any complex phase for $T_{ij}(ka)$ can be absorbed into a redefinition of the annihilation operators. We have also dropped any dependence on $\nu = \pm$ in the functions $\phi_{\uparrow\downarrow}(ka)$ since, based on symmetry arguments, the magnitude of zero-point fluctuations as quantifies by $\phi_{\uparrow\downarrow}(ka)$ must be equal for the left- and right moving fields.

In the limit $N \to \infty$, we have $\Delta ka \to 0$ and $k_l a$ becomes a real variable in the range $[0,\pi)$ for $l = 0,\ldots,N/2-1$. The above sum thus approaches an integral

$$\hat{\phi}_n = \sum_{\nu=\pm} \int_0^\pi d(ka) \left[\phi_\downarrow(ka)\hat{a}_\nu(ka)e^{\nu i(ka)n} + \phi_\uparrow(ka)\hat{c}_\nu(ka)e^{\nu i(ka)n}\right] + \text{H.c.}, \tag{27}$$

where we have defined operators $\hat{b}_\nu(k_l a) = \hat{b}_{\nu l}/\sqrt{\Delta ka}$ for $\hat{b} = \hat{a}, \hat{c}$, which in the $N \to \infty$ limit satisfy $[\hat{b}_\nu(ka), \hat{b}_\mu^\dagger(k'a)] = \delta_{\mu\nu}\delta(ka - k'a)$. Finally, changing to an integral over frequency we find

$$\hat{\phi}_n = \sum_{\nu=\pm} \int_{\Omega_\downarrow} d\omega\, \phi_\downarrow(\omega)\hat{a}_\nu(\omega)e^{\nu ik(\omega)x_n} + \int_{\Omega_\uparrow} d\omega\, \phi_\uparrow(\omega)\hat{c}_\nu(\omega)e^{\nu ik(\omega)x_n} + \text{H.c.}, \tag{28}$$

where

$$\Omega_\downarrow = [\omega_\downarrow(ka=\pi), \omega_\downarrow(ka=0)], \tag{29}$$
$$\Omega_\uparrow = [\omega_\uparrow(ka=0), \omega_\uparrow(ka=\pi)], \tag{30}$$

refers to the lower and upper frequency bands, respectively. We have moreover introduced $\hat{b}_\nu(\omega) = \hat{b}_\nu(ka)/\sqrt{v_g(\omega)}$ for $\hat{b} = \hat{a}, \hat{c}$ and $\phi_{\uparrow\downarrow}(\omega) = \phi_{\uparrow\downarrow}(ka)/\sqrt{v_g(\omega)}$, with the group velocity

$$v_g(\omega) = \left|\frac{dk}{d\omega}\right|^{-1}. \tag{31}$$

By introducing a density of left/right moving modes

$$\rho_\pm(\omega) = \frac{\rho(\omega)}{2} = \frac{Na}{2\pi v_g(\omega)}, \tag{32}$$

we can write

$$\phi_\downarrow(\omega) = \sqrt{\frac{\hbar|T_{10}[k(\omega)a]|^2 \rho(\omega)}{4\omega}}, \tag{33a}$$

$$\phi_\uparrow(\omega) = \sqrt{\frac{\hbar|T_{10}[k(\omega)a]|^2 \rho(\omega)}{4\omega}}, \tag{33b}$$

Analytical expressions for $\phi_{\uparrow\downarrow}(\omega)$ are in principle straight forward to find. However, the expressions are cumbersome and we have not been able to simplify them to a form that is useful to display. Instead we numerically compare $\phi_{\uparrow\downarrow}(\omega)$ to the expected expression Eq. (12). We have verified that $\phi_{\uparrow\downarrow}(\omega)$ approaches Eq. (12) for small $|ka|$, corresponding to the center of the band $\Omega_\downarrow \cup \Omega_\uparrow$. An example is shown in Fig. S3 for a choice of parameters corresponding to a balanced CRLH metamaterial ($\omega_n = \omega_g$). We thus formally recover the form Eq. (11) by defining $\hat{\phi}(x_n) = \hat{\phi}_n$ and taking the continuum limit $a \to 0$. In Eq. (11) then, $\Omega = \Omega_\downarrow \cup \Omega_\uparrow$ and $\hat{b}_\nu(\omega)$ refers to $\hat{a}_\nu(\omega)$ or $\hat{c}_\nu(\omega)$, depending on the frequency band. Note that $\nu = +$ refers to left-movers in the lower frequency band $\Omega_\downarrow$ and right-movers in the upper band $\Omega_\uparrow$, and vice versa for $\nu = -$.

6

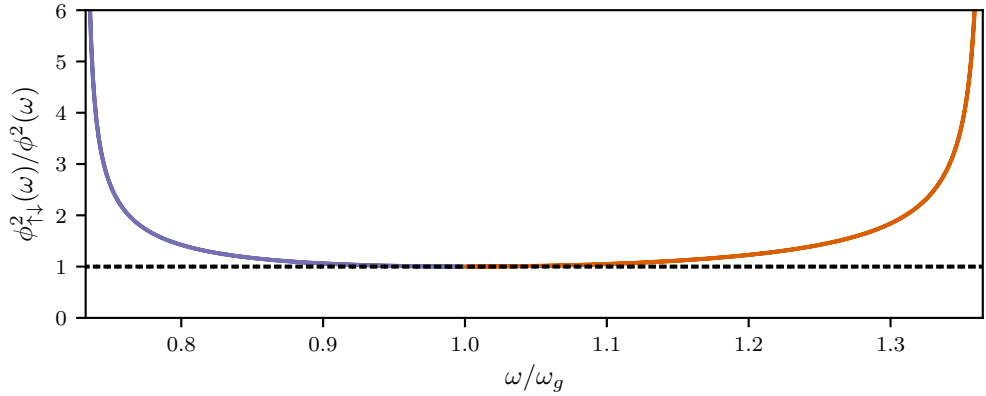

Figure S3: Comparison of $\phi_\downarrow^2(\omega)$ (purple) and $\phi_\uparrow^2(\omega)$ (orange) to $\phi^2(\omega) \sim Z_{\text{tml}}(\omega)$ in Eq. (12), for $c = C_n/C_g = 0.1$, $l = L_n/L_g = 10$. The center of the band $\omega \simeq \omega_g$ corresponds to long wavelengths where $|ka| \ll 1$.

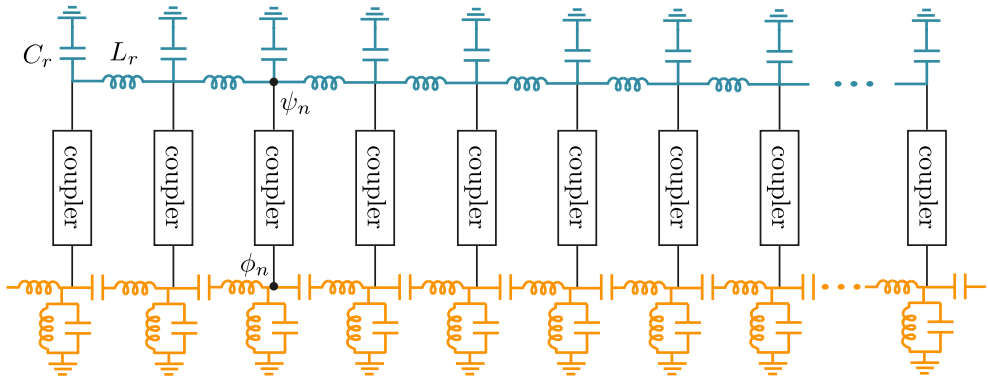

Figure S4: Simplified representation of the JTWPD where the transmission line resonator is represented by a telegrapher model with a finite number of unit cells. For clarity, the CRLH transmission line is shown in orange and the probe resonator in blue. In the numerics periodic boundary conditions are used for the CRLH transmission line.

## 2 CRLH coupled to a resonator

### 2.1 Nonlinear coupler element

The basis for our nonlinear coupler is a circuit which is identical to a flux qubit or a Superconducting Nonlinear Asymmetric Inductive eLement (SNAIL) [5], but used at a different operating point than these two devices. The nonlinear coupler is used to couple the CRLH metamaterial to a probe resonator. The resonator could be a 2D co-planar waveguide resonator or a 3D cavity, although we will focus mostly on a 2D architecture in the following. The setup is illustrated in Fig. S4, where the resonator is represented by a lumped element telegrapher model (blue).

The main original motivation for introducing the SNAIL in Ref. [5] was to realize a three-wave mixing element. The desired properties of our coupling element is rather different. The main feature we want is a purely nonlinear quartic interaction, with no (or minimal) quadratic contribution to the potential. The basic element of the coupler consists of a loop of $n_s$ large Josephson junctions with Josephson energy $E_J$ and a single smaller junction with energy $\beta E_J$, encircling an external flux $\Phi_x$, as illustrated in the following circuit diagram:

7

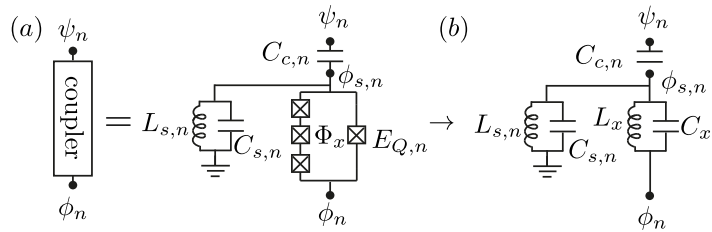

Figure S5: (a) Nonlinear coupler where $\phi_n$, $\psi_n$ refer to the CRLH metamaterial and transmission line resonator node flux at position $x_n$, respectively. (b) Linearized version of the coupler element, where $C_x$ and $L_x$ represent stray capacitance and inductance, respectively.

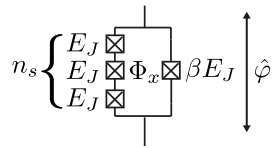

When $E_J/E_C \gg 1$ for each of the $n_s$ large junctions with $E_C$ the charging energy of a single junction due to the junction capacitance, we can assume that the phase drop $\hat{\varphi}$ across the entire chain is equally distributed across each junction, leading to a simplified expression for the potential energy [5]

$$\hat{U}(\hat{\varphi}) = -\beta E_J \cos(\hat{\varphi} - \varphi_x) - n_s E_J \cos\left(\frac{\hat{\varphi}}{n_s}\right), \tag{34}$$

where $\varphi_x = 2\pi\Phi_x/\Phi_0$. We focus on the choice $\varphi_x = \pi$ such that the two cosine terms in the previous equation have opposite signs. This can be used to cancel out the quadratic contribution to the potential. By expanding $\hat{U}(\hat{\varphi})$ around $\hat{\varphi} = 0$ we find

$$\hat{U}(\hat{\varphi}) = -\frac{E_J}{2}\left(\beta - \frac{1}{n_s}\right)\hat{\varphi}^2 + \frac{E_J}{24}\left(\beta - \frac{1}{n_s^2}\right)\hat{\varphi}^4 + \ldots, \quad \text{for } \phi_x = \pi, \tag{35}$$

where the ellipses refer to higher order terms in $\hat{\varphi}$ and we neglect a constant term. In particular, with the choice $\beta = 1/n_s$ (and $n_s > 1$) the quadratic term cancels out exactly leading to

$$\hat{U}(\hat{\varphi}) = \frac{E_Q}{24}\hat{\varphi}^4 + \ldots, \quad \phi_x = \pi, \ \beta = \frac{1}{n}, \tag{36}$$

where for notational convenience we have defined

$$E_Q = E_J \frac{n_s^2 - 1}{n_s^3}. \tag{37}$$

To construct a non-linear coupler element between the CRLH metamaterial and the resonator, we use the circuit illustrated in Fig. S5 (a), where $\phi_n$, $\psi_n$ refer to the CRLH metamaterial and transmission line resonator node flux at position $x_n$, respectively, while $\phi_{s,n}$ is a coupler mode which will typically remain in its ground state. The coupling capacitance $C_{c,n}$ is used to control the coupling to the resonator, while the shunt inductance $L_{s,n}$ and capacitance $C_{s,n}$ is used to control the frequency and characteristic impedance of the coupler mode $\phi_{s,n}$.

This coupling element has several advantageous features. The capacitance $C_{c,n}$ is used to control the participation of the resonator mode in the non-linear potential of the coupler. On the other hand, the CRLH metamaterial is galvanically coupled to the coupler. This allows us to maximize the cross-Kerr coupling between the two systems while minimizing the self-Kerr interaction of the resonator. Moreover, the use of a non-linear coupler without any quadratic coupling terms, c.f. Eq. (36), minimizes hybridization between the two systems.

Of course, in any realistic device, deviations from the ideal operating point for the nonlinear coupler, $\varphi_x = \pi$ and $\beta = 1/n_s$, will lead to some stray inductive coupling. In addition, the junction capacitances is another source of unwanted coupling between the two systems. We therefore investigate the robustness of the system to such stray linear couplings in the next section.

## 2.2 Eigenmodes of the linearized JTWPD

In this section we consider how the presence of the coupler elements influences the eigenmodes of the CRLH metamaterial and the probe resonator. To do this, we first consider a linearized version of the coupler element, as illustrated in Fig. S5 (b). In this circuit, $L_x$ represents stray inductive coupling between the resonator and the metamatrial due to deviations from the ideal operating point $\phi_x = \pi$ and $\beta = 1/n_s$ described in the previous section, and $C_x$ similarly represent stray capacitive coupling due to, e.g., junction capacitances. We for simplicity take all the coupler elements to be identical in the numerical results in this section, but this is not a necessary requirement as the detector is in fact highly robust to disorder.

The full circuit we consider in this section is illustrated in Fig. S4 (with the understanding that the couplers are replaced by their linearized versions). We use a lumped element telegrapher model to represent the probe resonator, with a finite number of unit cells with capacitance $C_r$ and inductance $L_r$, as illustrated in the figure. For simplicity, we use the same number of $N$ unit cells for both the transmission line resonator (blue) and the CRLH transmission line (orange), while we use periodic boundary conditions for the CRLH transmission line and open boundary conditions for the probe resonator.

With the couplers in Fig. S4 replaced by their linearized versions shown in Fig. S5 (b), the Lagrangian for the full system can be written in the form

$$L_{\text{lin}} = \frac{1}{2}\dot{\vec{\phi}} C \dot{\vec{\phi}} - \frac{1}{2}\vec{\phi} L^{-1} \vec{\phi}, \tag{38}$$

where $\vec{\phi} = [\phi_0, \phi_1, \ldots, \phi_{3N+1}]$ is a vector of node fluxes, and the capacitance matrix $C$ and inductance matrix $L^{-1}$ can be found using the standard circuit node approach [4]. There are in total $4N$ nodes in the circuit. We label the nodes such that the $n = 0, \ldots, 2N-1$ corresponds to the $N$ unit cells of the CRLH metamaterial (two nodes per unit cell), $n = 2N, \ldots, 3N-1$ are coupler nodes and the last $3N, \ldots, 4N-1$ nodes are for the probe resonator.

Next, introduce normal modes via $\vec{\phi} = C^{-1/2} P \vec{\xi}$, with $P$ an orthogonal matrix that diagonalizes $M \equiv C^{-1/2} L^{-1} C^{-1/2}$, i.e., $P^T M P = \text{diag}[\omega_0^2, \omega_1^2, \ldots, \omega_{4N}^2]$. In terms of these normal modes, the Lagrangian becomes

$$L_{\text{lin}} = \sum_{l=0}^{4N-1} \frac{\dot{\xi}_l^2}{2} - \frac{1}{2}\omega_l^2 \xi_l^2, \tag{39}$$

and the Hamiltonian is easily found

$$H_{\text{lin}} = \sum_{l=0}^{4N-1} \frac{q_{\xi_l}^2}{2} + \frac{1}{2}\omega_l^2 \xi_l^2, \tag{40}$$

where $q_{\xi_l} = \partial L_{\text{lin}} / \partial \dot{\xi}_l$.

The normal mode frequencies $\omega_l$ are found by numerical diagonalization. In Fig. S6 we show the numerically computed normal modes $\omega_l$ for metamaterial parameters given in table S1. We are restricted from using too many unit cells due to the correspondingly large size of the matrices involved in the diagonalization. We use a total of $N = 1000$ unit cells in the numerics, for illustration, but note that much larger numbers up to tens of thousands might be feasible in experiments.

## 2.3 Density of modes

The density of modes $\rho(\omega)$ in the CRLH band (orange in Fig. S6) can be approximated by binning the normal mode frequencies and using

$$\rho(\omega_i^*) \simeq \frac{n_i}{\delta \omega_i}, \tag{41}$$

with $\{\omega_i\}$ a partition of the frequency band, $\delta_i = \omega_i - \omega_{i-1}$, $\omega_i^* = (\omega_i + \omega_{i-1})/2$ and $n_i$ the number of modes found in the $i$th interval. We can compare this to the density of modes in the continuum

| Set | I | II | III | IV |
|---|---|---|---|---|
| | (high $Z_\text{tml}$ ideal) | (high $Z_\text{tml}$ non-ideal) | (low $Z_\text{tml}$ ideal) | (low $Z_\text{tml}$ non-ideal) |
| CRLH | | | | |
| $C_g$ | 1.59 pF | 1.59 pF | 6.37 pF | 6.37 pF |
| $C_n$ | 15.9 fF | 15.9 fF | 63.7 fF | 63.7 fF |
| $L_g$ | 0.637 nH | 0.637 nH | 0.159 nH | 0.159 nH |
| $L_n$ | 63.7 nH | 63.7 nH | 15.9 nH | 15.9 nH |
| resonator | | | | |
| $\omega_0/(2\pi)$ | 7.14 GHz | 7.14 GHz | 7.14 GHz | 7.14 GHz |
| $Z_r$ | 7.0 $\Omega$ | 7.0 $\Omega$ | 7.0 $\Omega$ | 7.0 $\Omega$ |
| couplers | | | | |
| $C_c$ | 10 fF | 10 fF | 10 fF | 10 fF |
| $\omega_s/(2\pi)$ | 7 GHz | 7 GHz | 7 GHz | 7 GHz |
| $Z_s$ | 50 $\Omega$ | 50 $\Omega$ | 200 $\Omega$ | 200 $\Omega$ |
| $I_s$ | 1.1 $\mu$A | 1.1 $\mu$A | 1.1 $\mu$A | 1.1 $\mu$A |
| $C_x$ | 0 | 1 fF | 0 | 1 fF |
| $L_x$ | $\infty$ | 2.0 nH | $\infty$ | 2.0 nH |
| derived params | | | | |
| $Z_\text{tml}$ | 200 $\Omega$ | 200 $\Omega$ | 50 $\Omega$ | 50 $\Omega$ |
| $\omega_g/(2\pi)$ | 5 GHz | 5 GHz | 5 GHz | 5 GHz |
| $\omega_n/(2\pi)$ | 5 GHz | 5 GHz | 5 GHz | 5 GHz |
| $\Omega_L/(2\pi)$ | 4.52 GHz | 4.52 GHz | 4.52 GHz | 4.52 GHz |
| $\Omega_H/(2\pi)$ | 5.52 GHz | 5.52 GHz | 5.52 GHz | 5.52 GHz |

Table S1: Example JTWPD parameter sets used in the numerics with $N = 1000$ unit cells. For the resonator, $\omega_0 = \pi/\sqrt{L_\Sigma C_\Sigma}$ and $Z_r = \sqrt{L_\Sigma/C_\Sigma}$ are the bare frequency and impedance of the fundamental mode, with $L_\Sigma = NL_r$ and $C_\Sigma = N(C_r + C_c)$. For the couplers $\omega_s = 1/\sqrt{L_s C_s}$ and $Z_s = \sqrt{L_s/C_s}$, and $I_s = E_Q/\varphi_0$, with $E_Q$ the effective nonlinearity of the coupler Eq. (37).

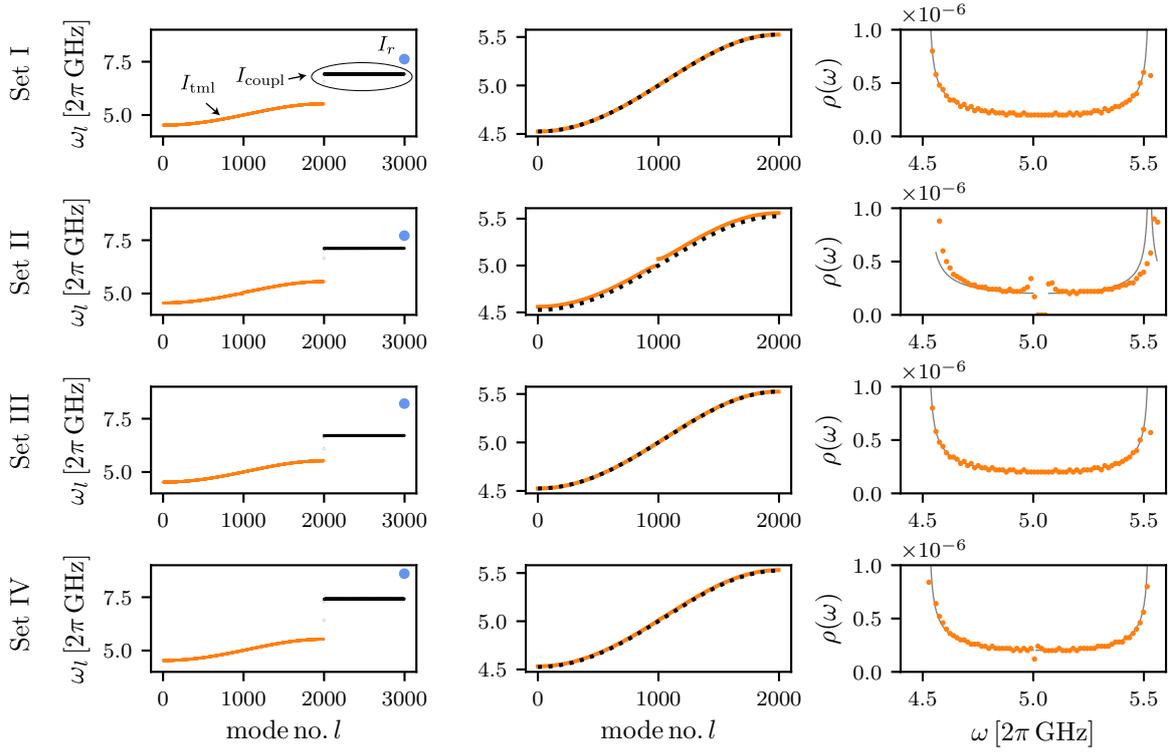

Figure S6: Normal modes of the JTWPD found by numerical diagonalization with $N = 1000$ unit cells, for the parameter sets in table S1. **Left column:** Normal modes in the CRLH band $I_{\mathrm{tml}}$ (orange), coupler modes $I_{\mathrm{coupl}}$ (black) and the lowest resonator mode $I_r$ (blue). **Middle column:** Normal modes in the CRLH band (orange) compared to the analytical expression for the uncoupled case Eq. (1) (black dotted). **Right column:** Density of modes $\rho(\omega)$ computed by numerically binning frequencies, Eq. (41) (orange), compared to the expected continuum limit form Eq. (42) (gray).

11

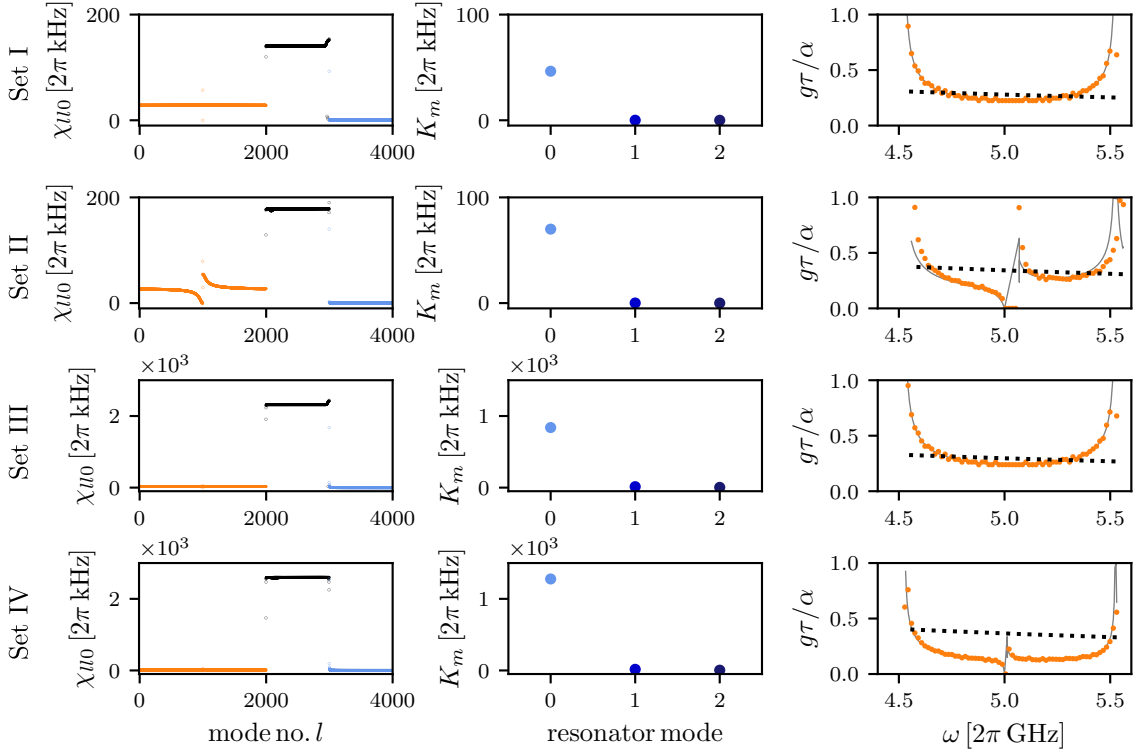

Figure S7: JTWPD nonlinearities for the parameter sets in table S1. **Left column:** Numerically computed cross-Kerr nonlinearity $\chi_{ll0} = \sum_n \chi_{ll0}^{(n)}$, Eq. (53), between the fundamental resonator mode and all other modes. **Middle column:** Resonator mode self-Kerr nonlinearities $K_m$, Eq. (52). **Right column:** JTWPD coupling strength $g\tau/\alpha$ in the narrow photon approximation, Eq. (75), computed using Eq. (41) (orange) and Eq. (42) (gray) for $\rho(\omega)$. The dashed line shows the approximation Eq. (80), which neglects the frequency dependence of the density of modes in the CRLH metamaterial and the spatial dependence of the resonator mode.

limit for the uncoupled CRLH metamaterial, given in Eq. (32),

$$\rho(\omega) \simeq 2 \times \frac{N}{2\pi} \left| \frac{d(ka)}{d\omega} \right|, \tag{42}$$

with $ka$ found from Eq. (1). The two estimates Eqs. (41) and (42) are compared in Fig. S6 for the parameter sets in table S1.

## 2.4 Self- and cross-Kerr nonlinearities

Having diagonalized the linear Hamiltonian $H_{\text{lin}}$, the non-linear contribution due to the couplers in Fig. S5 is approximated to fourth order as [6]

$$H_{\text{nl}} \simeq \sum_{n=0}^{N-1} E_{Q,n} \frac{\varphi_{J,n}^4}{24}, \tag{43}$$

where $\varphi_{J,n} = \phi_{J,n}/\phi_0$ is the dimensionless branch flux across the $n$th coupler. If we denote the corresponding node fluxes by $\phi_n$ and $\phi'_n$ such that $\phi_{J,n} = \phi_n - \phi'_n$ then

$$\varphi_{J,n} = \frac{1}{\phi_0}(\phi_n - \phi'_n) = \frac{1}{\phi_0} \sum_k (R_{n,k} - R_{n',k}) \xi_k. \tag{44}$$

Here we have related bare modes to the normal modes through

$$\vec{\phi} = R\vec{\xi}, \tag{45}$$

where $R = C^{-1/2}P$, with $C$ the capacitance matrix and $P$ the orthogonal matrix found in Sect. 2.2, as before.

Quantization follows by promoting the normal modes to operators

$$\hat{\xi}_l = \sqrt{\frac{\hbar}{2\omega_l}}(\hat{a}_l^\dagger + \hat{a}_l), \tag{46a}$$

$$\hat{q}_{\xi_l} = i\sqrt{\frac{\hbar\omega_l}{2}}(\hat{a}_l^\dagger - \hat{a}_l), \tag{46b}$$

such that

$$\hat{\varphi}_{J,n} = \sum_k \varphi_{n,k}(\hat{a}_k + \hat{a}_k^\dagger). \tag{47}$$

with $\varphi_{n,k} = \sqrt{\frac{4\pi}{R_K \omega_k}}(R_{n_1,k} - R_{n_2,k})$, $R_K = h/e^2$ the quantum of resistance. Using this in Eq. (43) we find

$$\begin{aligned}
\hat{H}_{\text{nl}} &\simeq \sum_{n=0}^{N-1} \frac{E_{Q,n}}{24} \sum_{klpq} \varphi_{n,k}\varphi_{n,l}\varphi_{n,p}\varphi_{n,q}(\hat{a}_k + \hat{a}_k^\dagger)(\hat{a}_l + \hat{a}_l^\dagger)(\hat{a}_p + \hat{a}_p^\dagger)(\hat{a}_q + \hat{a}_q^\dagger) \\
&\simeq \sum_{n=0}^{N-1} \frac{E_{Q,n}}{4} \left[ \sum_{klpq} \varphi_{n,k}\varphi_{n,l}\varphi_{n,p}\varphi_{n,q}\hat{a}_k^\dagger \hat{a}_l^\dagger \hat{a}_p \hat{a}_q + 2\sum_{klp} \varphi_{n,k}\varphi_{n,l}\varphi_{n,p}^2 \hat{a}_k^\dagger \hat{a}_l \right],
\end{aligned} \tag{48}$$

where in the last line we have dropped all terms that do not conserve the overall excitation number as well as a constant term.

In analyzing the above Hamiltonian we can qualitatively distinguish between three sets of modes: dressed metamaterial modes indexed by the set $I_{\text{tml}}$, primarily localized in the CRLH metamaterial with dressed frequences in a dense band (orange in Fig. S6), dressed coupler modes indexed by $I_{\text{coupl}}$ at frequencies near the coupler resonance frequency $\omega_s = 1/\sqrt{C_s L_s}$ (black in Fig. S6), and dressed resonator modes indexed by $I_r$ (blue in Fig. S6). The coupler modes are far off resonance from both the resonator and the metamaterial modes, and are thus only virtually populated. We therefore assume these modes to be in the vacuum state and drop them from the Hamiltonian.

Moreover, as the population of any metamaterial mode is much smaller than one for a single incoming photon (or even a few incoming photons), we can neglect any cross and self-Kerr interaction between them. With these approximations we can write

$$\hat{H}_{\text{nl}} \simeq \sum_{m \in I_r} \left( \hbar \Delta_m \hat{a}_m^\dagger \hat{a}_m + \frac{\hbar K_m}{2} \hat{a}_m^{\dagger 2} \hat{a}_m^2 \right) + \sum_{kl \in I_{\text{tml}}} \hbar \Delta_{kl} \hat{a}_k^\dagger \hat{a}_l$$
$$+ \sum_{m > m' \in I_r} \hbar \chi_{mm'} \hat{a}_m^\dagger \hat{a}_m \hat{a}_{m'}^\dagger \hat{a}_{m'} \qquad (49)$$
$$+ \sum_{n=0}^{N-1} \sum_{m \in I_r} \sum_{kl \in I_{\text{tml}}} \chi_{klm}^{(n)} \hat{a}_k^\dagger \hat{a}_l \hat{a}_m^\dagger \hat{a}_m,$$

where we drop a constant term as well as any term that does not conserve the photon number of each resonator mode $m \in I_r$. The latter approximation is based on the assumption that all the resonator modes are well separated in frequency. We have moreover defined

$$\hbar \Delta_{kl} = \frac{1}{2} \sum_{n=0}^{N-1} \sum_p E_{Q,n} \varphi_{n,k} \varphi_{n,l} \varphi_{n,p}^2, \qquad (50)$$

$$\hbar \chi_{mm'} = \sum_n E_{Q,n} \varphi_{n,m}^2 \varphi_{n,m'}^2, \qquad (51)$$

$$\hbar K_m = \frac{1}{2} \sum_{n=0}^{N-1} E_{Q,n} \varphi_{n,m}^4 = \frac{\chi_{mm}}{2}, \qquad (52)$$

$$\hbar \chi_{klm}^{(n)} = E_{Q,n} \varphi_{n,k} \varphi_{n,l} \varphi_{n,m}^2 = \hbar \sqrt{\chi_{kkm}^{(n)}} \sqrt{\chi_{llm}^{(n)}}, \qquad (53)$$

and $\Delta_m = \Delta_{mm}$.

The terms proportional to $\Delta_{kl}$ can be absorbed into the linear part of the Hamiltonian $H_{\text{lin}}$ which can be diagonalized again with a new unitary transformation of the modes. However, for the parameter regimes we are interested in $\Delta_{kl}$ is very small, and we therefore drop this linear term for simplicity. The full Hamiltonian can then be written

$$\hat{H} = \hat{H}_{\text{lin}} + \hat{H}_{\text{nl}} \simeq \hat{H}_0 + \hat{H}_r + \hat{H}_{\text{int}}, \qquad (54)$$

where

$$\hat{H}_0 = \sum_{l \in I_{\text{tml}}} \hbar \omega_l \hat{a}_l^\dagger \hat{a}_l, \qquad (55)$$

$$\hat{H}_r = \sum_{m \in I_r} \hbar \omega_m \hat{a}_m^\dagger \hat{a}_m + \frac{\hbar K_m}{2} \hat{a}_m^{\dagger 2} \hat{a}_m^2, \qquad (56)$$

$$\hat{H}_{\text{int}} = \sum_{n=0}^{N-1} \sum_{m \in I_r} \sum_{kl \in I_{\text{tml}}} \chi_{klm}^{(n)} \hat{a}_k^\dagger \hat{a}_l \hat{a}_m^\dagger \hat{a}_m. \qquad (57)$$

For notational simplicity we have here absorbed the frequency shift $\Delta_m$ into a redefinition of $\omega_m$, i.e., $\omega_m + \Delta_m \to \omega_m$, such that $\omega_m$ now refers to the measurable dressed frequencies of the resonator, including shifts due to the non-linear part of the couplers. We show cross- and self-Kerr non-linearities estimated from Eqs. (52) and (53) in Fig. S7 for the parameter sets in table S1.

Following similar steps as in Sect. 1.2 we can consider a large $N$ limit where the CRLH frequency band becomes dense. As $N \to \infty$, the sums over $I_{\text{tml}}$ in the above expressions approach integrals, leading to

$$\hat{H}_0 = \sum_{\nu = \pm} \int_\Omega d\omega \, \hbar \omega \hat{b}_{\nu\omega}^\dagger \hat{b}_{\nu\omega}, \qquad (58)$$

$$\hat{H}_{\text{int}} = \sum_{n=0}^{N-1} \sum_{\nu,\mu = \pm} \sum_{m \in I_r} \int_\Omega d\omega d\omega' \chi_{\nu\mu m}^{(n)}(\omega, \omega') \hat{b}_\nu^\dagger(\omega) \hat{b}_\mu(\omega') \hat{a}_m^\dagger \hat{a}_m, \qquad (59)$$

14

where, as in Sect. 1.2, the range of integration $\Omega$ is the band of CRLH normal modes, we have introduced left- and right movers index by $\nu = \pm$, and defined

$$\chi^{(n)}_{\nu\mu m}(\omega_k, \omega_l) = E_{Q,n}\sqrt{\rho_\mu(\omega_k)\rho_\nu(\omega_l)}\varphi^*_{n,\mu k}\varphi_{n,\nu l}\varphi^2_{n,m}, \tag{60}$$

with $\rho_\mu(\omega) = \rho(\omega)/2$ the density of left/right moving modes. Finally taking a large wavelength limit where $a \to 0$ the sum over $n$ approaches an integral over space

$$\hat{H}_{\text{int}} = \sum_{\nu,\mu=\pm}\sum_{m\in I_r}\int_0^z dx \int_\Omega d\omega d\omega' \chi_{\nu\mu m}(x;\omega,\omega')\hat{b}^\dagger_\nu(\omega)\hat{b}_\mu(\omega')\hat{a}^\dagger_m \hat{a}_m, \tag{61}$$

where $\chi_{\nu\mu m}(x_n;\omega,\omega') = \chi^{(n)}_{\nu\mu m}(\omega,\omega')/a$ is a dispersive shift per unit cell. We postpone a discussion of numerically estimating the non-linear coupling between the resonator and the metamaterial until Sect. 2.7.

## 2.5 Blackbox quantization for a general geometry

In this section, we outline a blackbox quantization approach [6] for an arbitrary geometry, generalizing the results from the preceding sections. The JTWPD design illustrated in Fig. S8 is one possible realization of this device. Other realizations, such as using a 3D cavity, are also possible. In the following we outline a general formalism leading to the JTWPD Hamiltonian used in the main paper.

To find a Hamiltonian for the metamaterial, we divide the Hamiltonian into a linear and a non-linear contribution, as in the previous section

$$\hat{H} = \hat{H}_{\text{lin}} + \hat{H}_{\text{nl}}, \tag{62}$$

where $\hat{H}_{\text{lin}}$ is the Hamiltonian of the linearized system which includes the junction's electromagnetic environment as well as the capacitances and linear inductance of the junctions themselves [6]. On the other hand, $\hat{H}_{\text{nl}}$ describes the non-linear contribution of the couplers, which to lowest order can be expressed as in Eq. (43).

Whether we are considering a 2D or 3D setup, the Hamiltonian $\hat{H}_{\text{lin}}$ can be diagonalized using of a set of eigenmodes with, in general, both discrete modes and continuous bands [7]

$$\hat{H}_{\text{lin}} = \sum_m \hbar\omega_m \hat{a}^\dagger_m \hat{a}_m + \sum_\nu \int_\Omega d\omega \hbar\omega\, \hat{b}^\dagger_\nu(\omega)\hat{b}_\nu(\omega), \tag{63}$$

where $\Omega$ refers to the continuous part of the spectrum. The index $\nu$ is used to label degenerate modes, in particular left- and right-movers, as before. A generic eigenspectrum is illustrated in Fig. S9. In the case of interest, the continuum modes are associated to the metamaterial waveguide at the bottom of Fig. S8, treated here in the continuum limit, and the discrete spectrum is associated to the probe resonator.

In the lumped element limit, we separate the $n$th coupler element into capacitive, linear inductive and non-linear inductive elements as illustrated in Fig. S10, where the spider symbol is used to represent the purely non-linear potential of the coupler, while $C_{x,n}$ and $L_{x,n}$ represent any stray linear capacitance and inductance [6]. Note that in contrast to a conventional Josephson junction, the quartic nonlinearity represented by the spider symbol has a positive prefactor, Eq. (36).

To express the non-linear part of the Hamiltonian we first need to find the dimensionless flux across the $n$th nonlinear element located at $\mathbf{x}_n$, defined as $\hat{\varphi}_{s,n}(t) = (1/\phi_0)\int_{-\infty}^t dt' V_n(t')$ with $V_n(t)$ the branch voltage across the $n$th element, $V_n(t) = \int_{\mathbf{x}_n}^{\mathbf{x}'_n} d\mathbf{l}\cdot\hat{\mathbf{E}}(\mathbf{x})$, with $\{\mathbf{x}_n,\mathbf{x}'_n\}$ as indicated in Fig. S10. Here $\mathbf{E}(\mathbf{x})$ is the electric field in space [4]. Since $\mathbf{E}(\mathbf{x})$ can be expressed in a set of mode functions and bosonic creation and annihilation operators [7], it follows that the flux across the $n$th nonlinear element has the general form

$$\hat{\varphi}_{s,n} = \sum_m [\varphi_{n,m}\hat{a}_m + \text{H.c.}] + \sum_\nu \int_\Omega d\omega \left[\varphi_{n,\nu}(\omega)\hat{b}_\nu(\omega) + \text{H.c.}\right], \tag{64}$$

15

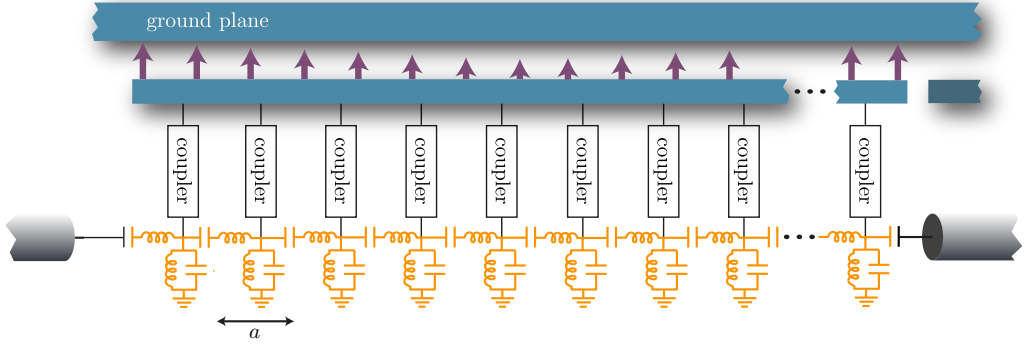

Figure S8: The detector consists of three parts. 1) A CRLH metamaterial transmission line forms the backbone of the detector, and serves effectively as a linear waveguide over some frequency range (orange). 2) A probe resonator, here shown as a transmission line resonator (blue). 3) The probe (resonator) is coupled to the waveguide (CRLH metamaterial) via an array of nonlinear couplers (black).

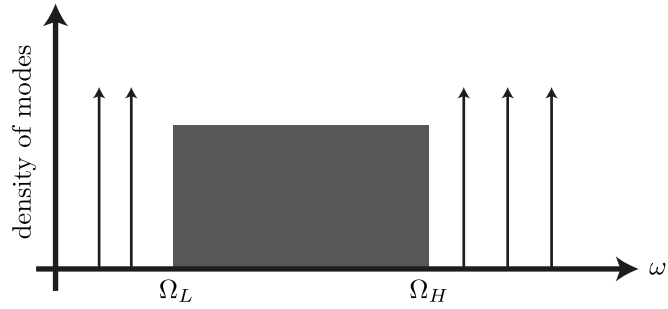

Figure S9: Schematic illustration of generic spectrum consisting of both discrete and continuum modes.

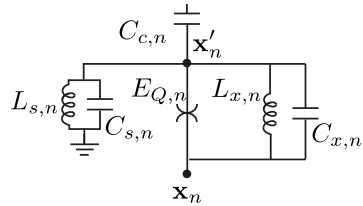

Figure S10: Representation of the coupler element in terms of capacitive, linear inductive, and and non-linear inductive elements. In this figure, the spider symbol represent the purely non-linear quartic potential, Eq. (36), and $C_{x,n}$, $L_{x,n}$ represent stray capacitance and inductance. The coordinates $\mathbf{x}_n$ and $\mathbf{x}'_n$ marks the physical location in space of the superconducting islands on the two respective sides of the purely nonlinear element.

where $\varphi_{n,m}$ and $\varphi_{n,\nu}(\omega)$ are zero-point fluctuations for discrete and continuum mode flux biases across the $n$th nonlinear element, respectively.

Using Eq. (64) we can express $\hat{H}_{\rm nl}$ in terms of boson operators $\hat{a}_m$, $\hat{b}_\nu(\omega)$. With the same set of approximations argued for in Eq. (49), we arrive at the general result

$$\hat{H}_{\rm nl} \simeq \sum_m \left[\hbar\Delta_m \hat{a}_m^\dagger \hat{a}_m + \frac{\hbar K}{2}\hat{a}_m^{\dagger 2}\hat{a}_m^2\right] + \sum_{m>n} \hbar\chi_{mn}\hat{a}_m^\dagger \hat{a}_m \hat{a}_n^\dagger \hat{a}_n + \hat{H}_{\rm int}, \qquad (65)$$

where

$$\hat{H}_{\rm int} = \sum_{n=0}^{N-1}\sum_m \sum_{\nu\mu}\int d\omega d\omega' \hbar\chi_{\nu\mu m}^{(n)}(\omega,\omega')\hat{b}_\nu^\dagger(\omega)\hat{b}_\mu(\omega')\hat{a}_m^\dagger \hat{a}_m, \qquad (66)$$

$\Delta_m = \frac{1}{2}\sum_{m'}\chi_{mm'} + \frac{1}{2}\sum_n\sum_\nu\int d\omega\chi_{\nu\nu m}^{(n)}(\omega,\omega)$, $K_m = \chi_{mm}/2$ and

$$\hbar\chi_{mm'} = \sum_n E_{Q,n}|\varphi_{n,m}|^2|\varphi_{n,m'}|^2, \qquad (67)$$

$$\hbar\chi_{\nu\mu m}^{(n)}(\omega,\omega') = E_{Q,n}\varphi_{n,\nu}^*(\omega)\varphi_{n,\mu}(\omega')|\varphi_{n,m}|^2. \qquad (68)$$

In Eq. (65), we have dropped terms that do not conserve the overall excitation number, fast-rotating terms for the discrete modes based on the assumption that they are well separated in frequency, and non-linear self- and cross-Kerr couplings between continuum modes. We have also ignored a linear term mixing different continuum modes. As argued for in Sect. 2.4, this term can be absorbed into $\hat{H}_{\rm lin}$, which can be re-diagonalized in a second step. This will lead to shifts of the parameters in Eq. (65), but preserves the general form.

## 2.6 Narrow photon approximations

At frequencies supporting traveling waves we assume a spatial dependence

$$\varphi_{n,\nu}(\omega) = |\varphi_n(\omega)|e^{\nu i k_\omega x_n + i\theta_{\nu\omega}}, \qquad (69)$$

for the flux field. The position-independent phase factor $\theta_{\nu\omega}$ can be absorbed into a redefinition of the annihilation operators $\hat{b}_\nu(\omega)$, and can therefore be ignored. A simplified expression for $\hat{H}_{\rm int}$ can be derived in the situation where the photon number distribution of the continuum is zero except in a reasonably narrow range of frequencies, as one would typically expect for a single incoming photon. We can then Taylor expand the wavevector around the center frequency $\bar{\omega}$

$$k_\omega \simeq k_{\bar{\omega}} + \frac{1}{v_g}(\omega - \bar{\omega}), \qquad (70)$$

where $v_g = v_g(\bar{\omega}) = \left|\frac{\partial k_\omega}{\partial \omega}\right|_{\bar{\omega}}^{-1}$. Moreover, it is reasonable to assume that $|\varphi_{n,\nu}(\omega)| \simeq |\varphi_{n,\nu}(\bar{\omega})| = |\varphi_n(\bar{\omega})|$ can be taken to be approximately independent of frequency (and $\nu$) over the relevant range. With these approximations in mind, it is convenient to introduce position-dependent annihilation operators

$$\hat{b}_\nu(x) = \sqrt{\frac{1}{2\pi v_g}}\int_\Omega d\omega e^{\nu i\omega x/v_g}\hat{b}_\nu(\omega), \qquad (71)$$

such that we can write in a continuum limit where $a \to 0$

$$\hat{H}_{\rm int} = \sum_m \sum_{\nu\mu}\int_0^z dx\hbar\chi_m(x_n)\hat{b}_\nu^\dagger(x)\hat{b}_\mu(x)\hat{a}_m^\dagger \hat{a}_m, \qquad (72)$$

where we have defined

$$\hbar\chi_m(x_n) = \frac{2\pi v_g}{a}E_{Q,n}|\varphi_n(\bar{\omega})|^2|\varphi_m(x_n)|^2 \qquad (73)$$

Finally, the Hamiltonian used in the main paper is found from Eqs. (65) and (72) by taking a single-mode approximation for the probe resonator, including only one term from the sum over $m$, and adding a drive term for this resonator mode.

## 2.7 Dimensionless JTWPD coupling strength $g\tau$

As discussed in the main paper, a key figure of merit for the JTWPD is the dimensionless number

$$g\tau \equiv \frac{2\alpha}{v_g} \times \int_0^z dx \chi_m(x) \simeq \frac{2\alpha \chi_m z}{v_g}, \tag{74}$$

where $\alpha$ is the displacement of the probe resonator, $m$ labels the resonator mode used as a probe, and the last approximation is for the simplification used in the main paper where the spatial dependence of $\chi_m(x)$ is neglected. Here we are assuming that the detector is used in reflection mode, effectively doubling $\tau$. Using Eq. (73) we can approximate

$$\frac{g\tau}{\alpha} = \frac{2}{v_g} \int_0^z dx \chi_m(x) = 4\pi \sum_{n=0}^{N-1} \frac{E_{Q,n}}{\hbar} \frac{\rho(\omega_k)}{2} |\varphi_{n,k}|^2 |\varphi_{n,m}|^2, \tag{75}$$

with $\bar{\omega} = \omega_k$ and $\rho(\omega) = 2\rho_\pm(\omega)$ the density of modes (counting both left- and right-movers), as before. To estimate this quantity numerically using the circuit model illustrated in Fig. S4, we use the approach in Sect. 2.2 to compute $\varphi_{n,k}$ and $\omega_k$. The density of modes $\rho(\omega)$ can be approximated by binning the normal mode frequencies in the CRLH band and using Eq. (41) or Eq. (42) with $ka$ found from Eq. (1). The last column of Fig. S7 shows the estimated $g\tau$ from Eq. (75) as a function of frequency for the parameter sets in table S1.

We can also find a simplified expression for $g\tau$ if we neglect any spatial dependence in $E_{Q,n} \equiv E_Q$, $\varphi_{n,k} \equiv \varphi(\bar{\omega})/\sqrt{\rho(\omega_k)}$ and $\varphi_m(x_n) \equiv \varphi_m$. This gives

$$\frac{g\tau}{\alpha} \simeq 4\pi \times |\varphi(\bar{\omega})|^2 \times \frac{N E_Q}{\hbar} |\varphi_m|^2. \tag{76}$$

Next using, from Eq. (67) that the resonator's Kerr nonlinearity is within this approximation given by

$$K_Q \simeq \frac{N E_Q}{2\hbar} |\varphi_m|^4, \tag{77}$$

we can write

$$\frac{g\tau}{\alpha} \simeq 4\pi \times |\varphi(\bar{\omega})|^2 \times \sqrt{N E_Q/\hbar} \sqrt{2 K_Q}. \tag{78}$$

In the last two expressions we have dropped the subscript $m$ for the resonator mode, for notational simplicity. Assuming the continuum density of modes of an uncoupled CRLH, we can moreover approximate [c.f. Eq. (12)]

$$\varphi(\bar{\omega}) \simeq \frac{1}{\phi_0} \sqrt{\frac{\hbar Z_{\text{tml}}}{4\pi \bar{\omega}}} = \sqrt{\frac{2 Z_{\text{tml}}}{R_K \bar{\omega}}}, \tag{79}$$

We thus get the result

$$\frac{g\tau}{\alpha} \simeq 8\pi \times \frac{Z_{\text{tml}}}{R_K \bar{\omega}} \times \sqrt{N E_Q/\hbar} \sqrt{2 K_Q}. \tag{80}$$

This approximation is shown as dotted black lines in Fig. S7 where the fundamental resonator mode is used as the probe mode.

Inverting this we find an expression for the number of unit cells

$$N \simeq \frac{1}{2} \left( \frac{g\tau}{\alpha} \times \frac{R_K}{8\pi Z_{\text{tml}}} \right)^2 \frac{\bar{\omega}^2}{K_Q E_Q/\hbar}. \tag{81}$$

This expression can be used to ball-park the number of unit cells needed to reach a certain value of $g\tau$. It also shows the inverse relation between $N$ and $K_Q$, such that by using a larger number of unit cells one can reduce the self-Kerr nonlinearity. In Fig. S11 we plot Eq. (81) as a function of $K_Q$ for representative choices of parameters.

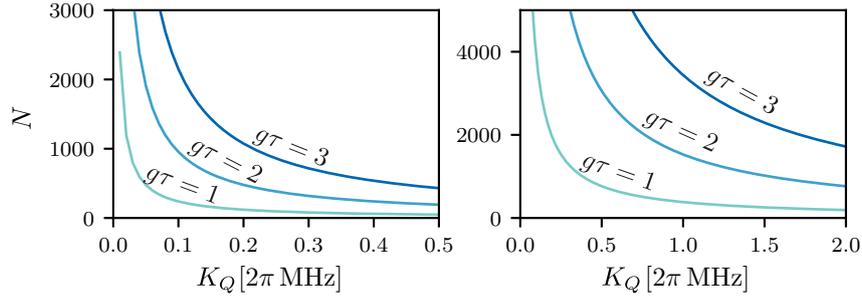

Figure S11: Estimated number of unit cells using Eq. (81) with coupler parameters $I_c = 5$ nA $n_s = 3$, drive strength $\alpha = 5$, and center frequency $\bar{\omega}/(2\pi) = 5.0$ GHz for three different $g\tau = 1, 2, 3$ and varying resonator self-Kerr nonlinearity $K_Q$. On the left $Z_{\text{tml}} = 200\,\Omega$ and on the right $Z_{\text{tml}} = 50\,\Omega$.

## 3 JTWPD characteristics

### 3.1 Controlling the coupling strength $g\tau$

As is clear from Eq. (80) the JTWPD coupling $g$ grows with $\sqrt{NE_Q}$. At a fixed $N$ and $E_Q$, the coupling can be increased by increasing $Z_{\text{tml}}$ and/or the resonator self-Kerr non-linearity $K_Q$. The latter is in turn controllable in two distinct ways. First, it depends on the coupling capacitance $C_c$ relative to the detuning of the resonator mode and the coupler modes. Second, at fixed $C_c$ and detuning, the self-Kerr can be increased or decreased by increasing or decreasing the characteristic impedance $Z_s$ of the coupler modes, respectively.

Since since $K_Q \sim Z_s^2 Z_r^2$ [6], we in fact see that

$$g \sim Z_{\text{tml}} Z_s Z_r. \tag{82}$$

It is generally more practical to increase $Z_s$ than $Z_r$, in particular since the additional coupling capacitances $C_{c,n}$ will tend to decrease the impedance of the resonator. For this reason we have used a low $Z_r = 7\,\Omega$ in our numerical examples above, c.f. table S1. Eq. (82) moreover explains why the results for Set I and Set III (Set II and Set VI) in Fig. S7 are practically identical, since the product $Z_{\text{tml}} \times Z_s = 50 \times 200\,\Omega^2$ is the same in both cases.

There are thus two options for increasing $g$ at fixed $N$, $E_Q$, $C_c$ and $Z_r$: Increasing $Z_{\text{tml}}$ or increasing $Z_s$. For many applications it is desirable to have $Z_{\text{tml}} = 50\,\Omega$, but impedance transformers can also be used, and some applications do not require a $50\,\Omega$ environment. On the other hand, the only downside to increasing $Z_s$ is that it increases $K_Q$, the self-Kerr non-linearity of the resonator. As exemplified in Fig. S7, we expect $K_Q/(2\pi)$ in the 10–100 kHz range for $Z_s = 50\,\Omega$ and in the MHz range for $Z_s = 200\,\Omega$. In the numerical results in the main paper we assumed $K/\kappa_a = 10^{-2}$ for $K$ the *total* Kerr nonlinearity of the probe mode. As discussed in more detail below, this will likely necessitate canceling the positive self-Kerr $K_Q$ arising from the nonlinear couplers by a negative self-Kerr.

### 3.2 The Kerr non-linear resonator

Adding a drive term and keeping only a single mode, the resonator Hamiltonian is

$$\hat{H}_r/\hbar = \omega_r \hat{a}^\dagger \hat{a} + \frac{K}{2}\left(\hat{a}^\dagger\right)^2 \hat{a}^2 + \left(i\varepsilon e^{-i\omega_d t}\hat{a}^\dagger + \text{H.c.}\right), \tag{83}$$

where $\varepsilon$ is the drive strength and $\omega_d$ the drive frequency. Here $K$ now refers to the *total* Kerr non-linearity of the resonator, which might have contributions from other sources than $K_Q$, the Kerr nonlinearity due to the metamaterial. Damping at a rate $\kappa_a$ is taken into account through the master equation

$$\dot{\rho} = -\frac{i}{\hbar}[\hat{H}, \rho] + \kappa_a \mathcal{D}[\hat{a}]\rho, \tag{84}$$

19

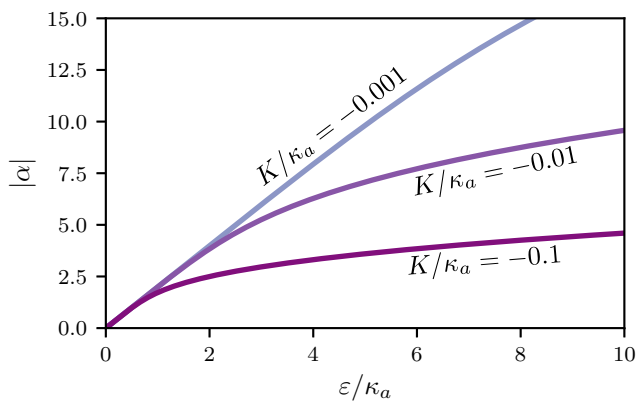

Figure S12: Displacement of a Kerr non-linear resonator driven at the shifted resonance $\omega_d = \omega_0 + 2K|\alpha|^2$ as a function of drive strength, for three different values of $K/\kappa_a$.

where $\mathcal{D}[\hat{a}]\rho \equiv \hat{a}\rho\hat{a}^\dagger - \frac{1}{2}\hat{a}^\dagger\hat{a}\rho - \frac{1}{2}\rho\hat{a}^\dagger\hat{a}$. Moving first to a frame rotating at the drive frequency $\omega_d$ and subsequently doing a displacement transformation $\hat{a} \to \hat{a} + \alpha$ we can write $\hat{H}_r$ in the new frame

$$\hat{H}'_r/\hbar \approx (\delta + 2K|\alpha|^2)\hat{a}^\dagger\hat{a} + \frac{K}{2}\left(\hat{a}^\dagger\right)^2\hat{a}^2, \tag{85}$$

where $\delta = \omega_r - \omega_d$, and we neglect fast rotating terms. The master equation retains its original form and $\alpha$ is chosen to satisfy

$$(\delta + K|\alpha|^2)\alpha - \frac{i\kappa_a}{2}\alpha + i\varepsilon = 0. \tag{86}$$

We wish to drive the resonator on resonance, taking the dynamic frequency shift due to the Kerr non-linearity into account, and consequently choose $\omega_d$ such that $\delta = -2K|\alpha|^2$. $\hat{H}'_r$ then reduces to

$$\hat{H}'_r = \frac{\hbar K}{2}\left(\hat{a}^\dagger\right)^2\hat{a}^2, \tag{87}$$

while the non-linear equation for $\alpha$ becomes

$$K|\alpha|^2\alpha + \frac{i\kappa_a}{2}\alpha = i\varepsilon. \tag{88}$$

For $K|\alpha|^2 \ll \kappa_a$ the solution is approximately $\alpha = 2\varepsilon/\kappa_a$ and the steady state of the resonator is to a good approximation the coherent state $|\alpha\rangle$. In the opposite limit, however, the solution goes as $|\alpha| \sim (\varepsilon/K)^{1/3}$, and the steady state is highly non-Gaussian due to the non-linear Hamiltonian $\hat{H}'_r$. The solution for $|\alpha|$ for different values of $K/\kappa_a$ is shown in Fig. S12.

### 3.2.1 Cancelling the Kerr non-linearity

Due to the large number of non-linear couplers in the JTWPD, the Kerr non-linearity of the resonator can be large and thus might reduce the fidelity of the detector. This effect can be mitigated by noting that the Kerr non-linearity $K_Q$ is always positive for the coupler in Fig. S5 [c.f. Eq. (67)]. One can therefore balance it out by introducing an additional negative Kerr non-linearity. This is straight forwardly achieved by galvanically or capacitively coupling the resonator to a Josephson junction, since the quartic potential of a Josephson junction

$$U_J(\hat{\varphi}) = -E_J \cos\hat{\varphi} \simeq -E_J + \frac{E_J}{2}\hat{\varphi}^2 - \frac{E_J}{24}\hat{\varphi}^4 + \ldots, \tag{89}$$

will give rise to a negative Kerr non-linearity, $K_J < 0$. The magnitude $|K_J|$ can easily be made large by galvanic coupling, or alternatively by using several junctions capacitively coupled along resonator (less than one per JTWPD unit cell is needed, since a stronger non-linearity can be used). Moreover, by using a flux-tunable $E_J$ realized using two junctions enclosing an external flux [8], the total self-Kerr can be tuned to zero in situ, $K = K_Q + K_J \simeq 0$.

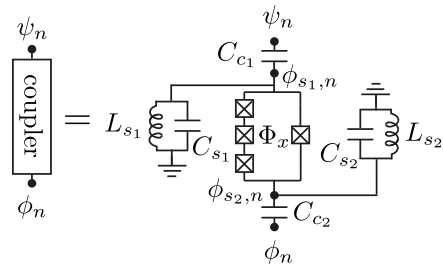

Figure S13: A modification of the coupler element where the detector bandwidth can be controlled via a coupling capacitance $C_{c_2}$, with a center frequency around $\omega_{s_2} = 1/\sqrt{L_{s_2} C_{s_2}}$. Maximizing the tradeoff between cross-Kerr coupling and resonator self-Kerr can be done by having a larger $Z_{s_2} = \sqrt{L_{s_2}/C_{s_2}}$ and smaller $Z_{s_1} = \sqrt{L_{s_1}/C_{s_1}}$

### 3.3 Controlling detector bandwidth

The large bandwidth distinguishes the JTWPD from other single-photon detector proposals in the microwave domain. The origin of this large bandwidth is the galvanic coupling of the coupler elements to the metamaterial, which ensures that all modes in the metamaterial frequency band couples strongly to the detector.

However, if the bandwidth is too large, thermal photons will dominate the detector count. Depending on application, limiting the bandwidth is therefore desirable. The use of a CRLH metamaterial is here practical, because the bandwidth can be adjusted by varying $c/l = C_n L_g / C_g L_n$. As discussed in Sect. 1.1, the bandwidth vanishes in the limit $c/l \to 0$. In the parameter sets in table S1 a value of $c/l = 10^{-2}$ was used, giving GHz bandwidths. In table S2 we show two alternative parameter sets, similar to Set III, where the center frequency is higher (sets V, VI) and $c/l = 10^{-3}$ (set VI), to alleviate thermal photon counts.

Another way to reduce the detector bandwidth, which might be simpler than realizing a CRLH with small $c/l$, is to change the coupler element. Changing from galvanic to capacitive coupling to the metamaterial as illustrated in Fig. S13 leads to a bandwidth that can tuned via the coupling capacitance $C_{c_2}$. The detector response will be maximal close to the frequency $\omega_{s_2} = 1/\sqrt{C_{s_2} L_{s_2}}$ of the coupler mode at the bottom of the circuit in Fig. S13 (given that this frequency lies in the metamaterial frequency band), and falls off away from this frequency with a bandwidth set by $C_{c_2}$. The price to pay is that the coupling to the metamaterial is in general weaker, but this can be compensated for by using a large impedance $Z_{s_2} = \sqrt{L_{s_2}/C_{s_2}}$. On the other hand, the impedance $Z_{s_1} = \sqrt{L_{s_1}/C_{s_1}}$ of the mode on the top side of the coupler in Fig. S13 should be kept lower to reduce the self-Kerr nonlinearity of the resonator mode. In analogy with Eq. (80), we can write an approximate form for the JTWPD coupling strength $g\tau$ in this case as

$$\frac{g\tau}{\alpha} \simeq 8\pi \times \frac{\eta(\bar{\omega}) Z_{s_2}}{R_K \bar{\omega}} \times \sqrt{N E_Q/\hbar} \sqrt{2K}, \qquad (90)$$

where $\eta(\omega) < 1$ is a parameter describing the bias of the nonlinear coupler due to the metamaterial field at frequency $\omega$. With this approach, the CRLH could also be replaced with a more standard lumped element telegrapher model transmission line, as used e.g. in JTWPAs [9] (the telegrapher model can be seen as a limit of a CRLH where $C_n, L_g \to \infty$). Such a realization does not have frequency cutoffs for the metamaterial, and it might therefore be necessary to use Purcell filters to avoid decay of resonator photons via the metamaterial.

### 3.4 Estimation of thermal noise

The response of the detector to a single incoming photon is in general non-Markovian and treated in Sect. 4 using a Keldysh path integral approach, and in Sect. 5 using Matrix Product States. The case of incoming thermal noise, however, can be treated as approximately Markovian. We follow the derivation of Carmichael, Chapter 2.2.4 in Ref. [10].

| Set | V | VI |
|---|---|---|
| CRLH | | |
| $C_g$ | 4.24 pF | 13.4 pF |
| $C_n$ | 42.4 fF | 13.4 fF |
| $L_g$ | 0.106 nH | 33.6 pH |
| $L_n$ | 10.6 nH | 33.6 nH |
| resonator | | |
| $\omega_0/(2\pi)$ | 9.43 GHz | 9.43 GHz |
| $Z_r$ | 5.3 $\Omega$ | 5.3 $\Omega$ |
| couplers | | |
| $C_c$ | 10 fF | 10 fF |
| $\omega_s/(2\pi)$ | 9.2 GHz | 9.2 GHz |
| $Z_s$ | 300 $\Omega$ | 300 $\Omega$ |
| $I_c$ | 5 $\mu$A | 5 $\mu$A |
| $C_x$ | 0 | 0 |
| $L_x$ | $\infty$ | $\infty$ |
| derived params | | |
| $Z_{\text{tml}}$ | 50 $\Omega$ | 50 $\Omega$ |
| $\omega_g/(2\pi)$ | 7.5 GHz | 7.5 GHz |
| $\omega_n/(2\pi)$ | 7.5 GHz | 7.5 GHz |
| $\Omega_L/(2\pi)$ | 6.79 GHz | 7.27 GHz |
| $\Omega_H/(2\pi)$ | 8.29 GHz | 7.74 GHz |

Table S2: Example JTWPD parameter sets similar to Set III in table S1 but with higher CRLH center frequency $\omega_g = \omega_n$. Set VI furthermore have a smaller value of $c/l$ giving a smaller CRLH frequency band.

The starting point is the interaction Hamiltonian

$$\hat{H}_{\text{int}} = \sum_{kl} \chi_{kl} \hat{b}_k^\dagger \hat{b}_l \hat{a}^\dagger \hat{a} \tag{91}$$

where

$$\chi_{kl} = \sum_{n=0}^{N-1} \chi_{kl}^{(n)}, \tag{92}$$

is the total cross-Kerr coupling due to all $N$ nonlinear couplers between metamaterial modes $k, l$ and the probe mode. We assume that each metamaterial mode is in a thermal state. Following Carmichael, we first move the average displacement

$$\delta_p = \sum_k \chi_{kk} \bar{n}_k, \tag{93}$$

with $\bar{n}_k = \langle \hat{b}_k^\dagger \hat{b}_k \rangle$ into the resonator Hamiltonian

$$H_r = (\omega_r + \delta_p)\hat{a}^\dagger \hat{a} + \frac{\hbar K}{2}\hat{a}^{\dagger 2}\hat{a}^2 + (i\varepsilon e^{-i\omega_d t}a^\dagger + \text{H.c.}), \tag{94}$$

such that the interaction becomes

$$\hat{H}_{\text{int}} = \sum_{kl} \chi_{kl}(\hat{b}_k^\dagger \hat{b}_l - \delta_{kl}\bar{n}_k)\hat{a}^\dagger \hat{a}. \tag{95}$$

Moving to a rotating frame with respect to the metamaterial Hamiltonian $\hat{H}_0$ and a rotating and displaced frame for the resonator as defined in Sect. 3.2 we have

$$\hat{H}_{\text{int}} = \hat{B}(t)\hat{A}, \tag{96}$$

22

where

$$\hat{B}(t) = \sum_{kl} \chi_{kl}(\hat{b}_k^\dagger \hat{b}_l - \delta_{kl}\bar{n}_i)e^{i(\omega_k-\omega_l)t}, \quad (97)$$

$$\hat{A} = \hat{a}^\dagger \hat{a} + \alpha(\hat{a} + \hat{a}^\dagger) + \alpha^2. \quad (98)$$

The derivation from here on follows that by Carmichael [10], leading to a Master equation for the resonator in the lab frame

$$\dot{\rho}_r = -i[\hat{H}_r + \Delta_p \hat{A}, \rho_r] + \kappa_a \mathcal{D}[\hat{a}]\rho_r + \Gamma_{\text{th}}\mathcal{D}[\hat{A}]\rho_r, \quad (99)$$

where we have added the usual resonator decay term. The influence of the resonator self-Kerr interaction $K$ on the dissipative terms is neglected in this equation. The frequency shift $\delta_p$ and noise rate $\Gamma_{\text{th}}$ are given by [10]

$$\delta_p = \int_0^\infty d\omega \chi(\omega,\omega)\bar{n}(\omega), \quad (100)$$

$$\Gamma_{\text{th}} = 2\pi \int_0^\infty d\omega \chi(\omega,\omega)^2 \Delta n(\omega)^2, \quad (101)$$

where $\Delta n(\omega)^2 = \bar{n}(\omega)[\bar{n}(\omega)+1]$ are fluctuations in the thermal photon number, and summation over metamaterial modes $\omega_k$ has been converted to an integral by introducing $\chi(\omega_k,\omega_l) = \sqrt{\rho(\omega_k)\rho(\omega_l)}\chi_{kl}$ with $\rho(\omega)$ the density of modes. Moreover,

$$\Delta_p = P \int_0^\infty \int_0^\infty d\omega d\omega' \frac{\rho(\omega)\rho(\omega')}{\omega - \omega'}|\chi(\omega,\omega')|^2 \bar{n}(\omega), \quad (102)$$

is a principal value part. To numerically estimate $\Gamma_{\text{th}}$ from a set of normal modes $\omega_k$ found using the approach outlined in Sect. 2.2 we use

$$\Gamma_{\text{th}} \simeq 2\pi \sum_{l \in I_{\text{tml}}} \rho(\omega_l)\chi_{ll}^2 \bar{n}(\omega_l)[\bar{n}(\omega_l)+1], \quad (103)$$

where the density of modes $\rho(\omega)$ can be estimated using Eq. (42). Numerical estimates of $\Gamma_{\text{th}}$ for the three different parameter sets are shown in Fig. S14, for temperatures in the range $T = 10$–$40$ mK. These results only include modes in the metamaterial band $I_{\text{tml}}$. The coupler modes also couple dispersively to the resonator, but the Markovian treatment used here is not valid for these higher quality modes. To ensure that the coupler modes have low thermal population it is useful to design high coupler frequencies $\omega_s$.

With a detection time $t_{\text{det}}$ of up to tens of $\mu$s (see Sect. 3.5), we should keep $\Gamma_{\text{th}}$ under $10^4$ to ensure $\Gamma_{\text{th}} t_{\text{det}} < 0.1$. As is clear from Fig. S14, parameter sets V and VI give lower thermal noise rates. This is due to higher center frequencies for the CRLH band, and smaller detector bandwidth in the case of set VI, as given in table S2.

## 3.5 Detection time and probe decay rate

Sine the dispersion relation of the CRLH metamaterial is only weakly modified by the coupling to the resonator, we can use that the time-of-flight through the detector is to a good approximation

$$\tau \simeq \frac{2z}{v_g} = 2N\left|\frac{d(ka)}{d\omega}\right|, \quad (104)$$

with $ka$ found from Eq. (1), and we are again assuming that the detector is operated in reflection mode. For the parameters studied in the previous sections and with $N = 2000$ we find $\tau$ to be in the range $1$–$10\,\mu$s, c.f. Fig. S2.

The numerical results of the main paper indicates a detection time of a few $\tau$, i.e., in the $\mu$s range. A typical dimensionless photon width of $\gamma\tau = 2$–$6$ as used in the numerical simulations in the main paper thus corresponds to $\gamma/(2\pi)$ under a MHz. We emphasize that the detection

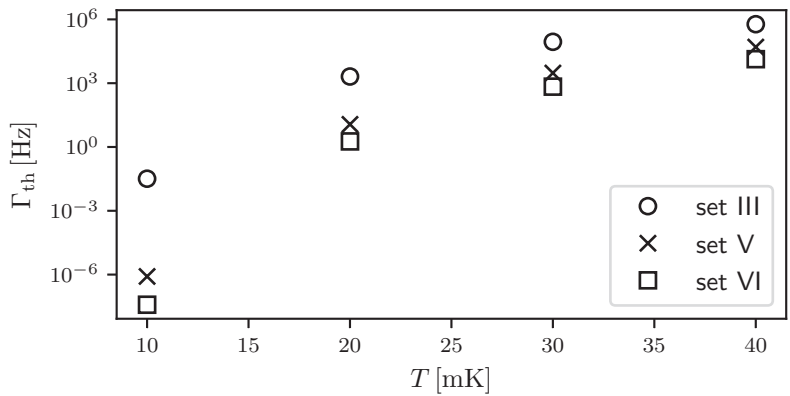

Figure S14: Estimates of thermal noise rate $\Gamma_{\text{th}}$ Eq. (103) for parameter set III from table S1 and sets V and VI from table S2.

fidelity improves with *increasing* $\gamma$. It is therefore encouraging that very high detection fidelities are found for such narrow photons.

In the numerical simulations presented in the main paper we used a resonator decay rate such that $\kappa_a \tau = 1.0$, which for $\tau$ in the range 1–10 $\mu$s corresponds to $\kappa_a/(2\pi)$ in the range 0.015–0.15 MHz. With such a small value for $\kappa_a$ it is necessary to reduce the total self-Kerr nonlinearity $K_{\text{tot}} = K + K_J \simeq 0$ of the resonator, e.g. using the approach discussed in Sect. 3.2.1. It might be beneficial to increase $\kappa_a$ to the range $\kappa_a/(2\pi) = 1$–10 MHz, thus correspondingly increasing $\kappa_a \tau$. Unfortunately we found it numerically too challenging to simulate this regime of larger $\kappa_a$, due to the correspondingly smaller time steps needed. Based on a simplified model, we expect the SNR to increase with $\kappa_a$ at short times, and go down as $1/\sqrt{\kappa_a}$ for large $\kappa_a \tau$ [11]. An order of magnitude increase in $\kappa_a$ might therefore lead to a decrease of about a factor of three in the SNR, which is comparable to reducing the JTWPD coupling strength $g$ by the same amount.

## 4 Keldysh path-integral treatment

To study the scaling of the detector back-action, we integrate out the waveguide part of the system using Keldysh field theory. This allows us to derive an effective Keldysh action describing the evolution of the measurement resonator when a photon travels through the nonlinear waveguide. As we show, this effective Keldysh action indicates that

1. The relevant small parameter for the measurement back-action is $g/\sigma$, with $\sigma^2$ the variance of an incoming Gaussian photon.

2. Back-scattering of the signal photons due to measurement back-action are suppressed when the local coupling strength is small compared to the photon carrier frequency, $g/\bar{\omega} \ll 1$.

Moreover, from the effective action, we derive an equivalent master equation that can be used to perform numerical simulations.

### 4.1 Keldysh action for the emitter-waveguide-probe system

Recall the Hamiltonian for the interacting system in the rotating frame for the resonator

$$\begin{aligned}
\hat{H} &= \hat{H}_0 + \hat{H}_{\text{int}}, \\
\hat{H}_0 &= \sum_\nu \int_\Omega d\omega\, \hbar\omega \hat{b}^\dagger_{\nu,\omega} \hat{b}_{\nu,\omega}, \\
\hat{H}_{\text{int}} &= \hbar g \sum_{\nu,\mu} \int_{-z/2}^{z/2} dx\, \hat{b}^\dagger_\nu(x) \hat{b}_\mu(x)(\hat{a}^\dagger + \hat{a}),
\end{aligned} \qquad (105)$$

24

where
$$\hat{b}_\nu(x) = \sqrt{\frac{\bar{\omega}}{2\pi v}} \int_\Omega \frac{d\omega}{\sqrt{\omega}} \hat{b}_{\nu\omega} e^{\nu i \omega x / v}. \tag{106}$$

In order to simplify the calculation and focus solely on the part of the interaction that induces a displacement in the measurement resonator, we have here set $K = \chi(x) = 0$ and $g(x) = g$ in the Hamiltonian above, as discussed in the main text. In this section, we will use the convention $\hbar = 1$ for brevity and, to avoid ambiguousness with the $\pm$ notation from the upper and lower branches of the Keldysh contour, use the $\nu = R, L$ subscripts for right- and left-moving fields, respectively.

Instead of directly considering a signal photon on the waveguide, we add the fictitious emitter of such a photon far away from the nonlinear waveguide at the initial position $x_0 \ll -z/2$ and consider that this emitter is only coupled to the right-moving modes. This is modeled by a combined emitter-waveguide-resonator Hamiltonian [12] in the rotating wave approximation

$$\begin{aligned}\hat{H} &= \hat{H}_0 + \hat{H}_\text{int} + \hat{H}_c, \\ \hat{H}_c &= \omega_c \hat{c}^\dagger \hat{c} + \sqrt{\kappa_c(t) v} \int_{-\infty}^{\infty} dx\, \delta(x - x_0) \left[ \hat{b}_R^\dagger(x) \hat{c} + \hat{c}^\dagger \hat{b}_R(x) \right], \end{aligned} \tag{107}$$

where $\kappa_c(t)$ is the coupling of the emitter to the waveguide and $\omega_c$ is the frequency of the emitter. In principle, any photon shape can be modeled this way and we follow the main text by considering a photon of carrier frequency $\omega_c = \bar{\omega}$ with a Gaussian envelope of linewidth $\sigma$,

$$\xi(t) = \left(\frac{2\sigma^2}{\pi}\right)^{1/4} e^{-i\omega_c t} e^{-\sigma^2 (t + x_0/v)^2}, \tag{108}$$

with $x_0$ the initial position of the signal photon. Here, we simulate such a photon wavepacket by initializing the emitter in the $|1\rangle$ photon Fock state and choosing [13]

$$\kappa_c(t) = \sqrt{\frac{8\sigma^2}{\pi}} \frac{e^{-2\sigma^2 t^2}}{\text{erfc}(\sqrt{2}\sigma t)}. \tag{109}$$

Note that Eq. (108) corresponds to the amplitude of the photon wavepacket as a function of time at the fixed waveguide position $x = 0$. Since we consider a waveguide without dispersion, the amplitude of the wavepacket as a function of both time and position is easily calculated, $\xi(x,t) = \xi(t - x/v)$.

Following Ref. [14], we write the Keldysh action for the combined emitter-waveguide-resonator system

$$\begin{aligned}S[a,b,c] = &\int dt\, a_+^* \partial_t a_+ - a_-^* \partial_t a_- + c_+^* \partial_t c_+ - c_-^* \partial_t c_- \\ &+ \int dt dx\, b_+(x)^* \partial_t b_+(x) - b_-(x)^* \partial_t b_-(x) - i\mathcal{L}(a,b,c), \end{aligned} \tag{110}$$

$$\mathcal{L}(a,b,c) = -i(H_+ - H_-) + \kappa_a \left[ a_+ a_-^* - \frac{1}{2}\left(a_+^* a_+ + a_-^* a_-\right) \right]$$

where the arguments of $S$ and $\mathcal{L}$ have been shortened to $\psi \leftrightarrow \psi_+^*, \psi_+, \psi_-^*, \psi_-$, with $\psi = a, b, c$. The Hamiltonian part $H_\pm$ is found by writing Eq. (107) in a normal-ordered form and replacing each operator by the corresponding variable $\hat{\psi} \to \psi_\pm$. Unless otherwise noted, we will keep the time dependence of variables implicit to make the notation more compact. Performing the Keldysh rotation $\psi_{cl} = (\psi_+ + \psi_-)/\sqrt{2}$ and $\psi_q = (\psi_+ - \psi_-)/\sqrt{2}$, we write the action as

$$S[a,b,c] = \int dt\, \mathbf{a}^\dagger G_a^{-1} \mathbf{a} + \mathbf{c}^\dagger G_c^{-1} \mathbf{c} + \int dt dx\, \mathbf{b}^\dagger \left[G_b^{-1} - V\right]\mathbf{b} + \mathbf{b}^\dagger \mathbf{j} + \mathbf{j}^\dagger \mathbf{b}, \tag{111}$$

where the dependence on $x$ was made implicit in the second part to make the notation more

compact and we defined the vectors

$$\mathbf{a} = \begin{pmatrix} a_{cl} \\ a_q \end{pmatrix},$$

$$\mathbf{c} = \begin{pmatrix} c_{cl} \\ c_q \end{pmatrix},$$

$$\mathbf{b} = \begin{pmatrix} b_{R,cl}(x) \\ b_{R,q}(x) \\ b_{L,cl}(x) \\ b_{L,q}(x) \end{pmatrix}, \quad (112)$$

$$\mathbf{j} = -\delta(x-x_0)\sqrt{\kappa_c(t)}v \begin{pmatrix} c_q \\ c_{cl} \\ 0 \\ 0 \end{pmatrix}.$$

We also defined

$$\begin{aligned} G_a^{-1} &= \begin{pmatrix} 0 & i\partial_t - i\kappa_a/2 \\ i\partial_t + i\kappa_a/2 & i\kappa_a \end{pmatrix}, \\ G_b^{-1} &= \begin{pmatrix} G_{b,R}^{-1} & 0_{2\times 2} \\ 0_{2\times 2} & G_{b,L}^{-1} \end{pmatrix}, \\ \tilde{G}_{b,\nu}^{-1}(k) &= \begin{pmatrix} 0 & i\partial_t - \nu v k - i\eta \\ i\partial_t - \nu v k + i\eta & i\eta \end{pmatrix}, \\ G_c^{-1} &= \begin{pmatrix} 0 & i\partial_t - \omega_c - i\eta \\ i\partial_t - \omega_c + i\eta & i\eta \end{pmatrix}, \\ V &= \begin{pmatrix} W & W \\ W & W \end{pmatrix}, \\ W &= g\,\theta(x+z/2)\theta(z/2-x)\begin{pmatrix} X_q & X_{cl} \\ X_{cl} & X_q \end{pmatrix}, \\ X_{cl/q} &\equiv (a_{cl/q} + a_{cl/q}^*)/\sqrt{2}, \end{aligned} \quad (113)$$

where $\eta$ is there for regularization and the limit $\eta \to 0$ is implicit. In Eq. (111), we expressed the waveguide's field inverse Green's function in position space instead of momentum as the interaction term V is local in space. However, it is easier to solve for the free field Green's function in momentum, so we will perform the Fourier transform back to position space after finding $\tilde{G}_{b,\nu}(k)$.

### 4.2 Tracing out the waveguide

The partition function associated with the action Eq. (111) is given by

$$Z = \int \mathcal{D}[a,b,c] e^{iS[a,b,c]}. \quad (114)$$

It is useful to point out two properties of the Keldysh action Eq. (111). First, it is quadratic in the waveguide field $b(x)$ and second, the coupling between the emitter and the waveguide acts as a source term for $b(x)$. Using these properties and the Gaussian integral identity,

$$\int D[z,z^*] e^{i\int_{t,t'} \psi^\dagger(t) G^{-1}(t,t')\psi(t') + \psi^\dagger(t)j(t) + j^\dagger(t)\psi(t)} = \frac{e^{-i\int_{t,t'} j^\dagger(t) G(t,t')j(t')}}{\det G^{-1}}, \quad (115)$$

we can integrate out exactly the waveguide field,

$$Z = \int \mathcal{D}[a,c] e^{iS[a,c] - i\int_{x_1 x_2} \mathbf{j}^\dagger(x_1)[G_b^{-1}-V]^{-1}(x_1,x_2)\mathbf{j}(x_2)}, \quad (116)$$

where the determinant denominator is canceled by a factor in the integration measure $\mathcal{D}[b]$. We also used the shorthand notation $\mathrm{x} \equiv x,t$ for compactness. From the partition function Eq. (116),

it is natural to define an effective action for the emitter-resonator system,

$$S_{\text{eff}}[a,c] = -\int d\mathrm{x}_1 \mathrm{x}_2\, \mathbf{j}^\dagger(\mathrm{x}_1)[\mathrm{G}_b^{-1} - \mathrm{V}]^{-1}(\mathrm{x}_1,\mathrm{x}_2)\mathbf{j}(\mathrm{x}_2) + \int dt\, \mathbf{a}^\dagger \mathrm{G}_a^{-1}\mathbf{a} + \mathbf{c}^\dagger \mathrm{G}_c^{-1}\mathbf{c}. \quad (117)$$

This effective action is not very useful in its present form as (1), it is nonlocal in time and (2), we don't have an exact expression for $[\mathrm{G}_b^{-1} - \mathrm{V}]^{-1}$. To remediate to this situation, we use perturbation theory and expand $(1 - \mathrm{G}_b \mathrm{V})^{-1}$ as a Taylor series in $\mathrm{G}_b \mathrm{V}$, assuming that the linewidth of the photon is large compared to the coupling strength $g$,

$$S_{\text{eff}}[a,c] \approx -\sum_n \int d\mathrm{x}_1 \mathrm{x}_2 \mathrm{x}_3\, \mathbf{j}^\dagger(\mathrm{x}_1)[\mathrm{G}_b \mathrm{V}]^n(\mathrm{x}_1,\mathrm{x}_2)\mathrm{G}_b(\mathrm{x}_2,\mathrm{x}_3)\mathbf{j}(\mathrm{x}_3) + \int dt\, \mathbf{a}^\dagger \mathrm{G}_a^{-1}\mathbf{a} + \mathbf{c}^\dagger \mathrm{G}_c^{-1}\mathbf{c},$$
$$\equiv \sum_n S_{\text{eff}}^{(n)} + \int dt\, \mathbf{a}^\dagger \mathrm{G}_a^{-1}\mathbf{a} + \mathbf{c}^\dagger \mathrm{G}_c^{-1}\mathbf{c}. \quad (118)$$

More precisely, any point in the waveguide will interact for a time $\sim 1/\sigma$ with the photon, and, consequently, the expansion is valid as long as $g/\sigma < 1$. Note that this expansion relies on the fact that the signal photon has a finite width and Eq. (118) is thus not valid for a general incoming signal. Before evaluating the effective action at various orders, we compute the waveguide Green's function in momentum space

$$\tilde{G}_{b,\nu}(k,t_1,t_2) = e^{-i\nu v k(t_1-t_2)} \begin{pmatrix} 1 & \theta(t_1-t_2) \\ -\theta(t_2-t_1) & 0 \end{pmatrix}. \quad (119)$$

Formally, the right-moving (left-moving) fields are defined only on the positive (negative) wavevectors. Here, we will make the assumption that the coupling to the waveguide is small, $\kappa_c \ll \omega_c$ and, consequently, only wavevectors around $k_c = \omega_c/v$ will contribute to the effective action. Note that this assumption was already used earlier to write the Hamiltonian in the rotating wave approximation for the emitter-waveguide coupling, Eq. (107). In practice, this means that for both right- and left-moving fields we extend the wavevector integral to the whole real line. Using the above approximation and Eq. (119), we can write the Green's function in position space

$$\begin{aligned}
\mathrm{G}_{b,\nu}(x_1,x_2,t_1,t_2) &= \frac{-i}{2\pi}\int_0^\infty dk\, e^{i\nu k(x_1-x_2)} e^{-i\nu k(t_1-t_2)} \begin{pmatrix} 1 & \theta(t_1-t_2) \\ -\theta(t_2-t_1) & 0 \end{pmatrix} \\
&\approx \frac{-i}{2\pi}\int_{-\infty}^\infty dk\, e^{i\nu k(x_1-x_2)} e^{-i\nu k(t_1-t_2)} \begin{pmatrix} 1 & \theta(t_1-t_2) \\ -\theta(t_2-t_1) & 0 \end{pmatrix} \\
&= -i\delta[v(t_1-t_2) - \nu(x_1-x_2)] \begin{pmatrix} 1 & \theta(t_1-t_2) \\ -\theta(t_2-t_1) & 0 \end{pmatrix} \\
&\equiv \begin{pmatrix} \mathrm{G}_{b,\nu}^K & \mathrm{G}_{b,\nu}^R \\ \mathrm{G}_{b,\nu}^A & 0 \end{pmatrix}, \\
\mathrm{G}_b &= \begin{pmatrix} \mathrm{G}_{b,R} & 0_{2\times 2} \\ 0_{2\times 2} & \mathrm{G}_{b,L} \end{pmatrix},
\end{aligned} \quad (120)$$

with the useful identity $\mathrm{G}_{b,\nu}^K = \mathrm{G}_{b,\nu}^R - \mathrm{G}_{b,\nu}^A$ [14] and the convention $\theta(0) = 1/2$. Since the Green's function depends only on the space and time difference, we will use the shorter notation $\mathrm{G}_b(x_1,x_2,t_1,t_2) = \mathrm{G}_{b,\nu}(x_1-x_2,t_1-t_2)$.

## 4.3 Zeroth order

Using Eqs. (118) and (120), we compute the zeroth order term of the effective action,

$$
\begin{aligned}
S_{\text{eff}}^{(0)} &= -\int dx_1 dx_2 dt_1 dt_2\, \mathbf{j}^\dagger(x_1,t_1) \mathrm{G}_b(x_1-x_2,t_1-t_2)\mathbf{j}(x_2,t_2), \\
&= -v\int dx_1 dt_1 dx_2 dt_2\, \delta(x_1-x_0)\delta(x_2-x_0)\sqrt{\kappa_c(t_1)\kappa_c(t_2)} \\
&\quad \times \tilde{\mathbf{c}}^\dagger(t_1)\mathrm{G}_{b,R}(x_1-x_2,t_1-t_2)\tilde{\mathbf{c}}(t_2), \\
&= iv\int dt_1 dt_2\, \delta(vt_1-vt_2)\sqrt{\kappa_c(t_1)\kappa_c(t_2)}\times \tilde{\mathbf{c}}^\dagger(t_1)\begin{pmatrix}1 & \theta(t_1-t_2) \\ -\theta(t_2-t_1) & 0\end{pmatrix}\tilde{\mathbf{c}}(t_2), \\
&= i\int dt\, \kappa_c(t)\left[c_q^*(t)c_q(t) - \tfrac{1}{2}c_q^*(t)c_{cl}(t) + \tfrac{1}{2}c_{cl}^*(t)c_q(t)\right]
\end{aligned}
\tag{121}
$$

where we defined $\tilde{\mathbf{c}}^\dagger = (c_q^*\ c_{cl}^*)$ and used the identity $\delta(\alpha t) = \delta(t)/|\alpha|$. $S_{\text{eff}}^{(0)}$ corresponds to the decay of the emitter into the waveguide and we include this term into the bare action for the emitter,

$$
\mathrm{G}_c^{-1} \to \begin{pmatrix} 0 & i\partial_t - \omega_c - i\kappa_c/2 \\ i\partial_t - \omega_c + i\kappa_c/2 & i\kappa_c \end{pmatrix}.
\tag{122}
$$

Note that keeping the full Green's function for the waveguide field instead of extending the wavevector integral in Eq. (120) would have resulted in an additional renormalization of the emitter frequency $\omega_c$.

## 4.4 First order

The first term of the effective action is given by

$$
\begin{aligned}
S_{\text{eff}}^{(1)} &= -\int dx_1 dx_2 dx_3 dt_1 dt_2 dt_3\, \mathbf{j}^\dagger(x_1,t_1)\mathrm{G}_b(x_1-x_2,t_1-t_2)\mathrm{V}(x_2,t_2)\mathrm{G}_b(x_2-x_3,t_2-t_3)\mathbf{j}(x_3,t_3), \\
&= \frac{1}{v}\int dx_2 dt_2\, \kappa_c\!\left(t_2+\frac{x_0-x_2}{v}\right)\tilde{\mathbf{c}}^\dagger\!\left(t_2+\frac{x_0-x_2}{v}\right)\begin{pmatrix}1 & \theta(x_0-x_2) \\ -\theta(x_2-x_0) & 0\end{pmatrix}\mathrm{W}(x_2,t_2) \\
&\quad \times \begin{pmatrix}1 & \theta(x_2-x_0) \\ -\theta(x_0-x_2) & 0\end{pmatrix}\tilde{\mathbf{c}}\!\left(t_2+\frac{x_0-x_2}{v}\right),
\end{aligned}
\tag{123}
$$

where we used the identity $\theta(\alpha t) = \theta(t)$ for $\alpha > 0$. Here, we know that the integral in position is non-zero only when $x_2 > x_0$ since we placed the emitter before the nonlinear waveguide, $x_0 < -z/2$. As a result, we can evaluate the Heaviside functions (in position) in the equation above,

$$
S_{\text{eff}}^{(1)} = -\frac{2g}{v}\int dt \int_{-z/2}^{z/2} dx\, \kappa_c\!\left(t+\frac{x_0-x}{v}\right) \times c_-^*\!\left(t+\frac{x_0-x}{v}\right) X_q(t) c_+\!\left(t+\frac{x_0-x}{v}\right).
\tag{124}
$$

We then use the fact that we know, by construction, the solution for the time evolution of the emitter, $\sqrt{\kappa_c(t)}c(t) = \xi(t-x_0/v)$. Using this, we replace $\sqrt{\kappa_c(t+\Delta t)}c(t+\Delta t) = \sqrt{\kappa_c(t)}c(t)\times \xi(t-x_0/v+\Delta t)/\xi(t-x_0/v)$ in the equation above. We carry out the integration in position and arrive at a first order effective action that is local in time

$$
S_{\text{eff}}^{(1)} = -g\int dt\, c_-^*(t)\, X_q(t) c_+(t)\, \frac{\kappa_c(t)}{\sigma}\sqrt{\frac{\pi}{2}}e^{\tilde{t}^2}\left\{\mathrm{erf}(\tilde{t}+\tilde{t}_0+\tilde{z}/2) - \mathrm{erf}(\tilde{t}+\tilde{t}_0-\tilde{z}/2)\right\},
\tag{125}
$$

where we defined $\tilde{t} \equiv \sqrt{2}\sigma t$, $\tilde{t}_0 \equiv \sqrt{2}\sigma x_0/v$ and $\tilde{z} \equiv \sqrt{2}\sigma z/v$. The above action is equivalent to a term in the master equation [14]

$$
S_{\text{eff}}^{(1)} \to \mathcal{L}^{(1)}(\rho) = -ig\frac{\mathrm{erf}(\tilde{t}+\tilde{t}_0+\tilde{z}/2) - \mathrm{erf}(\tilde{t}+\tilde{t}_0-\tilde{z}/2)}{\mathrm{erfc}(\tilde{t})}\left[(\hat{a}+\hat{a}^\dagger)\hat{c}\rho\hat{c}^\dagger - \hat{c}\rho\hat{c}^\dagger(\hat{a}+\hat{a}^\dagger)\right],
\tag{126}
$$

which corresponds to a drive term on the measurement resonator with an amplitude proportional to the probability of the photon being in the nonlinear waveguide at time $t$. We can write the above equation as a Hamiltonian-like term in the master equation

$$\mathcal{L}^{(1)}(\rho) = -i \left[ g n_{\text{det}}(t)(\hat{a} + \hat{a}^\dagger), \frac{2\hat{c}\rho\hat{c}^\dagger}{\text{erfc}(\tilde{t})} \right],$$

$$n_{\text{det}}(t) \equiv \frac{1}{2} \left[ \text{erf}(\tilde{t}_+) - \text{erf}(\tilde{t}_-) \right],$$
(127)

where we defined $\tilde{t}_\pm \equiv \tilde{t} + \tilde{t}_0 \pm \tilde{z}/2$ to further simplify the notation.

## 4.5 Second order

We now evaluate the second order term of the effective action,

$$S_{\text{eff}}^{(2)} = -\int dx_1 dx_2 dx_3 dx_4 \, \tilde{\mathbf{c}}^\dagger(x_1) G_{b,R}(x_1 - x_2) W(x_2)$$
$$\times [G_{b,R}(x_2 - x_3) + G_{b,L}(x_2 - x_3)] W(x_3) G_{b,R}(x_3 - x_4) \tilde{\mathbf{c}}(x_4).$$
(128)

In this second order term, we see that there is the possibility of backscattering of the photon due to the appearance of the left-moving field Green's function, $G_{b,L}$. In order to make the equations more manageable, we split the effective action in two, $S_{\text{eff}}^{(2)} \equiv S_{\text{eff},R}^{(2)} + S_{\text{eff},L}^{(2)}$, and evaluate the terms one at a time. First, we consider the "forward-scattering" part, $S_{\text{eff},R}^{(2)}$,

$$S_{\text{eff},R}^{(2)} = -\int dx_1 dx_2 dx_3 dx_4 \, \tilde{\mathbf{c}}^\dagger(x_1) G_{b,R}(x_1 - x_2) W(x_2) G_{b,R}(x_2 - x_3) W(x_3) G_{b,R}(x_3 - x_4) \tilde{\mathbf{c}}(x_4),$$

$$= \frac{-2ig^2}{v} \int dx_2 dx_3 \sqrt{\kappa_c\left(t_2 + \frac{x_0 - x_2}{v}\right) \kappa_c\left(t_3 + \frac{x_0 - x_3}{v}\right)} c_-^*\left(t_2 + \frac{x_0 - x_2}{v}\right) c_+\left(t_3 + \frac{x_0 - x_3}{v}\right)$$
$$\times \delta[v(t_2 - t_3) - (x_2 - x_3)] \begin{pmatrix} X_q(t_2) & X_{cl}(t_2) \end{pmatrix} \begin{pmatrix} \theta(t_2 - t_3) + \theta(t_3 - t_2) & \theta(t_2 - t_3) \\ -\theta(t_3 - t_2) & 0 \end{pmatrix} \begin{pmatrix} X_q(t_3) \\ X_{cl}(t_3) \end{pmatrix},$$

$$= \frac{-2ig^2\sqrt{2}}{v} \int dx_2 dx_3 \sqrt{\kappa_c\left(t_2 + \frac{x_0 - x_2}{v}\right) \kappa_c\left(t_3 + \frac{x_0 - x_3}{v}\right)} c_-^*\left(t_2 + \frac{x_0 - x_2}{v}\right) c_+\left(t_3 + \frac{x_0 - x_3}{v}\right)$$
$$\times \delta[v(t_2 - t_3) - (x_2 - x_3)] [X_q(t_2) X_+(t_3)\theta(t_2 - t_3) - X_-(t_2) X_q(t_3)\theta(t_3 - t_2)],$$

$$= \frac{-2ig^2\sqrt{2}}{v^2} \int_{-z/2}^{z/2} dx_2 dx_3$$
$$\times \left\{ \int dt_2 \, \kappa_c\left(t_2 + \frac{x_0 - x_2}{v}\right) c_-^* c_+\left(t_2 + \frac{x_0 + x_2}{v}\right) X_q(t_2) X_+\left(t_2 - \frac{x_2 - x_3}{v}\right) \theta(x_2 - x_3) \right.$$
$$\left. - \int dt_3 \, \kappa_c\left(t_3 + \frac{x_0 - x_3}{v}\right) c_-^* c_+\left(t_3 + \frac{x_0 - x_3}{v}\right) X_-\left(t_3 + \frac{x_2 - x_3}{v}\right) X_q(t_3) \theta(x_3 - x_2) \right\}.$$
(129)

Similar as above, we use $\sqrt{\kappa_c(t + \Delta t)} c(t + \Delta t) = \sqrt{\kappa_c(t)} c(t) \times \xi(t - x_0/v + \Delta t)/\xi(t - x_0/v)$ to replace the time dependance of the emitter variable. Moreover, because of the simple form of Eq. (107), the correlations in the resonator quadrature coupled to the waveguide are given by $X(t)X(t + \Delta t) = X^2(t)e^{-\kappa_a|\Delta t|/2}$. We remark that this simplification is possible only because we assumed $\chi = K = 0$ in the starting point of this derivation, Eq. (105). However, our result remains approximately correct in the limit of small spurious nonlinearities where $\chi, K \ll g, \sigma$. Using these properties and relabeling the integration variables, we find

$$S_{\text{eff},R}^{(2)} = \frac{-4ig^2}{v^2} \int dt \, \kappa_c(t) c_-^* c_+(t) X_q^2(t) \int_{-z/2}^{z/2} dx_2 dx_3 \, e^{-2\sigma^2 \left[ \frac{(x_0 - x_2)^2}{v^2} + 2\frac{(x_0 - x_2)}{v} t \right]} e^{\frac{-\kappa_a(x_2 - x_3)}{2v}} \theta(x_2 - x_3),$$

$$= \frac{-4ig^2}{\kappa_a} \int dt \, c_-^* c_+(t) X_q^2(t) \frac{\kappa_c(t)}{\sigma} \sqrt{\frac{\pi}{2}}$$
$$\times \left\{ e^{\tilde{t}^2} \left[ \text{erf}(\tilde{t}_+) - \text{erf}(\tilde{t}_-) \right] + e^{(\tilde{\kappa}_a - \tilde{t})^2 - \tilde{\kappa}_a(2\tilde{t}_0 + \tilde{z})} \left[ \text{erf}(\tilde{\kappa}_a - \tilde{t}_+) - \text{erf}(\tilde{\kappa}_a - \tilde{t}_-) \right] \right\},$$
(130)

where we defined $\tilde{\kappa}_a \equiv \kappa_a/(4\sqrt{2}\sigma)$. We can write the effective action Eq. (130) as an equivalent term in the master equation

$$S_{\text{eff},R}^{(2)} \to \mathcal{L}^{(2)}(\rho) = \frac{2\Gamma(t)}{\text{erfc}(\tilde{t})} \left[ (\hat{a}+\hat{a}^\dagger)\hat{c}\rho\hat{c}^\dagger(\hat{a}+\hat{a}^\dagger) - \frac{1}{2}\hat{c}\rho\hat{c}^\dagger(\hat{a}+\hat{a}^\dagger)^2 - \frac{1}{2}(\hat{a}+\hat{a}^\dagger)^2\hat{c}\rho\hat{c}^\dagger \right], \quad (131)$$

where

$$\Gamma(t) \equiv \frac{2g^2}{\kappa_a} \left\{ \left[ \text{erf}(\tilde{t}_+) - \text{erf}(\tilde{t}_-) \right] + e^{(\tilde{\kappa}_a - \tilde{t})^2 - \tilde{t}^2 - \tilde{\kappa}_a(2\tilde{t}_0+\tilde{z})} \left[ \text{erf}(\tilde{\kappa}_a - \tilde{t}_+) - \text{erf}(\tilde{\kappa}_a - \tilde{t}_-) \right] \right\}. \quad (132)$$

We can write the above equation as a dissipator-like term

$$\mathcal{L}^{(2)}(\rho) = \Gamma(t)\mathcal{D}[\hat{a}+\hat{a}^\dagger]\frac{2\hat{c}\rho\hat{c}^\dagger}{\text{erfc}(\tilde{t})}. \quad (133)$$

We now turn to the "backscattering" part of the effective second order action, $S_{\text{eff},L}^{(2)}$. Using a similar procedure as above, we obtain

$$\begin{aligned}
S_{\text{eff},L}^{(2)} &= -\int dx_1 dx_2 dx_3 dx_4 \, \tilde{\mathbf{c}}^\dagger(x_1) G_{b,R}(x_1-x_2) W(x_2) G_{b,L}(x_2-x_3) W(x_3) G_{b,R}(x_3-x_4) \tilde{\mathbf{c}}(x_4), \\
&= \frac{ig^2\sqrt{2}}{v} \int dx_2 dx_3 \sqrt{\kappa_c\left(t_2+\frac{x_0-x_2}{v}\right)\kappa_c\left(t_3+\frac{x_0-x_3}{v}\right)} c_-^*\left(t_2+\frac{x_0-x_2}{v}\right) c_+\left(t_3+\frac{x_0-x_3}{v}\right) \\
&\quad \times \delta[v(t_2-t_3)+(x_2-x_3)]\left[X_q(t_2)X_+(t_3)\theta(t_2-t_3) - X_-(t_2)X_q(t_3)\theta(t_3-t_2)\right], \\
&= \frac{-ig^2\sqrt{2}}{v^2} \int_{-z/2}^{z/2} dx_2 dx_3 \left\{ \int dt_3 \sqrt{\kappa_c\left(t_3-\frac{2x_2-x_3-x_0}{v}\right)\kappa_c\left(t_3+\frac{x_0-x_3}{v}\right)} \theta(x_2-x_3) \right. \\
&\quad \times c_-^*\left(t_3-\frac{2x_2-x_3-x_0}{v}\right) c_+\left(t_3+\frac{x_0-x_3}{v}\right) X_q(t_3) X_+\left(t_3-\frac{x_2-x_3}{v}\right) \\
&\quad - \int dt_2 \sqrt{\kappa_c\left(t_2+\frac{x_0-x_2}{v}\right)\kappa_c\left(t_2-\frac{2x_3-x_2-x_0}{v}\right)} \theta(x_3-x_2) \\
&\quad \left. \times c_-^*\left(t_2+\frac{x_0-x_2}{v}\right) c_+\left(t_2-\frac{2x_3-x_2-x_0}{v}\right) X_-\left(t_2+\frac{x_2-x_3}{v}\right) X_q(t_2) \right\}, \\
&= \frac{-ig^2\sqrt{2}}{v^2} \int dt \, X_q(t) c_-^*(t) c_+(t) \kappa_c(t) \int_{-z/2}^{z/2} dx_2 dx_3 \, \theta(x_2-x_3) e^{-2\sigma^2\left[\frac{(x_3-x_0)^2}{v^2}+\frac{(2x_2-x_3-x_0)^2}{v^2}-4\frac{(x_2-x_0)}{v}t\right]} \\
&\quad \times e^{-\kappa_a(x_2-x_3)/2v}\left\{ X_+(t)e^{-2i\omega_c(x_2-x_3)/v} - X_-(t)e^{2i\omega_c(x_2-x_3)/v} \right\},
\end{aligned} \quad (134)$$

The two integrals in position can easily be evaluated, but the exact form is lengthy and does not yield much insight, so we will not report it here. However, due to the fast-oscillating integrand, we see that

$$S_{\text{eff},L}^{(2)} \propto \frac{g}{2\omega_c}, \quad (135)$$

and, consequently, we can safely neglect backscattering for small couplings, $g \ll \omega_c$. This is easily understood if we consider that a backscattering event creates a momentum shift in the photon of $2\omega_c/v$, something that must be compensated by the interaction, $g$.

### 4.6 Effective emitter-probe master equation

To summarize, we can write an effective master equation for the emitter-resonator system in a Linblad-like form,

$$\begin{aligned}
\dot{\rho} &= \kappa_a \mathcal{D}[\hat{a}]\rho + \kappa_c(t)\mathcal{D}[\hat{c}]\rho - i\left[g\eta_{\text{det}}(t)(\hat{a}+\hat{a}^\dagger), \frac{\hat{c}\rho\hat{c}^\dagger}{\langle\hat{c}^\dagger\hat{c}\rangle}\right] + \Gamma(t)\mathcal{D}[\hat{a}+\hat{a}^\dagger]\frac{\hat{c}\rho\hat{c}^\dagger}{\langle\hat{c}^\dagger\hat{c}\rangle} \\
&\quad + \mathcal{O}\left(\frac{g^3}{\sigma^3}\right) + \mathcal{O}\left(\frac{g}{2\omega_c}\right).
\end{aligned} \quad (136)$$

Note that small parameter in the expansion is $g/\sigma$ only in the limit $\kappa_a \to 0$. Otherwise, the small parameter in the expansion is a non-trivial combination of $g$, $\kappa_a$ and $\sigma$.

In writing Eq. (136), we have used that, by construction, $\kappa_c(t)\langle \hat{c}^\dagger \hat{c} \rangle(t) = |\xi(t-t_0)|^2$, so that using Eq. (109) we have $\langle \hat{c}^\dagger \hat{c} \rangle(t) = \frac{1}{2}\operatorname{erfc}(\tilde{t})$. We recognize the state $\hat{c}\rho\hat{c}^\dagger/\langle \hat{c}^\dagger \hat{c} \rangle$ as the normalized state of the system, conditioned on the photon having left the emitter. We, moreover, have the following more intuitive forms for the coefficients $n_{\text{det}}$ and $\Gamma(t)$:

$$n_{\text{det}}(t) = \frac{1}{v} \int_{-z/2}^{z/2} dx \left| \xi\left(t - \frac{x}{v}\right) \right|^2, \tag{137}$$

$$\Gamma(t) = \frac{4g^2}{\kappa_a v} \int_{-z/2}^{z/2} dx \left[ 1 - e^{-\frac{\kappa_a}{2v}(x+z/2)} \right] \left| \xi\left(t - \frac{x}{v}\right) \right|^2. \tag{138}$$

In the limit $\kappa_a \to 0$ the coefficient $\Gamma(t)$ reduces to

$$\begin{aligned}
\lim_{\kappa_a \to 0} \Gamma(t) &= \frac{2g^2}{v} \int_{-z/2}^{z/2} \frac{dx}{v} (x + z/2) \left| \xi\left(t - \frac{x}{v}\right) \right|^2, \\
&= \frac{g^2}{\sqrt{2\pi}\sigma} \left\{ e^{-\tilde{t}_+^2} - e^{-\tilde{t}_-^2} + \pi \tilde{t}_+ \left[\operatorname{erf}(\tilde{t}_+) - \operatorname{erf}(\tilde{t}_-)\right] \right\}.
\end{aligned} \tag{139}$$

# 5 Matrix Product State simulations

To validate the performance of the detector in numerical simulations, we represent the state of the metamaterial as a Matrix Product State (MPS) [15]. Recall that the interacting system is described by a Hamiltonian

$$\hat{H} = \hat{H}_r + \hat{H}_0 + \hat{H}_{\text{int}}, \tag{140}$$

where we are now working in the displaced and rotating frame for the probe resonator, defined in the main text, such that

$$\hat{H}_r = \frac{\hbar K}{2} \left(\hat{a}^\dagger\right)^2 \hat{a}^2, \tag{141}$$

$$\hat{H}_0 = \sum_\nu \int_\Omega d\omega \hbar\omega \hat{b}_{\nu\omega}^\dagger \hat{b}_{\nu\omega}, \tag{142}$$

and

$$\begin{aligned}
\hat{H}_{\text{int}} = &\hbar \sum_{\nu\mu} \int_{-z/2}^{z/2} dx \chi(x) \hat{b}_\nu^\dagger(x) \hat{b}_\mu(x) \left(\hat{a}^\dagger \hat{a} + \alpha^2\right) \\
&+ \hbar \sum_{\nu\mu} \int_{-z/2}^{z/2} dx g(x) \hat{b}_\nu^\dagger(x) \hat{b}_\mu(x) \left(\hat{a}^\dagger + \hat{a}\right),
\end{aligned} \tag{143}$$

where

$$\hat{b}_\nu(x) = \sqrt{\frac{\bar{\omega}}{2\pi v}} \int_\Omega \frac{d\omega}{\sqrt{\omega}} \hat{b}_{\nu\omega} e^{\nu i\omega x/v}. \tag{144}$$

In the following we present an approach based on discretizing both space and time, to efficiently represent the state of the system as an MPS. A main additional simplification that is made in this section is that we only treat the right-moving field $\nu = +$ for the waveguide. This simplification is justified based on the Keldysh analysis which shows that backscattering is suppressed when $g/\bar{\omega} \ll 1$.

Returning to Eq. (143), we note that in a rotating frame defined with respect to $\hat{H}_0$ the interaction becomes

$$\hat{H}_{\text{int}}(t) = \hbar \int_0^z dx \hat{b}_+^\dagger(x - vt) \hat{b}_+(x - vt) \left[\chi(x)\left(\hat{a}^\dagger \hat{a} + \alpha^2\right) + g(x)\left(\hat{a}^\dagger + \hat{a}\right)\right], \tag{145}$$

where we now only treat the right-moving $\nu = +$ field and we shift the $x$-axis by $z/2$ in this section for later notational convenience. Following [16, 17], we trotterize the time-evolution

$$U(T) = \mathcal{T}e^{-\frac{i}{\hbar}\int_0^T dt \hat{H}(t)} = \lim_{N_t \to \infty} \hat{U}_{N_t-1} \ldots \hat{U}_1 \hat{U}_0, \tag{146}$$

where

$$\hat{U}_i = e^{-\frac{i}{\hbar}\int_{t_i}^{t_i+\Delta t} dt \hat{H}(t)}, \quad (147)$$

for $i = 0, \ldots, N_t - 1$, $\Delta t = T/N_t$ and $t_i = i\Delta t$. We can, moreover, do a similar discretization of the spatial integral for each $\hat{U}_i$

$$\hat{U}_i = \lim_{N_x \to \infty} \hat{U}_{i,N_x-1} \ldots \hat{U}_{i,1} \hat{U}_{i,0}, \quad (148)$$

where

$$\hat{U}_{i,n} = e^{-\frac{i}{\hbar}\int_{t_i}^{t_i+\Delta t} dt \int_{x_n}^{x_n+\Delta x} dx \hat{\mathcal{H}}_{\text{int}}(x,t) - \frac{i}{\hbar}\Delta t \hat{H}_r/N_x}, \quad (149)$$

where $n = 0, \ldots N_x - 1$, we choose $\Delta x = z/N_x = v\Delta t$, and $x_n = n\Delta x$ and the Hamiltonian density $\hat{\mathcal{H}}_{\text{int}}(x,t)$ is

$$\hat{\mathcal{H}}_{\text{int}}(x,t) = \hat{b}_+^\dagger(x - vt)\hat{b}_+(x - vt)\hat{A}(x), \quad (150)$$

where we have defined $\hat{A}(x) = \chi(x)\left(\hat{a}^\dagger \hat{a} + \alpha^2\right) + g(x)\left(\hat{a}^\dagger + \hat{a}\right)$ for notational convenience. For sufficiently small $\Delta t$ and $\Delta x$, we now make the approximations

$$\int_{t_i}^{t_i+\Delta t} dt \int_{x_n}^{x_n+\Delta x} dx \hat{b}_+^\dagger(x - vt)\hat{b}_+(x - vt)\hat{A}(x)$$
$$\simeq \int_{t_i}^{t_i+\Delta t} dt \hat{b}_+^\dagger(x_n - vt) \int_{x_n}^{x_n+\Delta x} dx \hat{b}_+(x - vt_i)\hat{A}(x_n) \quad (151)$$
$$= -\Delta t \hat{b}_{n-i} \hat{b}_{n-i} \hat{A}(x_n).$$

where in the last line we have defined

$$\hat{b}_n = \frac{1}{\sqrt{\Delta x}} \int_{x_n}^{x_n+\Delta x} dx \hat{b}_+(x). \quad (152)$$

When the incoming photon has a narrow width concentrated around $\omega = \bar{\omega}$ we can approximate

$$\hat{b}_+(x) = \sqrt{\frac{1}{2\pi v}} \int_{-\infty}^{\infty} d\omega \hat{b}_{\nu\omega} e^{i\omega x/v}, \quad (153)$$

such that $[\hat{b}_+(x), \hat{b}_+(y)^\dagger] = \delta(x - y)$, and consequently $[\hat{b}_n, \hat{b}_m^\dagger] = \delta_{mn}$.

The interaction Eq. (151) suggests the following picture, illustrated in Fig. S15: At time $t = 0$, the probe resonator (denoted $\hat{a}$ in the figure) interacts with a subset of oscillators $\hat{b}_n$ with $n = 0, \ldots, N_x - 1$, for a short time $\Delta t$. In the next time step, the interaction is instead with oscillators labeled $n = -1, \ldots, N_x - 2$, which we can think of as resulting from shifting all the waveguide oscillators one unit cell to the right. The waveguide thus appears as a "conveyor belt" of oscillators moving past the probe, where for each time step the probe interacts with the subset $-i, \ldots, N_x - 1 - i$. We now furthermore have a natural representation of the system as an MPS. Each oscillator $b_n$ represents a site in the MPS, and the probe resonator is a special site which needs to be swapped along the MPS interacting with sites $n = -i, \ldots, N_x - 1 - i$ for each time step $i$. To simulate an incoming photon we also include an emitter at a site $l = l_0 - i - 1$ with $l_0 \leq 0$ (i.e., the emitter is to the left of the detector), interacting with site $l_0 - i$ through an interaction

$$\hat{U}_{\text{emitter}} = e^{\sqrt{\Delta t}\sqrt{\kappa_c(t_i)}(\hat{c}\hat{b}_{l_0-i}^\dagger - \text{H.c.})}, \quad (154)$$

where $\hat{c}$ is the annihilation operator for the emitter, initialized in Fock state $|1\rangle$, and $\kappa_c(t)$ the decay rate determining the shape of the incoming photon (see Refs. [16, 17] for a derivation). Note that the emitter is swapped one site to the left in the MPS for each time step.

To model a non-zero measurement rate $\kappa_a > 0$, we must include the interaction to a bath in the resonator Hamiltonian $\hat{H}_r$. The total time-evolution for the resonator for a small time step $\Delta t$ can be written

$$\hat{U}_{i,n} \simeq \hat{U}_{i,n}^{\text{int}}(\Delta t/2) \hat{U}_r(\Delta t) \hat{U}_{i,n}^{\text{int}}(\Delta t/2), \quad (155)$$



%% file: photodet.bbl
\begin{thebibliography}{10}
\expandafter\ifx\csname url\endcsname\relax
  \def\url#1{\texttt{#1}}\fi
\expandafter\ifx\csname urlprefix\endcsname\relax\def\urlprefix{URL }\fi
\providecommand{\bibinfo}[2]{#2}
\providecommand{\eprint}[2][]{\url{#2}}

\bibitem{Helmer09}
\bibinfo{author}{Helmer, F.}, \bibinfo{author}{Mariantoni, M.},
  \bibinfo{author}{Solano, E.} \& \bibinfo{author}{Marquardt, F.}
\newblock \bibinfo{title}{Quantum nondemolition photon detection in circuit qed
  and the quantum zeno effect}.
\newblock \emph{\bibinfo{journal}{Phys. Rev. A}} \textbf{\bibinfo{volume}{79}},
  \bibinfo{pages}{052115} (\bibinfo{year}{2009}).

\bibitem{Chen11}
\bibinfo{author}{Chen, Y.-F.} \emph{et~al.}
\newblock \bibinfo{title}{Microwave photon counter based on josephson
  junctions}.
\newblock \emph{\bibinfo{journal}{Phys. Rev. Lett.}}
  \textbf{\bibinfo{volume}{107}}, \bibinfo{pages}{217401}
  (\bibinfo{year}{2011}).

\bibitem{Sathyamoorthy14}
\bibinfo{author}{Sathyamoorthy, S.~R.} \emph{et~al.}
\newblock \bibinfo{title}{Quantum nondemolition detection of a propagating
  microwave photon}.
\newblock \emph{\bibinfo{journal}{Phys. Rev. Lett.}}
  \textbf{\bibinfo{volume}{112}}, \bibinfo{pages}{093601}
  (\bibinfo{year}{2014}).

\bibitem{Fan14}
\bibinfo{author}{Fan, B.}, \bibinfo{author}{Johansson, G.},
  \bibinfo{author}{Combes, J.}, \bibinfo{author}{Milburn, G.} \&
  \bibinfo{author}{Stace, T.~M.}
\newblock \bibinfo{title}{Nonabsorbing high-efficiency counter for itinerant
  microwave photons}.
\newblock \emph{\bibinfo{journal}{Phys. Rev. B}} \textbf{\bibinfo{volume}{90}},
  \bibinfo{pages}{035132} (\bibinfo{year}{2014}).

\bibitem{Koshino16}
\bibinfo{author}{Koshino, K.}, \bibinfo{author}{Lin, Z.},
  \bibinfo{author}{Inomata, K.}, \bibinfo{author}{Yamamoto, T.} \&
  \bibinfo{author}{Nakamura, Y.}
\newblock \bibinfo{title}{Dressed-state engineering for continuous detection of
  itinerant microwave photons}.
\newblock \emph{\bibinfo{journal}{Phys. Rev. A}} \textbf{\bibinfo{volume}{93}},
  \bibinfo{pages}{023824} (\bibinfo{year}{2016}).

\bibitem{Inomata16}
\bibinfo{author}{Inomata, K.} \emph{et~al.}
\newblock \bibinfo{title}{Single microwave-photon detector using an artificial
  $\lambda$-type three-level system}.
\newblock \emph{\bibinfo{journal}{Nature Comm.}} \textbf{\bibinfo{volume}{7}},
  \bibinfo{pages}{12303} (\bibinfo{year}{2016}).

\bibitem{Narla16}
\bibinfo{author}{Narla, A.} \emph{et~al.}
\newblock \bibinfo{title}{Robust concurrent remote entanglement between two
  superconducting qubits}.
\newblock \emph{\bibinfo{journal}{Phys. Rev. X}} \textbf{\bibinfo{volume}{6}},
  \bibinfo{pages}{031036} (\bibinfo{year}{2016}).

\bibitem{Kyriienko16}
\bibinfo{author}{Kyriienko, O.} \& \bibinfo{author}{S{\o}rensen, A.~S.}
\newblock \bibinfo{title}{Continuous-wave single-photon transistor based on a
  superconducting circuit}.
\newblock \emph{\bibinfo{journal}{Phys. Rev. Lett.}}
  \textbf{\bibinfo{volume}{117}}, \bibinfo{pages}{140503}
  (\bibinfo{year}{2016}).

\bibitem{leppakangas2018multiplying}
\bibinfo{author}{Lepp{\"a}kangas, J.} \emph{et~al.}
\newblock \bibinfo{title}{Multiplying and detecting propagating microwave
  photons using inelastic cooper-pair tunneling}.
\newblock \emph{\bibinfo{journal}{Phys. Rev. A}} \textbf{\bibinfo{volume}{97}},
  \bibinfo{pages}{013855} (\bibinfo{year}{2018}).

\bibitem{Kono:18a}
\bibinfo{author}{Kono, S.}, \bibinfo{author}{Koshino, K.},
  \bibinfo{author}{Tabuchi, Y.}, \bibinfo{author}{Noguchi, A.} \&
  \bibinfo{author}{Nakamura, Y.}
\newblock \bibinfo{title}{Quantum non-demolition detection of an itinerant
  microwave photon}.
\newblock \emph{\bibinfo{journal}{Nature Physics}}
  \textbf{\bibinfo{volume}{14}}, \bibinfo{pages}{546--549}
  (\bibinfo{year}{2018}).

\bibitem{Besse:18a}
\bibinfo{author}{Besse, J.-C.} \emph{et~al.}
\newblock \bibinfo{title}{Single-shot quantum nondemolition detection of
  individual itinerant microwave photons}.
\newblock \emph{\bibinfo{journal}{Phys. Rev. X}} \textbf{\bibinfo{volume}{8}},
  \bibinfo{pages}{021003} (\bibinfo{year}{2018}).

\bibitem{Royer18}
\bibinfo{author}{Royer, B.}, \bibinfo{author}{Grimsmo, A.~L.},
  \bibinfo{author}{Choquette-Poitevin, A.} \& \bibinfo{author}{Blais, A.}
\newblock \bibinfo{title}{Itinerant microwave photon detector}.
\newblock \emph{\bibinfo{journal}{Phys. Rev. Lett.}}
  \textbf{\bibinfo{volume}{120}}, \bibinfo{pages}{203602}
  (\bibinfo{year}{2018}).

\bibitem{lescanne2019detecting}
\bibinfo{author}{Lescanne, R.} \emph{et~al.}
\newblock \bibinfo{title}{Detecting itinerant microwave photons with engineered
  non-linear dissipation}.
\newblock \emph{\bibinfo{journal}{arXiv:1902.05102}}  (\bibinfo{year}{2019}).

\bibitem{blais2020quantum}
\bibinfo{author}{Blais, A.}, \bibinfo{author}{Girvin, S.~M.} \&
  \bibinfo{author}{Oliver, W.~D.}
\newblock \bibinfo{title}{Quantum information processing and quantum optics
  with circuit quantum electrodynamics}.
\newblock \emph{\bibinfo{journal}{Nature Physics}} \bibinfo{pages}{1--10}
  (\bibinfo{year}{2020}).

\bibitem{Kubo10}
\bibinfo{author}{Kubo, Y.} \emph{et~al.}
\newblock \bibinfo{title}{Strong coupling of a spin ensemble to a
  superconducting resonator}.
\newblock \emph{\bibinfo{journal}{Phys. Rev. Lett.}}
  \textbf{\bibinfo{volume}{105}}, \bibinfo{pages}{140502}
  (\bibinfo{year}{2010}).

\bibitem{Burkard2020}
\bibinfo{author}{Burkard, G.}, \bibinfo{author}{Gullans, M.~J.},
  \bibinfo{author}{Mi, X.} \& \bibinfo{author}{Petta, J.~R.}
\newblock \bibinfo{title}{Superconductor--semiconductor hybrid-circuit quantum
  electrodynamics}.
\newblock \emph{\bibinfo{journal}{Nature Reviews Physics}}
  \textbf{\bibinfo{volume}{2}}, \bibinfo{pages}{129--140}
  (\bibinfo{year}{2020}).

\bibitem{Spring13}
\bibinfo{author}{Spring, J.~B.} \emph{et~al.}
\newblock \bibinfo{title}{Boson sampling on a photonic chip}.
\newblock \emph{\bibinfo{journal}{Science}} \textbf{\bibinfo{volume}{339}},
  \bibinfo{pages}{798--801} (\bibinfo{year}{2013}).

\bibitem{Nickerson14}
\bibinfo{author}{Nickerson, N.~H.}, \bibinfo{author}{Fitzsimons, J.~F.} \&
  \bibinfo{author}{Benjamin, S.~C.}
\newblock \bibinfo{title}{Freely scalable quantum technologies using cells of
  5-to-50 qubits with very lossy and noisy photonic links}.
\newblock \emph{\bibinfo{journal}{Phys. Rev. X}} \textbf{\bibinfo{volume}{4}},
  \bibinfo{pages}{041041} (\bibinfo{year}{2014}).

\bibitem{Lloyd08}
\bibinfo{author}{Lloyd, S.}
\newblock \bibinfo{title}{Enhanced sensitivity of photodetection via quantum
  illumination}.
\newblock \emph{\bibinfo{journal}{Science}} \textbf{\bibinfo{volume}{321}},
  \bibinfo{pages}{1463--1465} (\bibinfo{year}{2008}).

\bibitem{Lamoreaux:13}
\bibinfo{author}{Lamoreaux, S.~K.}, \bibinfo{author}{van Bibber, K.~A.},
  \bibinfo{author}{Lehnert, K.~W.} \& \bibinfo{author}{Carosi, G.}
\newblock \bibinfo{title}{Analysis of single-photon and linear amplifier
  detectors for microwave cavity dark matter axion searches}.
\newblock \emph{\bibinfo{journal}{Phys. Rev. D}} \textbf{\bibinfo{volume}{88}},
  \bibinfo{pages}{035020} (\bibinfo{year}{2013}).

\bibitem{Hadfield09}
\bibinfo{author}{Hadfield, R.~H.}
\newblock \bibinfo{title}{Single-photon detectors for optical quantum
  information applications}.
\newblock \emph{\bibinfo{journal}{Nature Photonics}}
  \textbf{\bibinfo{volume}{3}}, \bibinfo{pages}{696--705}
  (\bibinfo{year}{2009}).

\bibitem{Macklin15}
\bibinfo{author}{Macklin, C.} \emph{et~al.}
\newblock \bibinfo{title}{A near--quantum-limited josephson traveling-wave
  parametric amplifier}.
\newblock \emph{\bibinfo{journal}{Science}} \textbf{\bibinfo{volume}{350}},
  \bibinfo{pages}{307--310} (\bibinfo{year}{2015}).

\bibitem{Guo17}
\bibinfo{author}{Guo, L.}, \bibinfo{author}{Grimsmo, A.},
  \bibinfo{author}{Kockum, A.~F.}, \bibinfo{author}{Pletyukhov, M.} \&
  \bibinfo{author}{Johansson, G.}
\newblock \bibinfo{title}{Giant acoustic atom: A single quantum system with a
  deterministic time delay}.
\newblock \emph{\bibinfo{journal}{Phys. Rev. A}} \textbf{\bibinfo{volume}{95}},
  \bibinfo{pages}{053821} (\bibinfo{year}{2017}).

\bibitem{Shapiro06a}
\bibinfo{author}{Shapiro, J.~H.}
\newblock \bibinfo{title}{Single-photon kerr nonlinearities do not help quantum
  computation}.
\newblock \emph{\bibinfo{journal}{Phys. Rev. A}} \textbf{\bibinfo{volume}{73}},
  \bibinfo{pages}{062305} (\bibinfo{year}{2006}).

\bibitem{Shapiro07a}
\bibinfo{author}{Shapiro, J.~H.} \& \bibinfo{author}{Razavi, M.}
\newblock \bibinfo{title}{Continuous-time cross-phase modulation and quantum
  computation}.
\newblock \emph{\bibinfo{journal}{New J. Phys.}} \textbf{\bibinfo{volume}{9}},
  \bibinfo{pages}{16--16} (\bibinfo{year}{2007}).

\bibitem{Gea-Banacloche10a}
\bibinfo{author}{Gea-Banacloche, J.}
\newblock \bibinfo{title}{Impossibility of large phase shifts via the giant
  kerr effect with single-photon wave packets}.
\newblock \emph{\bibinfo{journal}{Phys. Rev. A}} \textbf{\bibinfo{volume}{81}},
  \bibinfo{pages}{043823} (\bibinfo{year}{2010}).

\bibitem{Fan13}
\bibinfo{author}{Fan, B.} \emph{et~al.}
\newblock \bibinfo{title}{Breakdown of the cross-kerr scheme for photon
  counting}.
\newblock \emph{\bibinfo{journal}{Phys. Rev. Lett.}}
  \textbf{\bibinfo{volume}{110}}, \bibinfo{pages}{053601}
  (\bibinfo{year}{2013}).

\bibitem{Caloz2004novel}
\bibinfo{author}{Caloz, C.}, \bibinfo{author}{Sanada, A.} \&
  \bibinfo{author}{Itoh, T.}
\newblock \bibinfo{title}{A novel composite right-/left-handed coupled-line
  directional coupler with arbitrary coupling level and broad bandwidth}.
\newblock \emph{\bibinfo{journal}{IEEE Trans. Microw. Theory. Tech.}}
  \textbf{\bibinfo{volume}{52}}, \bibinfo{pages}{980--992}
  (\bibinfo{year}{2004}).

\bibitem{Didier15}
\bibinfo{author}{Didier, N.}, \bibinfo{author}{Bourassa, J.} \&
  \bibinfo{author}{Blais, A.}
\newblock \bibinfo{title}{Fast quantum nondemolition readout by parametric
  modulation of longitudinal qubit-oscillator interaction}.
\newblock \emph{\bibinfo{journal}{Phys. Rev. Lett.}}
  \textbf{\bibinfo{volume}{115}}, \bibinfo{pages}{203601}
  (\bibinfo{year}{2015}).

\bibitem{Grimsmo15}
\bibinfo{author}{Grimsmo, A.~L.}
\newblock \bibinfo{title}{Time-delayed quantum feedback control}.
\newblock \emph{\bibinfo{journal}{Phys. Rev. Lett.}}
  \textbf{\bibinfo{volume}{115}}, \bibinfo{pages}{060402}
  (\bibinfo{year}{2015}).

\bibitem{Pichler16}
\bibinfo{author}{Pichler, H.} \& \bibinfo{author}{Zoller, P.}
\newblock \bibinfo{title}{Photonic circuits with time delays and quantum
  feedback}.
\newblock \emph{\bibinfo{journal}{Phys. Rev. Lett.}}
  \textbf{\bibinfo{volume}{116}}, \bibinfo{pages}{093601}
  (\bibinfo{year}{2016}).

\bibitem{Wiseman09}
\bibinfo{author}{Wiseman, H.~M.} \& \bibinfo{author}{Milburn, G.~J.}
\newblock \emph{\bibinfo{title}{Quantum measurement and control}}
  (\bibinfo{publisher}{Cambridge university press}, \bibinfo{year}{2009}).

\bibitem{White15}
\bibinfo{author}{White, T.} \emph{et~al.}
\newblock \bibinfo{title}{Traveling wave parametric amplifier with josephson
  junctions using minimal resonator phase matching}.
\newblock \emph{\bibinfo{journal}{Appl. Phys. Lett.}}
  \textbf{\bibinfo{volume}{106}}, \bibinfo{pages}{242601}
  (\bibinfo{year}{2015}).

\bibitem{Planat2020}
\bibinfo{author}{Planat, L.} \emph{et~al.}
\newblock \bibinfo{title}{Photonic-crystal josephson traveling-wave parametric
  amplifier}.
\newblock \emph{\bibinfo{journal}{Phys. Rev. X}} \textbf{\bibinfo{volume}{10}},
  \bibinfo{pages}{021021} (\bibinfo{year}{2020}).

\bibitem{Nigg12}
\bibinfo{author}{Nigg, S.~E.} \emph{et~al.}
\newblock \bibinfo{title}{Black-box superconducting circuit quantization}.
\newblock \emph{\bibinfo{journal}{Phys. Rev. Lett}}
  \textbf{\bibinfo{volume}{108}}, \bibinfo{pages}{240502}
  (\bibinfo{year}{2012}).

\bibitem{Zhang2018}
\bibinfo{author}{Mundada, P.}, \bibinfo{author}{Zhang, G.},
  \bibinfo{author}{Hazard, T.} \& \bibinfo{author}{Houck, A.}
\newblock \bibinfo{title}{Suppression of qubit crosstalk in a tunable coupling
  superconducting circuit}.
\newblock \emph{\bibinfo{journal}{Phys. Rev. Applied}}
  \textbf{\bibinfo{volume}{12}}, \bibinfo{pages}{054023}
  (\bibinfo{year}{2019}).

\bibitem{Frattini17}
\bibinfo{author}{Frattini, N.} \emph{et~al.}
\newblock \bibinfo{title}{3-wave mixing josephson dipole element}.
\newblock \emph{\bibinfo{journal}{Appl. Phys. Lett.}}
  \textbf{\bibinfo{volume}{110}}, \bibinfo{pages}{222603}
  (\bibinfo{year}{2017}).

\bibitem{Ye2020}
\bibinfo{author}{Ye, Y.} \& \bibinfo{author}{et~al.}
\newblock \emph{\bibinfo{journal}{in preparation}}  (\bibinfo{year}{2020}).

\bibitem{Gough:12a}
\bibinfo{author}{Gough, J.~E.}, \bibinfo{author}{James, M.~R.},
  \bibinfo{author}{Nurdin, H.~I.} \& \bibinfo{author}{Combes, J.}
\newblock \bibinfo{title}{Quantum filtering for systems driven by fields in
  single-photon states or superposition of coherent states}.
\newblock \emph{\bibinfo{journal}{Phys. Rev. A}} \textbf{\bibinfo{volume}{86}},
  \bibinfo{pages}{043819} (\bibinfo{year}{2012}).

\bibitem{Sieberer:16a}
\bibinfo{author}{Sieberer, L.~M.}, \bibinfo{author}{Buchhold, M.} \&
  \bibinfo{author}{Diehl, S.}
\newblock \bibinfo{title}{Keldysh field theory for driven open quantum
  systems}.
\newblock \emph{\bibinfo{journal}{Rep. Prog. in Phys.}}
  \textbf{\bibinfo{volume}{79}}, \bibinfo{pages}{096001}
  (\bibinfo{year}{2016}).

\bibitem{Wall16}
\bibinfo{author}{Wall, M.~L.}, \bibinfo{author}{Safavi-Naini, A.} \&
  \bibinfo{author}{Rey, A.~M.}
\newblock \bibinfo{title}{Simulating generic spin-boson models with matrix
  product states}.
\newblock \emph{\bibinfo{journal}{Phys. Rev. A}} \textbf{\bibinfo{volume}{94}},
  \bibinfo{pages}{053637} (\bibinfo{year}{2016}).

\bibitem{Kloeden92}
\bibinfo{author}{Kloeden, P.~E.} \& \bibinfo{author}{Platen, E.}
\newblock \emph{\bibinfo{title}{Numerical Solution of Stochastic Differential
  Equations}} (\bibinfo{publisher}{Springer-Verlag Berlin Heidelberg},
  \bibinfo{year}{1992}).

\bibitem{Grimsmo17}
\bibinfo{author}{Grimsmo, A.~L.} \& \bibinfo{author}{Blais, A.}
\newblock \bibinfo{title}{Squeezing and quantum state engineering with
  josephson travelling wave amplifiers}.
\newblock \emph{\bibinfo{journal}{npj Quantum Information}}
  \textbf{\bibinfo{volume}{3}}, \bibinfo{pages}{20} (\bibinfo{year}{2017}).

\bibitem{wang2018mode}
\bibinfo{author}{Wang, H.} \emph{et~al.}
\newblock \bibinfo{title}{Mode structure in superconducting metamaterial
  transmission-line resonators}.
\newblock \emph{\bibinfo{journal}{Phys. Rev. Appl.}}
  \textbf{\bibinfo{volume}{11}}, \bibinfo{pages}{054062}
  (\bibinfo{year}{2019}).

\bibitem{Vool17}
\bibinfo{author}{Vool, U.} \& \bibinfo{author}{Devoret, M.}
\newblock \bibinfo{title}{Introduction to quantum electromagnetic circuits}.
\newblock \emph{\bibinfo{journal}{Int. J. Circ. Theor. App.}}
  \textbf{\bibinfo{volume}{45}}, \bibinfo{pages}{897--934}
  (\bibinfo{year}{2017}).

\bibitem{Bhat06}
\bibinfo{author}{Bhat, N. A.~R.} \& \bibinfo{author}{Sipe, J.~E.}
\newblock \bibinfo{title}{Hamiltonian treatment of the electromagnetic field in
  dispersive and absorptive structured media}.
\newblock \emph{\bibinfo{journal}{Phys. Rev. A}} \textbf{\bibinfo{volume}{73}},
  \bibinfo{pages}{063808} (\bibinfo{year}{2006}).

\bibitem{Carmichael2013}
\bibinfo{author}{Carmichael, H.~J.}
\newblock \emph{\bibinfo{title}{Statistical methods in quantum optics 1}}
  (\bibinfo{publisher}{Springer-Verlag Berlin Heidelberg},
  \bibinfo{year}{2013}).

\bibitem{Schollwock11}
\bibinfo{author}{Schollw{\"o}ck, U.}
\newblock \bibinfo{title}{The density-matrix renormalization group in the age
  of matrix product states}.
\newblock \emph{\bibinfo{journal}{Ann. Phys.}} \textbf{\bibinfo{volume}{326}},
  \bibinfo{pages}{96--192} (\bibinfo{year}{2011}).

\bibitem{Gardiner:00}
\bibinfo{author}{Gardiner, C.} \& \bibinfo{author}{Zoller, P.}
\newblock \emph{\bibinfo{title}{Quantum Noise: A Handbook of Markovian and
  Non-Markovian Quantum Stochastic Methods with Applications to Quantum
  Optics}}.
\newblock Springer Series in Synergetics (\bibinfo{publisher}{Springer},
  \bibinfo{year}{2000}), \bibinfo{edition}{2nd enlarged ed.} edn.

\end{thebibliography}
